\documentclass[aps,prd,preprint,superscriptaddress,tightenlines,%
nofootinbib]{revtex4}
\newcommand{\PRE}[1]{{#1}}   
\usepackage{amsmath,amssymb,latexsym}
\usepackage{bm}
\usepackage{epsfig}

\usepackage{lineno}

\newcommand{\MET}{\mbox{${\hbox{$E$\kern-0.6em\lower-.1ex\hbox{/}}}_T$}} 

\newcommand{\postscript}[2]{\setlength{\epsfxsize}{#2\hsize}
   \centerline{\epsfbox{#1}}}

\newcommand{\md}{M_D}
\newcommand{\mbh}{M_{\text{BH}}}

\newcommand{\xmin}{x_{\text{min}}}

\newcommand{\tev}{\text{TeV}}
\newcommand{\pb}{\text{pb}}

\newcommand{\km}{\text{km}}

\newcommand{\ns}{\text{ns}}

\newcommand{\kmwe}{\text{kmwe}}

\newcommand{\etal}{{\em et al.}}

\def\sh{\sqrt{\hat s}}

\begin{document}


\rightline{\small{\tt NUB-3245/Th-04}}
\rightline{\small{\tt La Plata-Th-04-03}}
\rightline{\tt hep-ph/0407020}


\title{
\PRE{\vspace*{1.in}}

High Energy Physics in the Atmosphere:\\
 Phenomenology of Cosmic Ray Air Showers

\PRE{\vspace*{0.3in}}
}

\author{Luis Anchordoqui}
\affiliation{
Department of Physics,\\
Northeastern University, Boston, MA 02115, USA
\PRE{\vspace*{.1in}}
}

\author{Mar\'{\i}a Teresa Dova}
\affiliation{
Departamento de F\'{\i}sica,\\ 
Universidad Nacional de La Plata, CC67 La Plata (1900), Argentina
\PRE{\vspace*{.1in}}
}

\author{Analisa Mariazzi}
\affiliation{
Departamento de F\'{\i}sica,\\ 
Universidad Nacional de La Plata, CC67 La Plata (1900), Argentina
\PRE{\vspace*{.1in}}
}

\author{\mbox{Thomas McCauley}}
\affiliation{
Department of Physics,\\
Northeastern University, Boston, MA 02115, USA
\PRE{\vspace*{.1in}}
}

\author{Thomas Paul}
\affiliation{
Department of Physics,\\
Northeastern University, Boston, MA 02115, USA
\PRE{\vspace*{.1in}}
}

\author{Stephen Reucroft}
\affiliation{
Department of Physics,\\
Northeastern University, Boston, MA 02115, USA
\PRE{\vspace*{.1in}}
}

\author{John Swain}
\affiliation{
Department of Physics,\\
Northeastern University, Boston, MA 02115, USA
\PRE{\vspace*{.1in}}
}

\setcounter{footnote}{0}

\date{4 July 2004}


\begin{abstract}

\PRE{\vspace*{.5in}} 

\noindent The properties of cosmic rays with energies above $10^{6}$~GeV
have to be deduced from the spacetime structure and particle content of the air showers which
they initiate. In this review we summarize the phenomenology of these
giant air showers. We describe the  hadronic interaction  
models used to extrapolate results from collider data to ultra high energies, and
discuss the prospects for insights into forward physics at the LHC.  We also 
describe the main electromagnetic processes that govern the longitudinal shower evolution, 
as well as the lateral spread of particles. Armed with these two principal 
shower ingredients and motivation from the underlying physics, we provide an overview of 
some of the different methods proposed to distinguish primary species. The properties of 
neutrino interactions and the potential of forthcoming 
experiments to isolate deeply penetrating showers from baryonic cascades are also discussed. We finally  
venture into a {\it terra incognita} endowed with TeV-scale gravity and explore anomalous 
neutrino-induced showers.
\end{abstract}



\maketitle


\tableofcontents

\newpage
\section{Introduction}  

\subsection{Why another cosmic ray review?}

Some 40 years after the discovery of ultra high energy cosmic rays~\cite{Linsley:prl63}, 
fundamental questions regarding their origin and nature lack definitive answers. 
The highest primary energy measured thus far is $E \sim 10^{11.5}$~GeV~\cite{Bird:1994uy}, 
corresponding to a nucleon-nucleon center-of-mass energy 
$\sqrt{s} \sim  10^{5.9}$~GeV/$\sqrt{A},$ where $A$ is the mass number
of the primary particle. The existence of these particles, the most energetic known in the universe, 
challenges our current understanding of physics. From the perspective of astrophysics, one must 
identify where and how in the 
universe these particles obtain such high energies. A failure to uncover such mechanisms
may lead one to postulate new physics to explain the phenomenon. From the perspective 
of particle physics, ultra high energy cosmic ray interactions are orders of magnitude 
beyond what can be achieved in current (and future) terrestrial collider experiments and may 
open a window to energy and kinematic regions previously unexplored in the study of fundamental 
interactions. From both perspectives, the tantalizing possibility of new physics that may be 
found in the study of ultra high energy cosmic rays continues to motivate current and future 
cosmic ray experiments.

The literature abounds in reviews of experimental techniques for detection
of cosmic ray air showers~\cite{Sokolsky:rz,Yoshida:1998jr,Cronin:fr,Nagano:ve,Bertou:2000ip,Watson:kk,Cronin:2004ye}, as well as overviews of the physics of cosmic ray propagation and possible astrophysical and exotic
origins ~\cite{Hillas:is,Biermann:tr,Bhattacharjee:1998qc,Kuzmin:1999zk,Olinto:2000sa,Anchordoqui:2002hs,Sigl:2002yk,Stecker:2003wm,Torres:2004hk}.  This review follows a somewhat different path,
focusing exclusively on cosmic ray phenomenology from the top of the atmosphere to the
Earth's surface.  The topics covered are viewed from the perspective of particle physics, 
and the reader is assumed only to possess a basic background in this field.  
We hope this article can provide a sort of bridge for high energy physicists interested in 
exploring some of the challenges facing upcoming ground- and space--based cosmic ray observatories.

\subsection{Cosmic ray observations}

For primary energy $E \agt 1$~GeV the observed cosmic ray flux can be described by a series of 
power laws with the flux falling about three orders of magnitude for each decade increase in energy.
In the decade centered at $\left. \sqrt{s} \right|_{_{\rm knee}} \sim 10^{3.4}~{\rm GeV}/\sqrt{A},$ 
the spectrum steepens from $E^{-2.7}$ to $E^{-3.0}$, forming the feature
commonly known as ``the knee''.  The spectrum steepens further to $E^{-3.3}$ above 
$\left. \sqrt{s}\right|_{_{\rm dip}} \sim 10^{4.5}~{\rm GeV}/\sqrt{A},$  
and then flattens to  $E^{-2.7}$ at $\left. \sqrt{s}\right|_{_{\rm ankle}} \sim 
10^{5.1}~{\rm GeV}/\sqrt{A},$ forming a feature known as ``the ankle''~\cite{Nagano:1991jz}. 
Within the statistical 
uncertainty of existing data, which is large 
for $E> 10^{11}$~GeV, the tail of the spectrum is 
consistent with a simple extrapolation at that slope to the highest 
energies. Thus far, for Earth-based accelerators, the record holder  for collisions with the 
highest energy per nucleon is the Tevatron, which countercirculates
protons and antiprotons with $\sqrt{s} \simeq 1.8$~TeV.  This center-of-mass
energy corresponds closely to that at the knee.
The Large Hadron Collider (LHC), now under construction at CERN, will 
collide protons with protons at $\sqrt{s} \simeq 14$~TeV.  This impressive energy is still about a factor 
of 50 smaller than the center-of-mass energy of the highest energy cosmic ray so 
far observed, assuming $A=1$.

For primary cosmic ray energies above $10^{5}$~GeV, the flux becomes so low that direct detection
of the primary using devices in or above the upper atmosphere is, for all practical purposes, impossible. 
Fortunately, in such cases the primary 
particle has enough energy to initiate a particle cascade in the atmosphere  
large enough that the products are detectable at ground. 
There are several techniques which can be employed in detecting these extensive air showers (EAS), 
ranging from sampling of particles in the cascade
to measurements of fluorescence, \v{C}erenkov or radio emissions produced
by the shower.

The most commonly used detection method involves sampling the shower 
front at a given altitude using an array of sensors spread over a 
large area.  Sensors, such as plastic
scintillators or \v{C}erenkov detectors are used to infer the 
particle density and the relative arrival times of the shower 
front at different locations; from this, one can
estimate the energy and direction of the
primary cosmic ray. The spacing between stations determines the 
energy threshold for a vertical shower. The muon content is usually 
sought either by exploiting the signal timing in the 
surface sensors or by employing dedicated detectors which are shielded from the 
electromagnetic shower component.
Inferring the primary energy from energy deposits at the ground is 
not completely straightforward, and involves proper modeling of both the 
detector response and the physics of the first few cascade generations.
This second point is particularly subtle and will be the main subject of Sec.~\ref{hadronic}.

Another highly successful air shower detection method involves measurement of the longitudinal
development of the cascade by sensing the fluorescence light produced
via interactions of the charged particles in the atmosphere.  
As an extensive air shower develops, it dissipates much of its energy by
exciting and ionizing air molecules along its path.  Excited nitrogen
molecules fluoresce producing radiation in the 300 - 400 nm ultraviolet
range, to which the atmosphere is quite transparent. Under favorable
atmospheric conditions EAS can be detected at distances as large as
20~km, about 2 attenuation lengths in a standard desert atmosphere at ground
level, though observations can only be made on clear moonless nights, yielding a 
duty cycle of about 10\%.  The shower development appears as a rapidly moving spot of light
whose angular motion depends on both the distance and the orientation of the shower axis.
The integrated light signal is proportional to the total energy deposited in the 
atmosphere.  Systematic errors can arise from a variety of 
sources, including uncertainties in 
the nitrogen fluorescence induced by the particle beam~\cite{Bunner,Nagano:2004am}, as well as uncertainties
in the atmospheric conditions at the time the fluorescence measurements are taken~\cite{Keilhauer:2004jr}.

The first measurements of ultra high energy cosmic rays
were carried out by Linsley at Volcano Ranch 
($35^\circ 09'$N, $106^\circ47'$W) in
the late 1950's~\cite{Linsley:prl} using an array of 
scintillation counters.  More recent experiments using surface 
detection techniques include 
Haverah Park in England ($53^\circ 58'$N, $1^\circ38'$W)~\cite{Lawrence:cc}, 
Yakutsk in Russia ($62^\circ$N, $130^\circ$E)~\cite{Khristiansen}, the Sydney
University Giant  
Airshower Recorder (SUGAR) in  
Australia ($30^\circ 32'$S, $149^\circ 43'$E)~\cite{Winn:un}, and 
the Akeno Giant Air Shower Array (AGASA), about 100~km 
west of Tokyo ($38^\circ47'$N, $138^\circ 30'$E)~\cite{Chiba:1991nf,Ohoka:ww}.  
The fluorescence method has been used by the 
Fly's Eye experiment~\cite{Baltrusaitis:mx,Baltrusaitis:ce}, as well as its up-scoped descendant
High Resolution Fly's Eye experiment (HiRes)~\cite{Abu-Zayyad:uu}, operating 
at the Dugway proving ground in the Utah desert ($40^\circ$N, $112^\circ$W).

Over the next few years, the best observations of the extreme end
of the cosmic ray spectrum will be made by 
the Pierre Auger Observatory (PAO)~\cite{Abraham},
which is currently operational in Malarg\"ue, Argentina ($35^\circ 12'$S, $69^\circ 12'$W) 
and is in the process of growing to 
its final size of $3000$~km$^2$.  A twin site is pending for the Northern hemisphere, and
together the two observatories will have an acceptance of 14000~km$^2$ sr
above $10^{10}$~GeV for zenith angles below $60^\circ$.  
The PAO works in 
a hybrid mode, and when complete will comprise 24 fluorescence detectors overlooking 
a ground array of 1600 water \v{C}erenkov detectors. 
During clear, dark nights,  events 
are simultaneously observed by 
fluorescence light and particle detectors, allowing powerful reconstruction
and cross-calibration techniques.  Simultaneous observations of showers using
two distinct detector methods will also help to control the systematic 
errors that have plagued cosmic ray experiments to date. Moreover, each site of PAO reaches 
$\sim 15$~km$^3$we~sr of target mass
around $10^{10}$~GeV~\cite{Capelle:1998zz}, which is comparable to dedicated
neutrino detectors being planned.\footnote{we $\equiv$ water equivalent.} 
This renders PAO a neutrino telescope operating in an energy regime complimentary
to existing and upcoming facilities.  The characteristics of neutrino-induced
showers are discussed in Sec.~\ref{nus}.

Space--based experiments are also in the offing, the most thoroughly planned being 
the Extreme Universe Space Observatory (EUSO)~\cite{Catalano:mm,Scarsi:fy}.  EUSO comprises
a single fluorescence eye, and is scheduled to fly aboard the  
International Space Station for more than three years.  After taking account of the 
10\% duty cycle, this experiment will image a vast volume of 750~km$^{3}$we~sr.

\begin{figure} [t]
\postscript{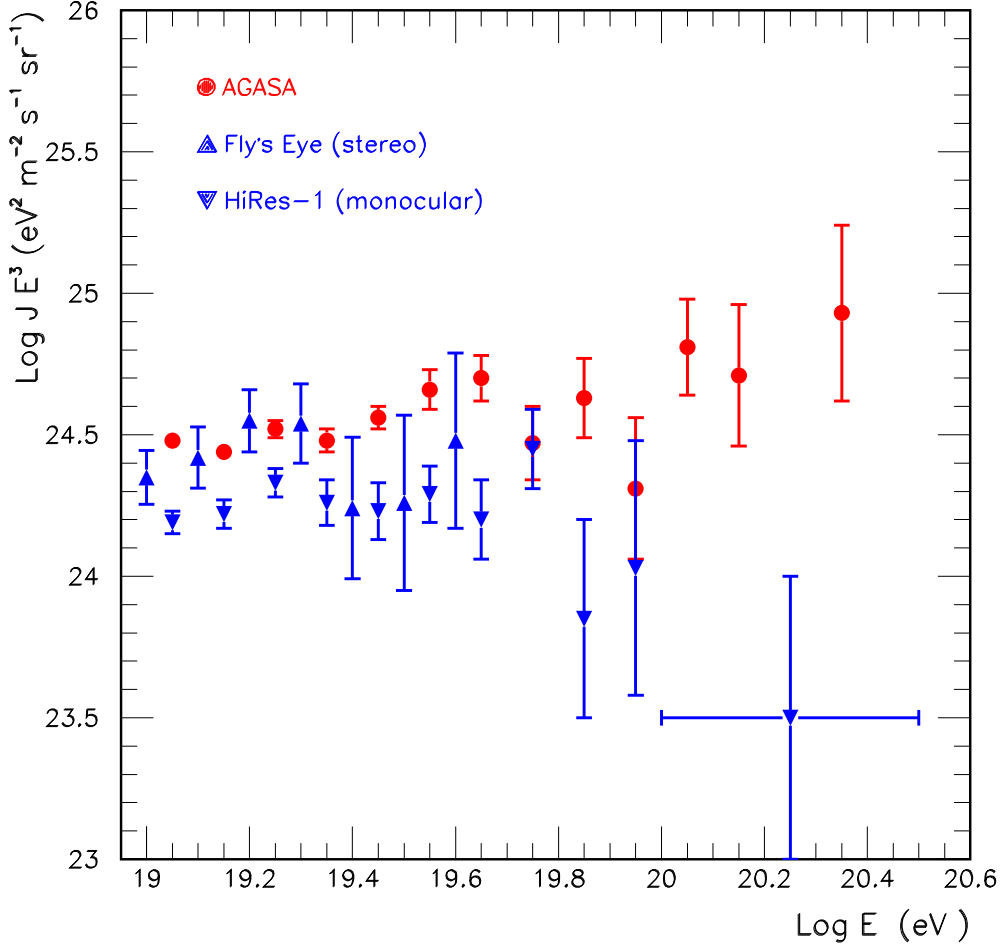}{0.80}
\caption{Upper end of the cosmic ray energy spectrum as observed by
 AGASA~\cite{Takeda:2002at} and HiRes~\cite{Abu-Zayyad:2002sf}/Fly's Eye~\cite{Bird:wp}.} 
\label{spectrum}
\end{figure}

In recent years, a somewhat confused picture {\it vis--{\`a}--vis} the energy spectrum and arrival 
direction distribution has been emerging.  Since 1998, the AGASA Collaboration has consistently 
reported~\cite{Takeda:1998ps,Takeda:2002at} a continuation of the spectrum beyond the 
expected Greisen--Zatsepin--Kuzmin (GZK) cutoff~\cite{Greisen:1966jv,Zatsepin:1966jv}, 
which should arise at about $10^{10.9}$~GeV if cosmic ray sources are
at cosmological distances.  This theoretical feature of the spectrum is mainly a consequence
of interactions of the primary cosmic ray with the microwave background radiation. In contrast, 
the most recent results from 
HiRes~\cite{Abu-Zayyad:2002sf} describe a spectrum which is consistent with the 
expected GZK feature. The discrepancy between the 2 estimated fluxes 
is shown in Fig.~\ref{spectrum}. Several analyses were done in trying to understand this difference. In
particular, since the quoted systematic errors in the energy estimate are
in the neighbourhood of 30\%, it was argued~\cite{DeMarco:2003ig} that if the AGASA Collaboration 
overestimates their energies by 30\% and the HiRes Collaboration underestimates them by about the same 
amount, then the two spectra can be brought into reasonable agreement within statistical errors.

Although there seems to be a remarkable agreement among experiments on
predictions about isotropy on large scale
structure~\cite{Anchordoqui:2003bx,Abbasi:2003tk}, this is certainly not the case when
considering the two-point correlation function on a small angular
scale.  The AGASA Collaboration reports observations of event clusters
which have a chance probability smaller than 1\% to arise from a random
distribution~~\cite{Hayashida:bc,Hayashida:2000zr}. Far from
confirming the AGASA results, the recent analysis reported
by the HiRes Collaboration showed that their data are consistent with no
clustering among the highest energy events~\cite{Abbasi:2004ib,Abbasi:2004dx}.
The discovery of such clusters would be a tremendous breakthrough for the
field, but the case for them is not yet proved.  Special
care must be taken when computing the statistical significance in such an analysis.
In particular, it is important to define
the search procedure {\it a priori} in order to ensure one does not inadvertently perform
``trials'' by studying the data before deciding upon the cuts. Very recently, with the aim 
of avoiding accidental bias on the
number of trials performed in selecting the angular bin, the original
claim of the AGASA Collaboration~\cite{Hayashida:bc} was re-examined
considering
only those events observed after the original claim~\cite{Finley:2003ur}. This study
showed that the evidence for clustering in the AGASA data set is weaker than was
previously supposed, and is consistent with the hypothesis of isotropically
distributed arrival directions.

The confusing experimental situation regarding the GZK feature and the small-scale clustering in
the distribution of arrival direction should be resolved in the near future by 
the PAO, which will provide not only a data set of unprecedented size, but also the
machinery for controlling some of the more problematic systematic uncertainties.   
As we will discuss in this review, however, the task of
identifying primary species is more challenging.  
The rest of this article is organized as follows.  In Sec.~\ref{hadronic}, we describe the 
phenomenology of hadronic interactions with the goal of providing an overview of
the main systematic uncertainties hindering the determination of the primary energy from 
observations at the ground.  In the following section, we discuss the electromagnetic
processes responsible for generating the great majority of particles in the shower.  Armed with these
two principal shower ingredients, we then discuss observables that are accessible to experiment.
Next, in Sec.~\ref{discrimination}, we describe how these observables are used to infer the primary 
composition.  In Sec.~\ref{nus} we summarize properties of neutrino interactions and
discuss observables in the deeply penetrating showers they produce.  Since the expected rate of 
such events is very low, any enhancement of the cross section due to physics beyond the electroweak
scale should be evident in this channel.  At the end we allow ourselves to venture into speculative 
territory and discuss the experimental signatures of neutrino interactions 
in scenarios with TeV-scale gravity.  Before getting underway, we first briefly summarize the main features
of the atmospheric ``calorimeter.'' 

\subsection{Nature's calorimeter}

Unlike man-made calorimeters, the atmosphere is a calorimeter whose 
properties vary in a predictable way with altitude,  and in a relatively unpredictable
way with time.  Beginning with the easier of the two variations, we note
that the density and pressure depend strongly on the height, while the temperature
does not change by more than about 30\% over the range 0--100~km above sea level.
Therefore we can get a reasonable impression of the density variation by assuming
an isothermal atmosphere, in which case the density $\rho_{\rm atm} (h) \approx \rho_0 e^{-h/h_0}$,
where $\rho_0 \approx 1.225$~kg/m$^3$ and $h_0 = R\,T /(\mu\,g) \approx 8.4$~km 
is known as the scale-height of the 
atmosphere,  $R$ being the ideal gas constant, $\mu$ the average molecular weight of 
air, $g$ the acceleration due to gravity and $T \approx 288$~K.
Of course, reading out such a natural calorimeter is complicated by the 
effects of varying aerosol and molecular attenuation and scattering.

\begin{figure}[tbp]
\postscript{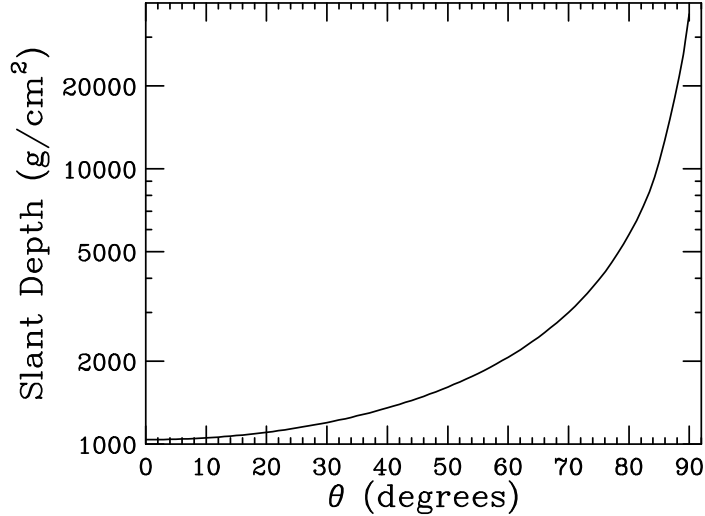}{0.56}
\caption{Slant depths corresponding to various zenith angles
$\theta$ considering the curvature of the Earth.}
\label{slantdepth}
\end{figure}

The quantity that most intuitively describes the varying density of the 
atmospheric medium is the vertical atmospheric depth,
$X_v (h) = \int_h^\infty \rho_{\rm atm}(z) \,\,dz$, where $z$ is the height. 
However, the quantity 
most relevant in air shower simulations is the 
slant depth, $X$, which defines the actual amount of air traversed by
the shower.  The variation of the 
slant depth with zenith angle is shown in Fig.~\ref{slantdepth}.
If the Earth curvature is not taken into account, then $X = X_v(h)/ \cos 
\theta$, where $\theta$ is the zenith angle of the shower axis. For 
$\theta \alt 80^\circ,$ the error associated with this  
approximation is less than 4\%.

The atmospheric 
medium is endowed with a magnetic field. In general, the geomagnetic field is 
described by 3 parameters, 
its strength $|\vec{B}|,$ its inclination $\iota,$ and its declination 
$\delta$. The 
inclination is defined as the angle between the local horizontal plane and the 
$\vec B$-field. The declination is defined as the angle between the horizontal 
component of the field $B_\perp$ (i.e., perpendicular to the arrival 
direction of the air shower) and the geographical North 
(direction of the local meridian). The angle $\iota$ is 
positive when $\vec B$ points downward and $\delta$ is positive 
when $B_\perp$ is inclined towards the East.

\section{Hadronic Processes}  
\label{hadronic}

 Uncertainties in hadronic interactions at ultra high energies
 constitute one of the most problematic sources of systematic error
 in analysis of air showers.  This section will explain the
 two principal schools of thought for extrapolating collider data to
 ultra high energies.  We start with a general description of $pp$ collisions 
 within the eikonal model. Next, we consider the specific case of hadronic interactions 
 in the atmosphere, and discuss the most widely used Monte Carlo codes. 
 Finally, we study the potential of present and future accelerators to produce 
 data valuable for understanding extensive air shower physics. Particular emphasis is placed on 
 measurements of interaction processes at extreme forward directions and the cross
 sections for diffractive interactions.

\subsection{Low-${\bm p}_{_{{\bm T}}}$ jet physics beyond collider energies}

Soft multiparticle production with small transverse momenta with respect to 
the collision axis is a dominant feature of most hadronic events at 
center-of-mass energies 
$10~{\rm GeV} < \sqrt{s} < 50~{\rm GeV}$ (see {\it e.g.},~\cite{Capella:yb,Predazzi:1998rp}). 
Despite the fact that strict calculations based on ordinary QCD perturbation 
theory are not feasible, there are some phenomenological models that 
successfully take into account the main properties of the soft diffractive 
processes. These models, inspired by $1/N$ QCD expansion are also 
supplemented with generally accepted theoretical principles like duality, 
unitarity, Regge behavior, and parton structure. The 
interactions are no longer described by single particle exchange, but by 
highly complicated modes known as Reggeons. Up to about 50~GeV, the slow growth of the cross 
section with $\sqrt{s}$ is driven by a dominant contribution of a special 
Reggeon, the Pomeron. 

At higher energies, semihard interactions arising from 
the hard scattering of partons that carry only a very small fraction of the 
momenta of their parent hadrons can compete successfully with soft 
processes~\cite{Cline:1973kv,Ellis:1973nb,Halzen:1974vh,Gaisser:1984pg,Pancheri:sr,Pancheri:ix,Pancheri:qg,Albajar:1988tt}. These semihard interactions lead 
to the minijet phenomenon, i.e., jets with transverse energy 
($E_T = |p_{_T}|$) 
much smaller than the total center-of-mass energy.  Unlike soft 
processes, this low-$p_{_T}$ jet physics can be computed in perturbative QCD. 
The parton-parton minijet cross section is given by
\begin{equation}
\sigma_{\rm QCD}(s,p_{{_T}}^{\rm cutoff}) = \sum_{i,j} \int 
\frac{dx_1}{x_1}\, \int \frac{dx_2}{x_2}\,
\int_{Q_{\rm min}^2}^{\hat{s}/2} \, d|\hat t|\,\, 
\frac{d\hat{\sigma}_{ij}}{d|\hat t|}\,\,
x_1 f_i(x_1, |\hat t|)\,\,\, x_2 f_j(x_2, |\hat t|) \,\,\,,
\label{sigmaminijet}
\end{equation}
where $x_1$ and $x_2$ are the fractions of the momenta of the parent hadrons 
carried by the partons which collide,
$d\hat{\sigma}_{ij}/d|\hat t|$ is the cross section for scattering of 
partons of types $i$ and $j$ according to elementary QCD diagrams, 
$f_i$ and $f_j$ are parton distribution functions (pdf's),
$\hat{s} = x_1\,x_2 s$ 
and $-\hat{t} = \hat{s}\, (1 - \cos \vartheta^*)/2 =  Q^2$ 
are the Mandelstam variables for this parton-parton process,
and the sum is over all parton species. Here,
\begin{equation}
p_{_T} = E_{\rm jet}^{\rm lab} \,\,\sin \vartheta_{\rm jet} = \frac{\sqrt{\hat s}}{2}\,\, 
\sin \vartheta^*\,,
\end{equation}
and
\begin{equation}
p_{_\parallel} = E_{\rm jet}^{\rm lab} \,\,\cos \vartheta_{\rm jet}\,
\end{equation}
where $E_{\rm jet}^{\rm lab}$ is the energy 
of the jet in the lab frame,   
$\vartheta_{\rm jet}$ the angle of the jet with respect to the beam 
direction in the lab frame, and $\vartheta^*$ is the angle of the jet with respect to the beam direction 
in the center-of-mass frame of the elastic parton-parton collision. This implies that for 
small $\vartheta^*,$ $p_{_T}^2 \approx Q^2$. The integration limits satisfy 
$Q_{\rm min}^2 < |\hat t| < \hat{s}/2,$ with $Q_{\rm min}$ the 
minimal momentum transfer.

\begin{figure}[tbp]
\postscript{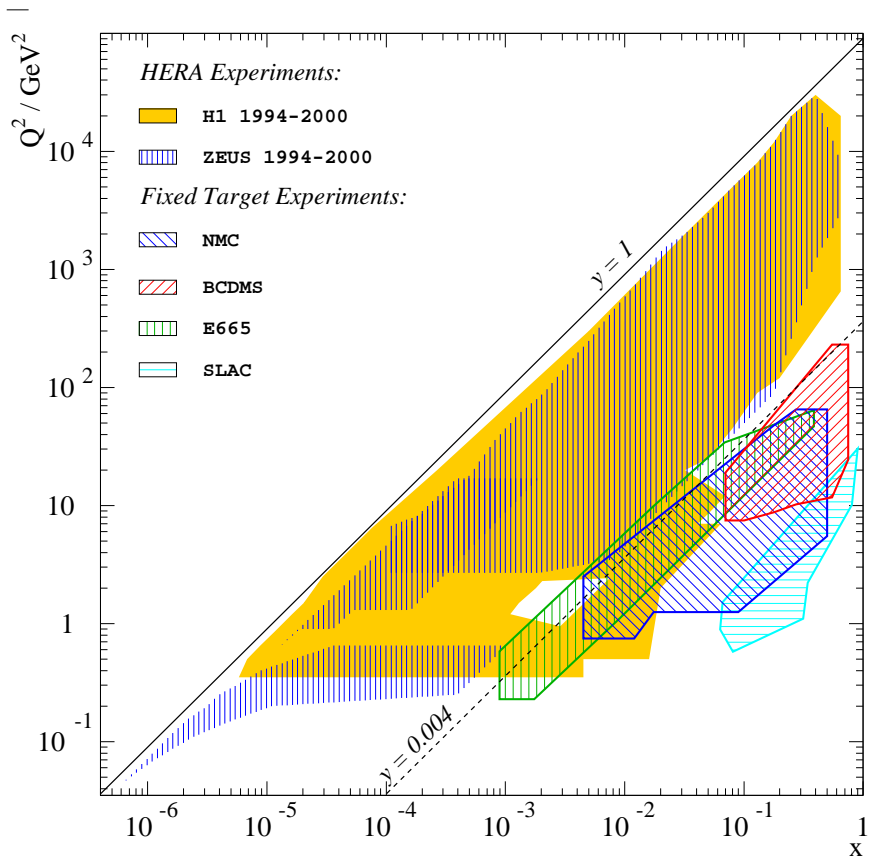}{0.70}
\caption{Kinematic $x$-$Q^2$ plane accessible to the H1 and ZEUS experiments at HERA 
and the region accessible to fixed-target experiments. The inelasticity $y = (1 -\cos \vartheta^*)/2$ is 
also shown. This figure is courtesy of Max Klein.}
\label{hera}
\end{figure}

A first source of  uncertainty in modeling cosmic ray interactions at 
ultra high energy is encoded in the extrapolation of the measured 
parton densities  several orders of magnitude down to low $x$. 
Primary protons that impact on the upper atmosphere with energy $\agt~10^{11}$~GeV, 
yield partons with $x \equiv 2 p^*_{_\parallel}/\sqrt{s} \agt m_\pi/\sqrt{s} \sim  10^{-7},$ whereas current data on quark and gluon 
densities are only available for $x \agt 10^{-4}$ to within an experimental accuracy of 3\% for 
$Q^2 \approx 20$~GeV$^2$~\cite{Adloff:2000qk}. In Fig.~\ref{hera} we show the region of the 
$x-Q^2$ plane probed by H1, ZEUS,\footnote{In the $pe^\pm$ storage ring 
HERA at DESY, 27.6~GeV $e^\pm$'s are collided on 820~GeV $p$'s, and 
data are recorded by two experiments, H1 and ZEUS.  These collisions
correspond to $\sqrt{s} \approx 300$~GeV, or equivalently a lepton energy 
$\approx 47$~TeV in the proton rest frame.} and fixed target experiments.
Moreover, application of HERA data to baryonic cosmic rays assumes universality
of the pdf's. The QCD factorization conjecture, which is essentially equivalent to the 
Ingleman-Schlein model~\cite{Ingelman:1984ns}, posits that the parton-parton minijet cross section of
Eq.~(\ref{sigmaminijet}) can always be written in a form which factorizes the parton
densities and the hard interaction processes irrespective of the order 
in perturbation theory and the particular hard process.  This conjecture holds in
the limit $Q^2 \gg \Lambda_{\rm QCD}$, where $\Lambda_{\rm QCD} \sim 200$~MeV is the 
QCD renormalization scale.  However, a severe breakdown of the 
factorization conjecture has been observed when using the pdf's obtained by
the HERA experiments 
to predict diffractive jet production in hadron-hadron
interactions at the Tevatron~\cite{Alvero:1998ta}.

\begin{figure}[tbp]
\postscript{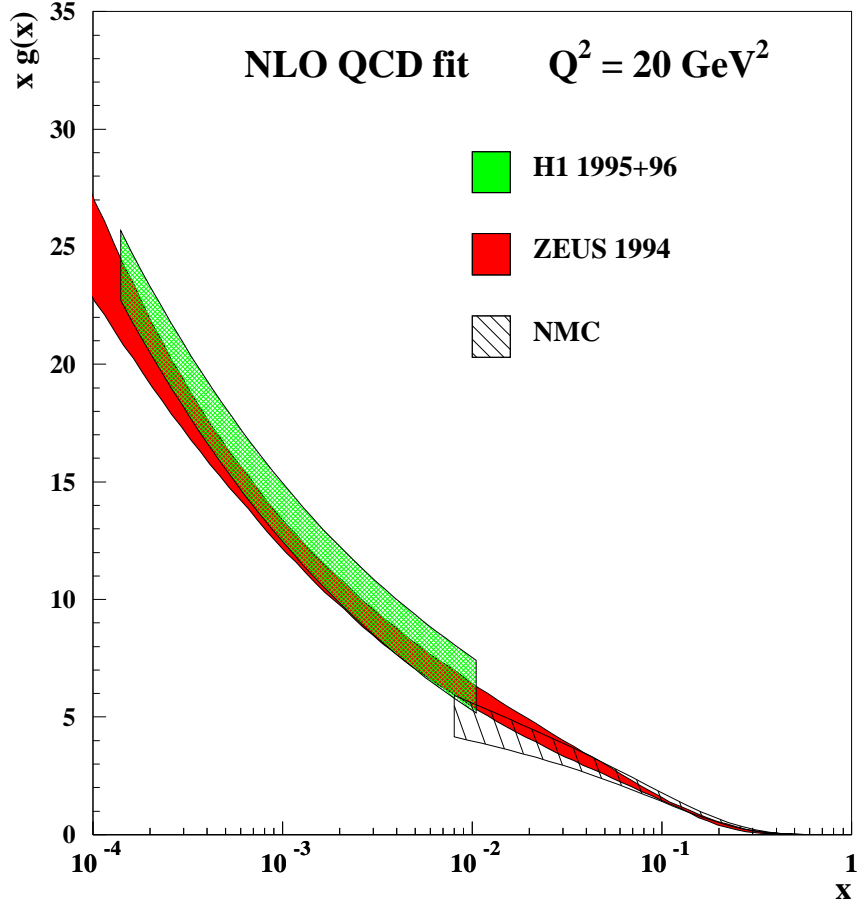}{0.70}
\caption{Gluon density for  $x > 10^{-4}$ and $Q^2 = 20$~GeV$^2,$ as measured by H1, ZEUS and NMC 
experiments.}
\label{gluonpdf}
\end{figure}

For large $Q^2$ and not too small $x$, the 
Dokshitzer-Gribov-Lipatov-Altarelli-Parisi (DGLAP) 
equations~\cite{Gribov:rt,Gribov:ri,Dokshitzer:sg,Altarelli:1977zs} 
\begin{equation}
\frac{\partial}{\partial \ln Q^2} {q(x, Q^2) \choose g(x,Q^2)}  =
\frac{\alpha_s(Q^2)}{2 \pi} {P_{qq} \,\,\,\, P_{qg} \choose P_{gq} 
\,\,\,\, P_{gg}}  \otimes  {q(x, Q^2) \choose g(x,Q^2)} 
\end{equation} 
successfully 
predict the $Q^2$ dependence of the quark and gluon densities ($q$ and $g,$ 
respectively). Here, $\alpha_s = g_s^2/(4\pi),$ with  $g_s$ the strong 
coupling constant. The splitting functions $P_{ij}$ indicate the probability
of finding a daughter parton $i$ in the parent parton $j$ with a given 
fraction of parton $j$ momentum. This probability will depend on the number 
of splittings allowed in the approximation. In the double--leading--logarithmic 
approximation, limit $[\ln(1/x)$, $\ln(Q^2/\Lambda_{\rm QCD}^2)] 
\rightarrow \infty,$ the 
DGLAP equations predict a steeply rising gluon density, $xg \sim x^{-0.4},$
which dominates the quark density at low $x$, 
in agreement with experimental results obtained with the HERA 
collider~\cite{Abramowicz:1998ii}.  Specifically, as can be seen in Fig.~\ref{gluonpdf}, 
HERA data are found to be consistent with a 
power law, $xg(x,Q^2) \sim x^{-\Delta_{\rm H}},$ with an exponent $\Delta_{\rm H}$ 
between 0.3 and 0.4~\cite{Engel:ac}.
However, it is easily seen using 
geometrical arguments that the rapid growth 
of the gluon density at low $x$ would eventually require corrections to the 
evolution equations~\cite{Gribov:tu}.

The high energy minijet cross section is then 
determined by the small-$x$ behavior of the parton distributions or, rather, 
by that of the dominant gluon distribution (via the lower limits of 
the $x_1$, $x_2$ 
integrations)
\begin{equation}
\sigma_{\rm QCD} (s,p_{{_T}}^{\rm cutoff})   
\approx \int \frac{dx_1}{x_1}\, \int \frac{dx_2}{x_2} \,
\int_{Q_{\rm min}^2}^{\hat{s}/2} \, d|\hat t|\,\,\, 
\frac{d\hat \sigma}{d|\hat t|}\,\,\,
x_1 g(x_1, |\hat t|)\,\,\, x_2 g(x_2, |\hat t|) \,\,.
\label{gluon}
\end{equation}
A na\"{i}ve estimate of the $\sigma_{\rm QCD}$ behavior at high energies can 
be obtained via extrapolation of 
$x g \propto x^{-\Delta_{\rm H}}$ to 
small $x$ in Eq.~(\ref{gluon}). Within this approximation it is 
sufficiently accurate to keep only the leading contribution  of the 
differential cross section for $gg$ scattering 
({\it i.e.}, $d\hat \sigma/d|\hat t|  \propto |\hat t|^{-2}$), and so  
Eq.~(\ref{gluon}) becomes~\cite{Kwiecinski:1990tb}
\begin{equation}
\sigma_{\rm QCD} (s)   
\propto \int_{2\,Q_{\rm min}^2/s}^{1} \frac{dx_1}{x_1}\,\,
x_1^{-\Delta_{\rm H}}\, \int_{2 \,Q_{\rm min}^2/s}^{1} \frac{dx_2}{x_2} 
\,\,x_2^{-\Delta_{\rm H}} \sim 
s^{\Delta_{\rm H}}\, \ln (s/s_0)\,,
\label{KH}
\end{equation}
where $s_0$ is a normalization constant. 
One caveat is that the inclusive QCD 
cross section given in Eq.~(\ref{KH}) is a Born approximation, and therefore 
automatically violates unitarity.

The procedure of calculating the inelastic cross section from inclusive cross 
sections is known as unitarization. In the eikonal 
model~\cite{Glauber:1970jm,L'Heureux:jk,Durand:prl,Durand:cr} of high energy 
hadron-hadron scattering, the unitarized (elastic, inelastic, and total) cross section, 
assuming a real eikonal function, is given by
\begin{equation}
\sigma_{\rm el}=\int d^2\vec b\,
\left\{1-\exp\left[ -\chi_{_{\rm soft}}(s,\vec b)
-\chi_{_{\rm hard}}(s,\vec b)\right]\right\}^2\ \,,
\label{elastic}
\end{equation}
\begin{equation}
\sigma_{\rm inel}=\int d^2\vec b\,
\left\{1-\exp\left[ -2\chi_{_{\rm soft}}(s,\vec b)
-2\chi_{_{\rm hard}}(s,\vec b)\right]\right\}\ ,
\label{inelastic}
\end{equation}
\begin{equation}
\sigma_{\rm tot}= 2 \int d^2\vec b\,
\left\{1-\exp\left[-\chi_{_{\rm soft}}(s,\vec b)
-\chi_{_{\rm hard}}(s,\vec b)\right]\right\}\ ,
\label{total}
\end{equation}
where the scattering is compounded as a sum of QCD ladders
via hard and soft processes through the
eikonals $\chi_{_{\rm hard}}$ and $\chi_{_{\rm soft}}$. It should be 
noted that we have ignored spin-dependent effects and the small real part of 
the scattering amplitude, both good approximations at high energies 
(see {\it e.g.},~\cite{Block:1984ru}). Now, if the eikonal function, 
$\chi (s,\vec b) \equiv \chi_{_{\rm soft}}(s,\vec b) + 
\chi_{_{\rm hard}}(s,\vec b) =\lambda/2,$
indicates the mean number of partonic interaction pairs at impact parameter 
$\vec b,$ the probability $p_n$ for having $n$ independent partonic 
interactions using Poisson statistics reads, 
$p_n = (\lambda^n/n!) \, e^{-\lambda}$.\footnote{This relation can be derived 
within a field theoretical context~\cite{Ter-Martirosyan:yn} using the 
Abramovski-Gribov-Kancheli (AGK) 
cutting rules~\cite{Abramovsky:fm}.}
Therefore, the factor $1-e^{-2\chi} = \sum_{n=1}^\infty p_n$ in Eq.~(\ref{inelastic}) 
can be interpreted semiclassically as the probability 
that at least 1 of the 2 protons is broken up in a collision at impact 
parameter $\vec b$.
With this in mind, the inelastic cross section is simply the integral
over all collision impact parameters of the probability of having at 
least 1 interaction, yielding a mean minijet multiplicity of 
\mbox{$\langle n_{\rm jet} \rangle \approx \sigma_{\rm QCD}/
\sigma_{\rm inel}$~\cite{Gaisser:1988ra}.} The leading
contenders to approximate the (unknown) cross sections at
cosmic ray energies, {\sc sibyll}~\cite{Fletcher:1994bd} and 
{\sc qgsjet}~\cite{Kalmykov:te}, share the eikonal
approximation but differ in their {\em ans\"atse} for the
eikonals. In both cases, the core of dominant scattering at
very high energies is the parton-parton minijet cross section given in
Eq.~(\ref{sigmaminijet}),
\begin{equation}
\chi_{_{\rm hard}} = \frac{1}{2} \, 
\sigma_{\rm QCD}(s,p_{{_T}}^{\rm cutoff})\,\, A(s,\vec b) \,,
\label{hard}
\end{equation}
where the normalized profile function, $\int d^2\vec b \,\,A(s,\vec b) = 1,$ 
indicates the distribution of partons in the plane transverse to the 
collision axis. 

In the {\sc qgsjet}-like 
models, the  core of the hard eikonal 
is dressed with a soft-pomeron pre-evolution factor. This amounts to
taking a parton distribution which is Gaussian in the transverse
coordinate distance $|\vec b|,$ 
\begin{equation}
A(s,\vec b) = \frac{e^{-|\vec b|^2/R^2(s)}}{\pi R^2(s)} \, ,
\label{a}
\end{equation}
where $R^2(s) \sim 4 R_0^2 + 4 \, \alpha^\prime_{\rm eff} \,\ln^2 (s/s_0)$, with 
$\alpha^\prime_{\rm eff} \approx 0.11$~GeV$^{-2}$. Fits to collider data have been 
carried out~\cite{Engel:icrc} using a Gaussian profile function with energy-independent width, $R_0$, 
which was allowed to vary in the fit. Under the assumption that
the partons contributing to jet production are uniformly distributed
in the transverse space all over the proton, one can obtain a reasonable
fit to the data with $R_0 = 3.5~{\rm GeV}^{-2}$ and $p_{_T}^{\rm cutoff} = 3.5$~GeV.
However, if one allows for the possibility of parton clustering, $R_0$ shrinks
to $1.5~{\rm GeV}^{-2}$ with $p_{_T}^{\rm cutoff} = 2.5$~GeV.  This leads to a 
smaller rise of the cross section with energy.  In fact, the CDF Collaboration has
reported~\cite{Abe:1997xk} measurements which may indicate that partons are distributed in clusters inside
the proton.  Specifically, measurements 
of the 4-jet to 2-jet ratio for a jet transverse energy cutoff
of 5~GeV when conveniently express in term of the effective cross section~\cite{Calucci:1999yz} lead to
\begin{equation} \label{cdf}
\sigma_{\rm eff} = \frac{1}{2} \, \frac{[\sigma_{\rm 2-jet}]^2}{\sigma_{\rm 4-jet}} = 
14.5 \pm 1.7^{+1.7}_{-2.3} \,\, {\rm mb}\,.
\end{equation}
Within the eikonal unitarization, this corresponds to 
$\sigma_{\rm eff} = 8 \pi R_0^2$.  From Eq.~(\ref{cdf}), 
$R_0 \approx 1.5~{\rm GeV}^{-2}$, which is consistent with the 
clustering hypothesis.

In  {\sc sibyll}-like models, the transverse density
distribution is taken as the Fourier transform of the proton electric
form factor, resulting in an energy-independent exponential 
(rather than Gaussian) fall-off of the parton density profile 
with $|\vec b|$. The main characteristics of the $pp$ cascade spectrum
resulting from these choices are readily predictable: the harder 
form of the {\sc sibyll}
form factor allows a greater retention of energy by the leading
particle, and hence less available for the ensuing 
shower. Consequently, on average {\sc sibyll}-like models predict a smaller  
multiplicity than {\sc qgsjet}-like models 
(see {\it e.g.}~\cite{Anchordoqui:1998nq,Anchordoqui:1999hn,Alvarez-Muniz:2002ne,Engel:is}).

At high energy, $\chi_{_{\rm soft}} \ll \chi_{_{\rm hard}},$ and so the 
inelastic cross section is dominated by the hard eikonal. 
For impact parameters larger than some threshold, $b_s$, where
$\chi_{_{\rm hard}}(s,b_s) \gg 1,$ the damping from the exponential 
term in the Gaussian profile function of Eq.~(\ref{a}) is so strong 
that any increase in $\sigma_{\rm QCD}$ does not significantly
alter the contribution to the inelastic cross section from the region
where $|\vec b| < b_s$. At high energy, with the appropriate choice of 
normalization, the cross section in Eq.~(\ref{KH}) can be well--approximated 
by a power law. Hence, by taking $\sigma_{\rm QCD} 
\sim s^{\Delta_{\rm H}},$ one fixes $b_s^2 \sim 4 \, \alpha^\prime_{\rm eff} 
\, \Delta_{\rm H}\,\ln^2 (s/s_0)$~\cite{Alvarez-Muniz:2002ne}. This implies 
that the growth of the inelastic  cross section according to {\sc qgsjet}-like models is given by
\begin{equation}
\sigma_{\rm inel} \sim \int d^2 \vec b \,\,\, \Theta (b_s - |\vec b|) = 
\pi b_s^2 \sim 4\pi \, \alpha^\prime_{\rm eff} \,\,\Delta_{\rm H}\,\,
\ln^2 (s/s_0) \sim 0.52 
\, \, \Delta_{\rm H} \,\, \ln^2 (s/s_0) \,\,\, {\rm mb}\,\,.
\label{qsig}
\end{equation}
For {\sc sibyll}-like models, where Eq.~(\ref{a}) is replaced by the 
 Fourier transform of the proton electric
form factor, the  growth of the inelastic
cross section also saturates the $\ln^2s$ Froissart 
bound~\cite{Froissart:ux},
but with a multiplicative constant which is larger 
than the one in {\sc qgsjet}-like models~\cite{Alvarez-Muniz:2002ne}. Namely,
\begin{equation}
\sigma_{\rm inel} \sim 3.2 \,\, \Delta_{\rm H}^2 \ln^2 (s/s_0) \,\,\, {\rm mb}\,\,.
\label{ssig}
\end{equation}

\begin{figure} [t]
\postscript{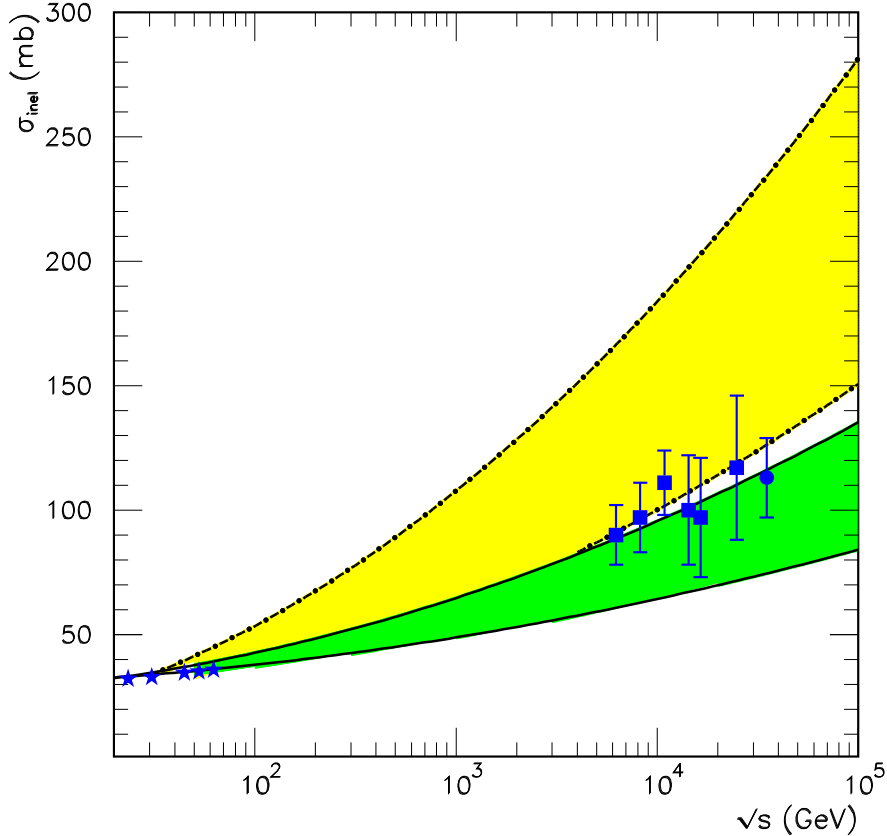}{0.70}
\caption{Energy dependence of the $pp$ inelastic cross section as predicted by 
Eqs.~(\ref{qsig}) and (\ref{ssig})  with  $0.3 < \Delta_{\rm H} < 0.4.$ The 
darkly shaded region between the solid lines corresponds to the model 
with Gaussian parton  distribution in $\vec b$. The region between the 
dashed-dotted lines 
corresponds to the model with exponential fall-off of the parton density in $\vec b$.  
In both cases the cross sections are normalized to reproduce the data $(\star)$~\cite{Amaldi:1979kd} 
from the CERN Intersecting Storage Ring (ISR) at 30~GeV. Also shown are 
estimates~\cite{Block:2000pg}  of the inelastic $pp$ cross section  as derived from 
measurements of the inelastic $p$-air cross section by the 
AGASA ({\tiny $\blacksquare$})~\cite{Honda:1992kv}  
and the Fly's Eye ($\bullet$)~\cite{Baltrusaitis:ka} experiments.}
\label{sigmahera}
\end{figure}

Figure~\ref{sigmahera} illustrates the 
large range of predictions for $pp$ inelastic cross section which remain 
consistent with HERA data. When the two leading order approximations discussed above are
extrapolated to higher energies, both are consistent
with existing cosmic ray data.  Note, however, that in both cases the range of
allowed cross-sections at high energy varies by a factor of about 2 to 3. A point worth noting 
at this juncture: a number of 
approaches have been used to extract the $pp$ cross section from cosmic ray shower 
data ~\cite{Gaisser:ha,Kopeliovich:iy,Nikolaev:mc,Block:1999ub}. The points in Fig.~\ref{sigmahera} 
correspond to the most up-to-date estimate~\cite{Block:2000pg}.

There are three event generators, {\sc sibyll}~\cite{Fletcher:1994bd}, 
{\sc qgsjet}~\cite{Kalmykov:te}, 
and {\sc dpmjet}~\cite{Ranft:fd} 
which are tailored specifically for simulation of hadronic interactions up to 
the highest cosmic ray energies.\footnote{Additionally, a new event
generator, ne{\sc x}us~\cite{Bossard:2000jh}, is available for simulation
of the region below $\sqrt{s}\sim 10^{4}$~GeV.} The latest versions of these packages are {\sc sibyll} 
2.1~\cite{Engel:db}, {\sc qgsjet} 01~\cite{Heck01}, and {\sc dpmjet III}~\cite{Roesler:2000he}; 
respectively.  In {\sc qgsjet}, both the soft and hard 
processes are formulated in terms of Pomeron exchanges. To describe the 
minijets, the soft Pomeron  mutates into a ``semihard Pomeron'', 
an ordinary soft Pomeron with the middle piece replaced by a QCD parton 
ladder, as sketched in the previous paragraph.  This is generally referred to 
as the ``quasi-eikonal'' model.  In
contrast, {\sc sibyll} and {\sc dpmjet} follow a ``two channel'' eikonal model, where the soft and  
the semi-hard regimes are demarcated by a sharp cut in the transverse momentum: 
{\sc sibyll} 2.1 uses a cutoff parametrization inspired in the double leading logarithmic approximation 
of the DGLAP equations,
\begin{equation}
p_{{_T}}^{\rm cutoff} (\sqrt{s}) =  p_{{_T}}^0 + 0.065~{\rm GeV}\, \exp[0.9\,\sqrt{\ln s}]\,,
\end{equation}
whereas {\sc dpmjet} III uses an {\it ad hoc} parametrization for the transverse momentum cutoff
\begin{equation}
p_{{_T}}^{\rm cutoff} (\sqrt{s}) =  p_{{_T}}^0 + 0.12~{\rm GeV}\, [\log_{10} (\sqrt{s}/50{\rm GeV})]^3\,,
\end{equation}
where  $p_{{_T}}^0 = 2.5$~GeV~\cite{Engel:ac}.

The transition process from asymptotically free partons to colour-neutral 
hadrons is described in all codes by string fragmentation 
models~\cite{Sjostrand:1987xj}.  Different choices of fragmentation functions 
can lead to some differences in the hadron multiplicities. 
However, the main difference in the predictions of {\sc qgsjet}-like and 
{\sc sibyll}-like models arises from different assumptions in extrapolation of the parton 
distribution function to low energy.

\subsection{Hadronic interactions in the Earth's atmosphere}
\label{HIEA}

Now we turn to nucleus-nucleus interactions, which 
cause additional headaches for event generators 
which must somehow extrapolate $pp$ interactions
in order to simulate the proton-air collisions of interest. 
All the event generators described above adopt the
Glauber formalism~\cite{Glauber:1970jm}, which is equivalent
to the eikonal approximation in nucleon-nucleon scattering, 
except that the nucleon density functions of the target nucleus are
folded with that of the nucleon.  The inelastic and production
cross sections read:
\begin{equation}
\widetilde\sigma_{\rm inel} \approx \int d^2\vec b\,
\left\{1-\exp\left[\sigma_{\rm tot} 
\,\, T_A(\vec b) \right]\right\}\ \,,
\label{inelasticn}
\end{equation}
\begin{equation}
\widetilde\sigma_{\rm prod} \approx \int d^2\vec b\,
\left\{1-\exp\left[\sigma_{\rm inel} 
\,\, T_A(\vec b) \right]\right\}\ \,,
\label{prod}
\end{equation}
where  $T_A(\vec b)$ is the transverse density of hadronic 
matter of the target nucleus folded with that of the projectile hadron. Here, 
$\sigma_{inel}$ and $\sigma_{tot}$ are given by Eqs.~(\ref{inelastic}) and 
(\ref{total}), respectively. The $p$-air inelastic cross section is the sum of the
``quasi-elastic'' cross section, which corresponds to 
cases where the target nucleus breaks up without production
of any new particles, and the production cross section,
in which at least one new particle is generated.  Clearly 
the development of EAS is mainly sensitive to the production
cross section.  Overall, the geometrically large size of 
nitrogen and oxygen nuclei dominates 
the inclusive proton-target cross section, and as a result
the disagreement from model-dependent
extrapolation is not more than about 15\%.  

The event generators also make different choices in their
handling of nucleus-air collisions~\cite{Engel:vf,Kaidalov:zf}.
Models of nucleus-nucleus interactions are particularly important to 
describe the first few generations of secondaries in cosmic ray showers 
produced by nuclei.  Measurements of proton-nucleus reactions at lower 
energies~\cite{Elias:1979cp} suggested that the charged multiplicity from a soft production mechanism 
should simply scale with the 
number of nucleons that participate in the collision~\cite{Bialas:1976ed}, thus allowing for comparison of 
different nuclear systems based on simple nucleon-nucleon superposition models. The particle densities 
are sensitive to the relative contributions of soft and hard 
processes~\cite{Wang:2000bf,Eskola:2000xq}. 
More recent experimental input suggests a simple superposition model is not completely realistic.
Specifically, RHIC~\footnote{The Relativistic Heavy Ion Collider (RHIC) collides ultra relativistic ions 
at energies up to 0.2 TeV/$N$. RHIC has two large detectors, STAR and PHENIX and two smaller experiments: 
BRAHMS and PHOBOS.} data have shown that the observed 
central particle densities~\cite{Bearden:2001qq,Adcox:2000sp,Klein:2002wn} are smaller than predictions 
from conventional 
multi-string models, with differences of 20\% -- 
30\%~\cite{DiasdeDeus:2000cg,DiasdeDeus:2000gf}.  To reduce the multiplicity in the models, 
the percolation process, which leads with increasing density to more and more fusion strings, 
has been proposed~\cite{Braun:1997ch,Braun:1999hv,Braun:2001us}. 
Very recently, the data from d-Au collisions collected by the PHOBOS 
Collaboration~\cite{Back:2003hx} were used to improve the event generator {\sc dpmjet}~III and
bring the predicted multiplicity in line with the data~\cite{Bopp:2004xn}.

So far the discussion has concerned $p-$air and nucleus-air interactions. 
Of course in air shower simulations, we are also concerned with $\pi$-air interactions.
Each approach discussed above handles $\pi p$ collisions using the same 
interaction model it uses for $p p$ collisions.  For energies of 
interest, both models predict, on average, 
a $\pi p$ inelastic cross section about 20\% smaller than the $p p$ cross 
section~\cite{Alvarez-Muniz:2002ne}.

\begin{figure}[tbp]
\begin{minipage}[t]{0.49\textwidth}
\postscript{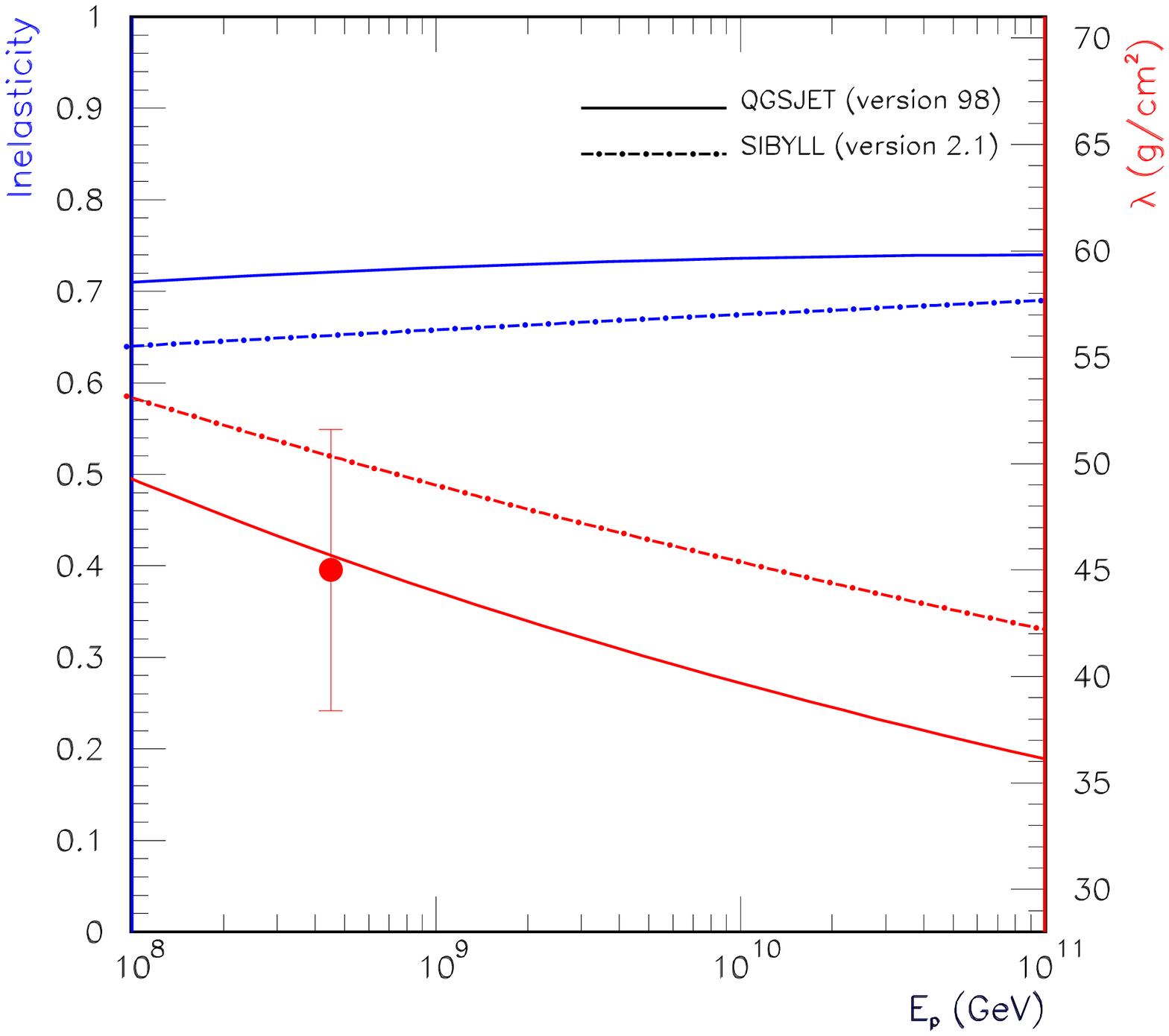}{0.99}
\end{minipage}
\hfill
\begin{minipage}[t]{0.49\textwidth}
\postscript{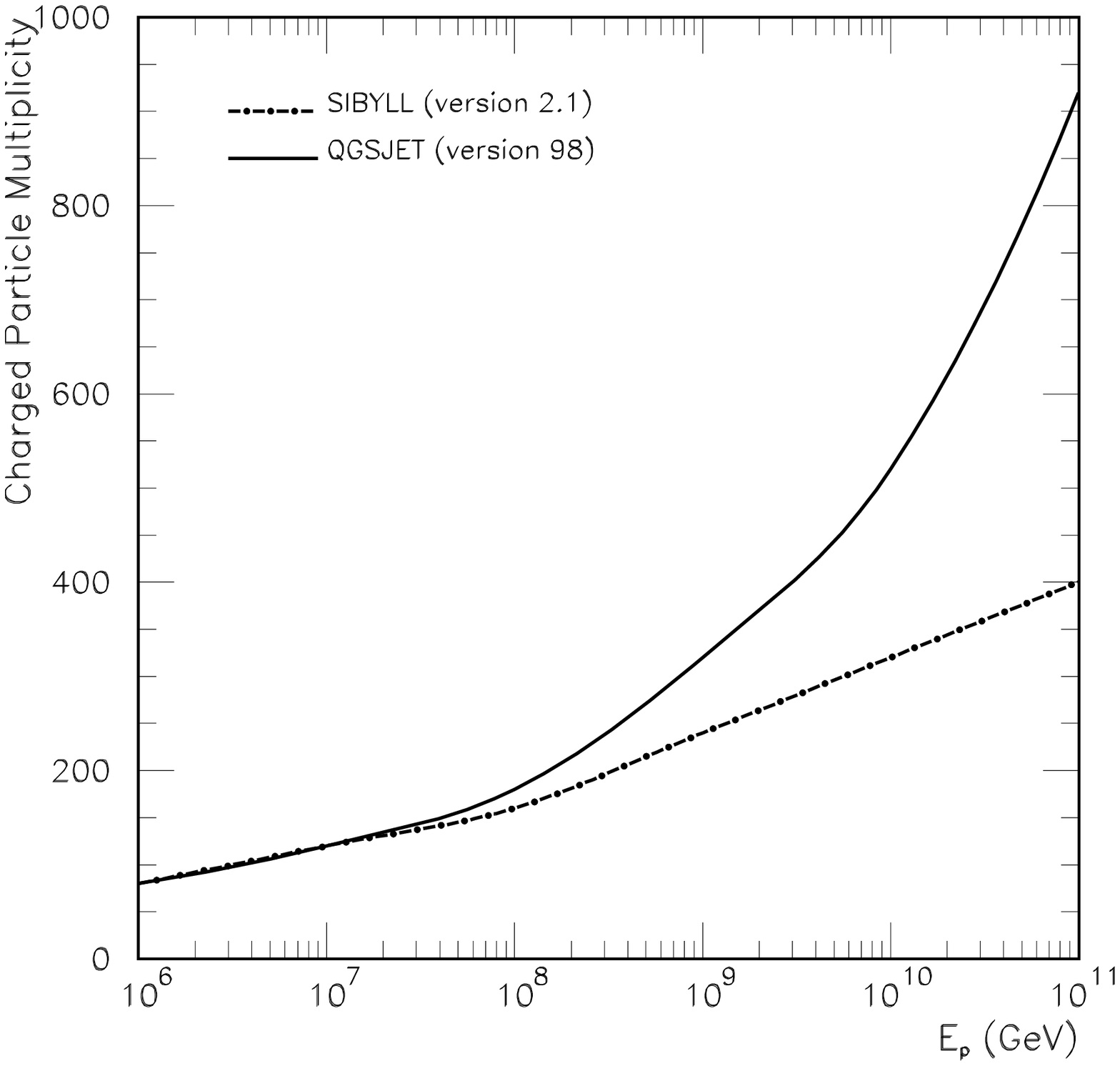}{0.95}
\end{minipage}
\caption{Left panel:
The slowly rising curves indicate the mean 
inelasticity in  proton air collisions as predicted 
by {\sc qgsjet} and {\sc sibyll}. The falling 
curves indicate the proton mean free path in the atmosphere. 
The data point is from Fly's Eye measurements~\cite{Baltrusaitis:ka}.
Right panel: Mean multiplicity of charged secondary particles produced in inelastic 
proton-air collisions processed with {\sc qgsjet} and 
{\sc sibyll}.}
\label{mfp_n}
\end{figure}

Since the codes described above are still being refined, the disparity between
them can vary even from version to version.
At the end of the day, however, the relevant parameters 
boil down to two: the mean free path, 
$\lambda = (\widetilde \sigma_{\rm prod} \,n)^{-1},$ and the inelasticity,  
$K = 1 - E_{\rm lead}/E_{\rm proj}$, where 
 $n$ is the number density of atmospheric target nucleons,
$E_{\rm lead}$ is the 
energy of the most energetic hadron with a long lifetime, and $E_{\rm proj}$ 
is the energy of the projectile particle. The first parameter characterizes 
the frequency of interactions, whereas the second one quantifies the energy 
lost per collision. Overall,
{\sc sibyll} has a shorter mean free path and a smaller inelasticity than 
{\sc qgsjet},
as indicated in Fig.~\ref{mfp_n}.  Since a shorter mean free path tends to 
compensate a smaller inelasticity, the two codes generate similar predictions 
for an air shower which has lived through several generations. The different 
predictions for the mean charged 
particle multiplicity in proton-air collisions are shown in Fig.~\ref{mfp_n}. 
Both models predict the same multiplicity below about $10^{7}$~GeV, but 
the predictions diverge above that energy. Such a divergence readily increases 
with rising energy. While {\sc qgsjet} predicts a 
power law-like increase of the number of secondaries up to the highest 
energy, {\sc sibyll} multiplicity exhibits a logarithmic growth. As it is 
extremely difficult to observe the first interactions experimentally,
it is not straightforward to determine which model is closer to reality.  
In Sec.~\ref{EMP}, however, we will discuss observables which may 
offer a hint of which model better predicts overall shower characteristics.

\subsection{Measurements of forward processes at the LHC}

Interpretation of cosmic ray data suffers from the lack 
of knowledge of high energy hadronic interaction models. 
Hard interactions with high momentum transfer are calculable in perturbation theory using QCD. 
At present, collider experiments have mostly concentrated on these hard processes in the central region,
thereby excluding soft processes in the far-forward direction.  
These low momentum transfer processes, which 
are of great interest in the development of cosmic ray EAS, are not calculable from the fundamental QCD 
Lagrangian.   

Some guidance towards understanding hadronic processes in the forward
direction may come directly from measurements of hadrons in airshowers~\cite{Antoni:2001qj}.
However, the most useful experimental input in the forseeable future will
likely come from the LHC.  This machine,
expected to become operational in 2007 or so,
will provide $pp$ collisions with $\sqrt{s}$ = 14~TeV and  luminosity up to 
$L \approx 10^{34}$~cm$^{-2}$ s$^{-1}$~\cite{Evans:1999dk}, as well as, a few years later, lead-lead ion 
collisions
with $\sqrt{s} = 1000$~TeV.  Two general-purpose experiments, ATLAS and CMS, presently cover
 up to $|\eta| < 5$.  A dedicated heavy ion detector, ALICE, will also operate at this 
collider.\footnote{A $b$-physics experiment, LHC-$b$, is also under construction at the LHC and 
will offer particle identification in the range $1.9<\eta<4.9$.}

\begin{figure}[tbp]
\postscript{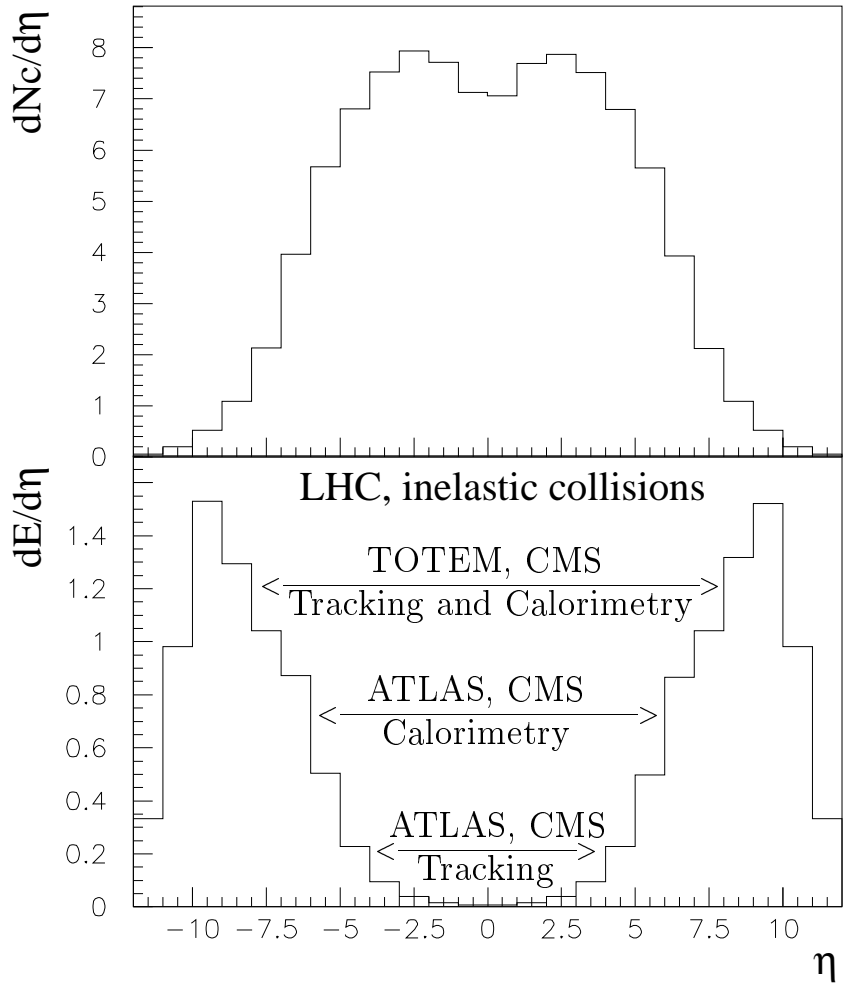}{0.70}
\caption{Pseudorapidity distributions of charged particles (upper panel) and of 
the energy flow (lower panel) for $pp$ collisions at LHC~\cite{Eggert:ca}.} 
\label{totem1}
\end{figure}

The interesting low momentum transfer processes tend to populate the 
region at very small angles $\vartheta$ with respect to the beam direction.  
The distribution of pseudorapidity, $\eta = -\ln \tan (\vartheta/2)$, and 
the energy flow distribution are shown in Fig.~\ref{totem1}.  While
the particle multiplicity is greatest in the low $| \eta |$ region, it  
is clearly seen that the energy flow is peaked at small production angles (large $| \eta |$).

A study of diffraction, in $pp$ as well as heavy ion collisions, must use 
detectors with excellent forward acceptance to allow for a comparison with cosmic ray data. Dedicated runs of the LHC with lower luminosity $(L = 10^{28} {\rm cm}^{-2} {\rm s}^{-1})$ and specially
tuned beam optics are planned to study these diffractive events.
At present the only approved experiment at LHC with a capability of measuring, to some extent, very forward 
particles is TOTEM~\cite{totemdr,Eggert:ca,TDR}, which will comprise Roman pots placed on each 
side of the CMS interaction region and forward trackers which cover the pseudorapidity range 
$3.0 \le \eta \le 6.8$.  It should be mentioned, however, that 
the fragmentation region that plays a crucial role in the development of EAS corresponds to pseudorapidity range 
$6\le |\eta|\le 10$. 

The main goal of TOTEM is the measurement of elastic and total cross-sections 
with an expected precision of about 1\%, in a luminosity independent manner.
To calculate the total cross section in terms of the number of 
elastic and inelastic events measured by TOTEM, we can resort to the well-known optical theorem
\begin{equation}
\sigma_{\rm tot} = {8 \pi \over \sqrt{s} } \; \Im {\rm m} [f(0)]\,\,,
\label{0}
\end{equation}
where $f(\vartheta)$ satisfies
\begin{equation}
{d\sigma_{\rm el}  \over dt} = 
{4 \pi \over s } \; 
{d\sigma_{\rm el}  \over d\Omega} =
{4 \pi \over s } \; |f(\vartheta)|^2 \,\,,
\end{equation}
with $\vartheta$ the angle of the scattered proton with respect to the beam direction.
Squaring Eq.(\ref{0}) we obtain
\begin{equation}
\sigma^2_{\rm tot} = {16 \pi \,\,\Im {\rm m}^2 [f(0)] \over  \Re  {\rm e}^2[f(0)] + 
\Im  {\rm m}^2[f(0)] } \; \left.{d\sigma_{\rm el}  \over dt}\right|_{t=0} = 
{16 \pi \over  1 + \rho^2}  \;\; \frac{[dN_{\rm el}/ dt]_{t=0}}{{\cal L}}\,,
\end{equation}
where ${\cal L}$ is the integrated luminosity. 
Now, following~\cite{Matthiae:gb,Lipari:2003es}, we can obtain 
the total cross section independently from ${\cal L}$, by using $\sigma_{\rm tot} =
(\sigma_{\rm el} + \sigma_{\rm inel})  = (N_{\rm el} + N_{\rm inel})/{\cal L},$
\begin{equation}
\sigma_{\rm tot} = {16 \pi \over 1 + \rho^2} \, { [ d N_{\rm el} / dt  ]_{t = 0} \over (N_{\rm el} + N_{\rm inel}) } \,. 
\end{equation}
Here, $N_{\rm el}$ and $N_{\rm inel}$ are the numbers of elastic and inelastic events, and
$\rho = 0.10 \pm 0.01$ is the ratio between the real and imaginary parts of the forward scattering 
amplitude~\cite{Augier:1993ta}.\footnote{Note that the quoted value of $\rho$ is an extrapolation to 
$\sqrt{s} = 14$~TeV, and may be measured by the LHC experiments. Otherwise, it will contribute to 
the uncertainty in $\sigma_{\rm tot}$.} The difficult aspect of this measurement is obtaining
a good extrapolation of the cross section for low momentum transfer. 
Recall that  $-t = s \,(1-\cos \vartheta) / 2 \simeq
s \, \vartheta^2 / 4$, and so $t \to 0$ implies a measurement in the
extreme forward direction.
The TOTEM experiment aims to measure down to values of $|t| \approx  \times 10^{-3}~\rm{GeV}^{2},$ which
corresponds to $\vartheta \approx 4.5~\mu$rad~\cite{TDR}.   The 
design for the pseudorapidity range $5.5 < |\eta| < 6.8$ is under discussion in a joint CMS/TOTEM working 
group. A tungsten \v{C}erenkov calorimeter known as CASTOR has also been proposed which 
would compliment the measurements of TOTEM and CMS for $|\eta| <6.8$ and facilitate
simultaneous measurements of particle flow in diffractive and non-diffractive events.

The ATLAS experiment is planning to implement additional detectors to cover the forward diffractive 
regions with tracking and/or calorimetry~\cite{Rijssenbeek:cb}, with proposed coverage for 
the region $|t| \approx 6 \times 10^{-4}$~GeV $^2$.

In summary, existing event generators rely on theoretical extrapolations of existing data up to 
the energies near the GZK energy.  There is a general 
consensus in the community that in order to understand the development of EAS at these 
extreme energies, new inputs from accelerator experiments are needed. A series of workshops have been 
organized to discuss what experimental inputs are most 
needed~\cite{Jones:bz,Engel:2002id,Schatz:2003we}; a preliminary list of the 
requirements includes~\cite{Heck:2003wc}: 
{\it (i)} measurements of total and inelastic cross sections for $pp$, $pA$, $AA$  
{\it (ii)} measurements of the ratio between soft diffractive and semi-hard processes, 
$\sigma_{\rm diff}/\sigma_{\rm inel}$, 
{\it (iii)} measurement of inclusive final state hadrons in the two momentum ranges, $ 0.8 < x < 1.0$ 
and $ 0.1 < x < 0.8$.

\section{Electromagnetic Processes}  
\label{EMP}

In this section we describe the electromagnetic interactions
of relevance in ultra high energy shower development.  
The most important processes are electron 
and muon bremsstrahlung and pair production. 
We also discuss the Landau-Pomeranchuk-Migdal (LPM)
effect, which suppresses the cross sections for pair production
and bremsstrahlung above roughly $10^{10}$~GeV, and photon
conversion in the geomagnetic field, which to a large
degree compensates for the LPM effect in terms of shower observables.
We comment further on shower observables such as the
age parameter, Moliere radius and shower size within the
context of the Nishimura, Kamata and Greisen (NKG) model.
Finally, we discuss the extension of the NKG formalism
to the corresponding shower parameters  describing the lateral
spread and the longitudinal development of EAS initiated by hadrons.

\subsection{The electromagnetic component}
\label{EM}

The evolution of an extensive air shower is dominated by electromagnetic
processes. The interaction of a 
baryonic cosmic ray  with an air nucleus high in the atmosphere leads to a 
cascade of secondary mesons and nucleons. The first few 
generations of charged pions interact again, producing a hadronic core, 
which continues to feed the electromagnetic and muonic components of the 
showers. Up to about $50$~km above sea level,
the density of atmospheric target nucleons is $n \sim 
10^{20}$~cm$^{-3},$ and so even for relatively
low energies, say  $E_{\pi^{\pm}}\approx 1$~TeV, the probability 
of decay before interaction falls below 10\%. 
Ultimately, the electromagnetic cascade dissipates around 90\%
of the primary particle's energy, and hence the total number of 
electromagnetic particles is very nearly proportional to the shower energy~\cite{Barbosa:2003dc}.

By the time a vertically incident 
$10^{11}$~GeV proton shower
reaches the ground, there are about $10^{11}$ secondaries with energy above
90~keV in the the annular region extending 8~m to 8~km from the shower core.
Of these, 99\%  are photons, electrons, and positrons, with a typical ratio 
of $\gamma$ to $e^+  e^-$ of 9 to 1. Their mean energy  is 
around 10~MeV and they transport 85\% of the total energy at ground level. Of course, 
photon-induced showers are even more dominated by the electromagnetic channel, 
as the only significant muon generation mechanism in this case is the decay of
charged pions and kaons produced in $\gamma$-air  interactions~\cite{Mccomb:tp}. 

It is worth mentioning that these figures dramatically change for the case of 
very inclined showers. For a primary zenith angle, $\theta > 70^{\circ},$ the 
electromagnetic component becomes attenuated exponentially with atmospheric 
depth, being almost completely absorbed at ground level. We remind the reader 
that the vertical atmosphere is $\approx 1000$ g/cm$^{2}$, and is about 36 
times deeper for completely horizontal showers (see Fig.~\ref{slantdepth}). 
As a result, most of the energy at ground level from an inclined shower is
carried by muons.

In contrast to  hadronic collisions, the electromagnetic 
interactions of shower particles can be calculated very accurately from 
quantum electrodynamics. Electromagnetic interactions are thus not a 
major source of systematic errors in shower simulations. The first 
comprehensive treatment of electromagnetic showers was elaborated by 
Rossi and Greissen~\cite{Rossi}.  This treatment was recently cast in a more 
pedagogical form by Gaisser~\cite{Gaisser:vg}, which we summarize in the 
subsequent paragraphs.

The generation of the electromagnetic component is 
driven by electron bremsstrahlung and pair production~\cite{Bethe:1934za}.
Eventually the average energy per particle drops below a critical energy,
$\epsilon_0$, at which point ionization takes over from bremsstrahlung and pair
production as the dominant energy loss mechanism. The $e^\pm$ energy loss rate due to
bremsstrahlung radiation is nearly proportional to their energy, whereas the 
ionization loss rate varies only logarithmically with the $e^\pm$ energy.
Though several different definitions of the critical energy appear in the 
literature~\cite{Hagiwara:fs}, throughout this review we take the critical energy 
to be that at which the ionization loss per radiation lenght
is equal to the electron energy, yielding $\epsilon_{0} = 710~{\rm MeV}/(Z_{\rm eff} +0.92) \sim 
$~86~MeV~\cite{Rossi:book}.\footnote{For altitudes up to 90~km above sea level, the air is a mixture 
of 78.09\% of N$_2$, 20.95\% of O$_2$, and 0.96\% of other gases~\cite{Weast}. 
Such a mixture is generally modeled as an homogeneous substance with atomic charge
and mass numbers $Z_{\rm eff} = 7.3$ and $A_{\rm eff} = 14.6,$ respectively.}  The changeover 
from radiation losses to ionization losses depopulates the shower.
One can thus categorize the shower development in three phases: the growth phase, in which all the particles 
have energy $> \epsilon_0$; the shower maximum, $X_{\rm max}$; and the shower 
tail, where the particles only lose energy, get absorbed or decay. 

The relevant quantities participating in the development
of the electromagnetic cascade are the probability for an electron of
energy $E$ to radiate a photon of energy $k=yE$ and
the probability for a photon to produce a pair $e^+e^-$
in which one of the particles (hereafter $e^-$) has energy $E=xk$.
These probabilities are determined by the properties of the air and 
the cross sections of the two processes. 

In the energy range of interest, the impact 
parameter of the electron or photon is larger than an atomic radius, so 
the nuclear field is screened by its electron cloud.  
In the case of complete screening, where the momentum transfer is small, 
the cross section for bremsstrahlung can be approximated by~\cite{Tsai:1973py}
\begin{equation}
\frac{d\sigma_{e \rightarrow \gamma}}{dk} \approx \frac{A_{\rm eff}}{X_0 N_A k}
\left(\frac{4}{3}-\frac{4}{3}y+y^2\right)\,\,,\label{brem}
\end{equation}
where $A_{\rm eff}$ is the effective mass number of the air, $X_0$ is a constant,
and $N_A$ is Avogadro's number. In the infrared limit ({\it i.e.}, $y \ll 1$) this approximation 
is inaccurate at the level of about 2.5\%, which is small compared to typical experimental 
errors associated with cosmic air shower detectors.  Of course, the approximation fails 
as $y \rightarrow 1$, when nuclear screening becomes incomplete, and as $y \rightarrow 0$, at which 
point the LPM and dielectric suppression effects become important, as we discuss below.

Using similar approximations, the cross section
for pair production can be written as~\cite{Tsai:1973py}
\begin{equation}
\frac{d\sigma_{\gamma \rightarrow e^+e^-}}{dE} \approx \frac{A_{\rm eff}}{X_0 N_A}
\left(1-\frac{4}{3}x+\frac{4}{3}x^2\right) \,. \label{pair}
\end{equation}
The similarities between this expression and Eq.~(\ref{brem}) are to be expected, 
as the Feynman diagrams for pair production and bremsstrahlung are variants of one another.

The probability for an electron to radiate a photon 
with energy in the range $(k,k+dk)$ in traversing $dt=dX/X_0$ of atmosphere is 
\begin{equation} \label{p_b}
\frac{d\sigma_{e \rightarrow \gamma}}{dk}\,\frac{X_0 N_A}{A_{\rm eff}}\,dk\,dt
\approx \left(y+\frac{4}{3}\,\,\frac{1-y}{y}\right)\,dy\,dt \,\,,
\end{equation}
whereas the corresponding probability density for a photon producing 
a pair, with electron energy in the 
range $(E,E+dE)$, is
\begin{equation} \label{p_pp}
\frac{d\sigma_{\gamma \rightarrow e^+e^-}}{dE}\,\frac{X_0 N_A}{A_{\rm eff}}\,dE\,dt
\approx \left(1-\frac{4}{3}x+\frac{4}{3}x^2\right)\,dx\,dt \,\,.
\end{equation}
The total probability for pair production per unit of $X_0$ follows from 
integration of Eq.~(\ref{p_pp}), 
\begin{equation}
\int \frac{d\sigma_{\gamma \rightarrow e^+e^-}}{dE}\,\,\,
\frac{X_0 N_A}{A_{\rm eff}}\, dE\, \approx \int_0^1 
\left(1-\frac{4}{3}x + \frac{4}{3}x^2\right)
\,dx = \frac{7}{9}\,.
\end{equation}

As can be seen from Eq.~(\ref{p_b}), the total probability for 
bremsstrahlung radiation is logarithmically divergent.  However, this 
infrared divergence is eliminated by the interference of 
bremsstrahlung amplitudes from multiple scattering centers.  
This collective effect of the electric potential of several
atoms is known as the Landau-Pomeranchuk-Migdal (LPM)
effect~\cite{Landau:um,Migdal:1956tc}. Of course,
the LPM suppression of the cross section results in an 
effective increase of the mean free path of electrons and photons. This 
effectively retards the development of the electromagnetic component of the shower. 
It is natural to introduce 
an energy scale, $E_{\rm LPM}$, at which the inelasticity is low enough that the LPM effect becomes 
significant~\cite{Stanev:au}.
Below $E_{\rm LPM}$, the energy loss rate due to bremsstrahlung is roughly
\begin{equation}
\frac{dE}{dX} \approx -\frac{1}{X_0} \, \int_0^1 y\ E\,  
\left(y+\frac{4}{3}\,\,\frac{1-y}{y}\right) \,dy \, = -\frac{E}{X_0} \,.
\end{equation}

With this in mind, we now identify the constant $X_0 \approx 36.7$~g cm$^{-2}$ 
with the radiation length in air, defined as the mean distance over which a high-energy 
electron loses $1/e$ of its energy, 
or equivalently $7/9$ of the mean free path for pair production by a high-energy 
photon~\cite{Hagiwara:fs}.

The experimental confirmation of the LPM effect at Stanford Linear 
Accelerator Center (SLAC)~\cite{Anthony:1997ed,Klein:1998du} has motivated new analyses of 
its consequences in cosmic ray 
physics~\cite{Alvarez-Muniz:1997sh,Alvarez-Muniz:1998px,Cillis:1998hf,Capdevielle:jt,Plyasheshnikov:2001xw}.
The most evident signatures of the LPM effect on shower development are  
a shift in the position of the shower maximum $X_{\rm max}$ and larger 
fluctuations in the shower development.

\begin{figure} [t]
\postscript{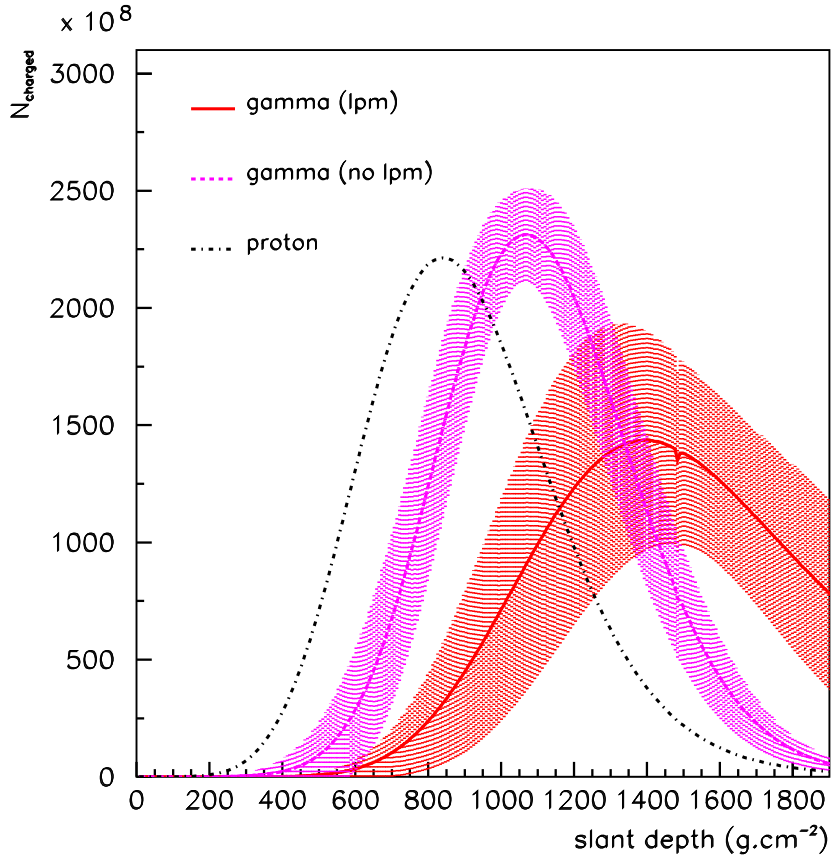}{0.70}
\caption{Average longitudinal shower developments 
of  $10^{11}$~GeV proton (dashed-dotted line) and $\gamma$-rays with 
and without the LPM effect (solid and dotted lines, respectively). The 
primary zenith angle was set to $\theta = 60^{\circ}$. 
The shadow area represents the intrinsic fluctuations of the showers. Larger 
fluctuations can be observed for $\gamma$-ray showers with the LPM effect, 
as expected.} 
\label{lpm} 
\end{figure}

When considering the LPM effect in the development of air showers produced by 
ultra high energy cosmic rays, one has to keep in mind that the 
suppression in the cross sections is strongly 
dependent on the atmospheric depth.\footnote{The same occurs for dielectric 
suppression, although the influence is not as important as 
for the LPM effect~\cite{Cillis:1998hf}.} Since the upper atmosphere is very thin the 
LPM effect becomes noticeable only for photons and electrons with energies above
$E_{\rm LPM} \sim 10^{10}$~GeV.  For baryonic primaries the LPM effect does not 
become important until the primary energy exceeds $10^{12}$GeV.  This is 
because the electromagnetic shower does not commence until after a significant
fraction of the primary energy has been dissipated through hadronic interactions.
To give a visual impression of how the LPM effect 
slows down the initial growth of high energy photon-induced showers,
we show the average longitudinal 
shower development of  $10^{10}$~GeV proton and $\gamma$-ray showers (generated using 
{\sc aires} 2.6.0~\cite{Sciutto:1999jh}) with and without the LPM effect in Fig.~\ref{lpm}.

At energies at which the LPM effect is important ({\it viz.}, $E > E_{\rm LPM}$), 
$\gamma$-ray showers will have already commenced in the geomagnetic field at almost 
all latitudes.  This reduces 
the energies of the primaries that reach the atmosphere, and thereby  
compensates the tendency of the LPM effect to retard the shower development.
The first description of photon interactions in the geomagnetic field 
dates back at least as far as 1966~\cite{Erber:1966vv},
with a punctuated revival of activity in the early 1980's~\cite{Mcbreen:yc}.
More recently, a rekindling of interest in the topic has led to 
refined calculations~\cite{Bertou,Vankov:2002cb,Homola:2003ru}.   
Primary photons with energies above  $10^{10}$~GeV convert into $e^{+}e^{-}$ 
pairs, which in turn emit synchrotron photons. Regardless of the primary energy, 
the spectrum of the resulting 
photon ``preshower'' entering the upper atmosphere 
extends over several decades below the primary photon energy, and is peaked 
at energies below  $10^{10}$~GeV~\cite{Bertou}.  The geomagnetic cooling
thus switches on at about the same energy at which the LPM effect does, and 
thereby preempts the LPM-related observables which would otherwise be evident.
Recent simulations~\cite{Risse:2004mx} which include photon preshowering indicate that 
above $\sim 10^{11}$~GeV this effect tends to accelerate the shower development,
shifting the $X_{\rm max}$ to a smaller value than previous calculation 
suggested~\cite{Halzen:1994gy,Anchordoqui:2000sm}, 
and into a range consistent with $X_{\rm max}$ typical of a proton primary.

\begin{figure} [t]
\postscript{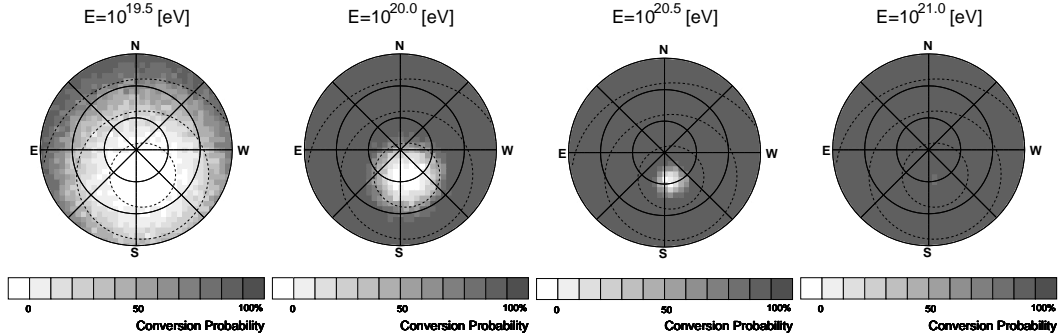}{0.85}
\caption{Maps of gamma ray conversion probability in the geomagnetic field for several primary 
energies. Azimuths are as labeled, ``N'' denotes true north. The inner circles correspond to 
zenith angles $30^\circ$, $60^\circ$ and 
horizon. Dashed curves indicate the opening angles of $30^\circ,$ $60^\circ$ and $90^\circ$ to the 
local magnetic field~\cite{Vankov:2002cb}.}
\label{map} 
\end{figure}

The relevant parameter to 
determine both conversion probability and synchrotron emission  is 
$E\times B_{\perp}$, where $E$ is the $\gamma$-ray energy and $B_{\perp}$ the transverse 
magnetic field. This leads to a large 
directional and geographical dependence of shower observables. Thus, each experiment has its own 
preferred direction for identifying primary gamma rays.  For instance,
Fig.~\ref{map} shows a map of the photon 
conversion probability in the geomagnetic field for 
all incident directions evaluated at the location of the 
HiRes experiment ($|\vec{B}| = 0.53$~G, $\iota = 25^{\circ},$ and 
$\delta = 14^{\circ}$)~\cite{Vankov:2002cb}.  The smallest probabilities for conversion are found, 
not surprisingly, around the direction parallel to the local geomagnetic field.  
Note that this conversion-free region shrinks rapidly with increasing primary energy.  
A similar evaluation for the Southern Site of the Pierre Auger 
Observatory  
($|\vec B| = 0.25$~G, $\iota = - 35^{\circ}$, and $\delta = 86^{\circ}$) can
be found in~\cite{Bertou}.

\subsubsection{Paper-and-pencil air shower modeling}
\label{ppm}

Most of the general features of an electromagnetic cascade can be understood in terms of the toy model 
due to Heitler~\cite{Heitler}. In this model, the shower is imagined to develop exclusively 
via bremsstrahlung and pair production, each of which results in the conversion of one particle into two.
As was previously discussed, these physical processes are characterized by an 
interaction length $X_0$. One can thus imagine the shower as a particle tree with branches that
bifurcate every $X_0$, until they fall below a critical energy, $\epsilon_0$, at which point
energy loss processes dominate. 
Up to $\epsilon_0$, the number of particles grows geometrically, so that after 
$n = X/X_0$ branchings, the total number of particles in the shower is 
$N \approx 2^n$.  At the depth of shower maximum $X_{\rm max}$, all particles are at 
the critical energy, $\epsilon_0$, and the energy of the primary particle, $E_0$, is 
split among all the $N_{\rm max} = E_0 / \epsilon_0$ particles.
Putting this together, we get:
\begin{equation} \label{heitler}
X_{\rm max} \approx X_0 \,\, \frac{\ln(E_0/\epsilon_0)}{\ln 2} \,\,.
\end{equation}
In real life, the combination of the LPM and geomagnetic effects introduces large fluctuations in the 
value of $X_{\rm max}$ for photon showers. The prediction of this toy model roughly 
lies within the range of these fluctuations. 

Even baryon-induced showers are dominated by electromagnetic processes, so 
this toy model is still enlightening for such cases.  In particular, for proton 
showers, Eq.~(\ref{heitler}) tells us that the $X_{\rm max}$ scales logarithmically
with primary energy, while $N_{\rm max}$ scales linearly. Moreover, to extend 
this discussion to heavy nuclei, we can apply the superposition principle as 
a reasonable first approximation. In this approximation, we pretend that the nucleus
comprises unbound nucleons, such that the point of first interaction of one nucleon is
independent of all the others.  Specifically, a shower produced by a nucleus with energy 
$E_{_A}$ and mass $A$ is modeled by a collection of $A$ proton showers, each with $A^{-1}$ of the
nucleus energy. Modifying Eq.~(\ref{heitler}) accordingly one easily obtains
$X_{\rm max} \propto \ln (E_0/A)$.
 
While the  Heitler model is very useful for imparting a first intuition regarding global shower properties, 
the details of shower evolution are far too complex to be fully described
by a simple analytical model. Full Monte Carlo simulation of interaction and transport of each individual
particle is required for precise modeling of the shower development. At present two Monte Carlo  
packages are available to simulate EAS: {\sc corsika} (COsmic Ray SImulation for 
KAscade)~\cite{Heck:1998vt} and {\sc aires} (AIR shower Extended Simulation)~\cite{Sciutto:1999jh}.  
Both programs provide fully 4-dimensional simulations of the air showers initiated by protons, photons, 
and nuclei. To simulate hadronic physics the programs make use of the event generators described in 
Sec.~\ref{hadronic}. Propagation of particles takes into account the Earth's curvature and geomagnetic field.
For further details on these codes the reader is referred to~\cite{Knapp:2002vs}.

As a bridge between the first order approximation just described and a full-blown Monte Carlo 
treatment of air shower cascades, a hybrid method has recently been presented~\cite{Drescher:2002cr}. 
The approach is as follows. The first few interactions are treated using Monte Carlo event 
generators. The second step in the approach utilizes one-dimensional cascade equations up to the 
point where the lateral spread of the particles becomes non-negligible, then the output of the cascade 
equations is treated again with Monte Carlo. The method shows a reasonable agreement when compared 
with results of the two detailed simulation packages~\cite{Drescher:2002vp}.

\subsubsection{Electron lateral distribution function}

The transverse development of electromagnetic showers is dominated by Coulomb 
scattering of charged particles off the nuclei in the atmosphere. The lateral 
development in electromagnetic cascades in different materials 
scales well with the Moli\`ere radius $r_{\rm M} = E_s\, X_0/\epsilon_0,$ which varies
inversely with the density of the medium,
\begin{equation} 
r_{\rm M}  = r_{\rm M}(h_{\rm OL}) \,\,\frac{\rho_{\rm atm}(h_{\rm OL})}{\rho_{\rm atm}(h)} \simeq 
\frac{9.0~{\rm g}/{\rm cm}^{2}}{\rho_{\rm atm}(h) } \,\,,\label{moliere}
\end{equation}
where $E_s \approx 21$~MeV~\cite{Hagiwara:fs} and the subscript ${\rm OL}$ indicates 
a quantity taken at a given observation level.

Approximate calculations of cascade equations in three dimensions to derive 
the lateral structure function for a pure electromagnetic cascade in vertical showers were 
obtained by Nishimura and Kamata~\cite{Kamata}, and later worked out by Greisen~\cite{Greisen} 
in the well-known NKG formula,
\begin{equation}
\rho_{}(r)  =  \frac{N_e}{r_{\rm M}^2} \, C \, \left( \frac{r}{r_{\rm M}}
\right)^{s_{_{\rm NKG}} -2} \left(1+\frac{r}{r_{\rm M}}\right) ^{s_{_{\rm NKG}}-4.5} \,\,,
 \label{nkg}
\end{equation}
where $N_e$ is the total number of electrons, $r$ is the distance from the shower axis, and
\begin{equation}
C  =  \frac{\Gamma (4.5 - s_{_{\rm NKG}})}{2\pi \,\,\Gamma(s_{_{\rm NKG}})\,\, 
\Gamma(4.5 -2 \, s_{_{\rm NKG}})} \,\, .
\end{equation}
For a primary of energy $E_0,$ the  so-called ``age parameter'', 
\begin{equation}
s_{_{\rm NKG}} = 3\, \left[1+\frac{2\, \ln(E_0/\epsilon_0)}{t} \right]^{-1} \,\,,
\end{equation} 
characterizes the stage of the shower development in terms of 
the depth of the shower in radiation lengths, {\it i.e.},
$t = \int_{z}^{\infty} \rho_{\rm atm}(z) \,\,dz/X_0.$

\begin{figure} [t]
\postscript{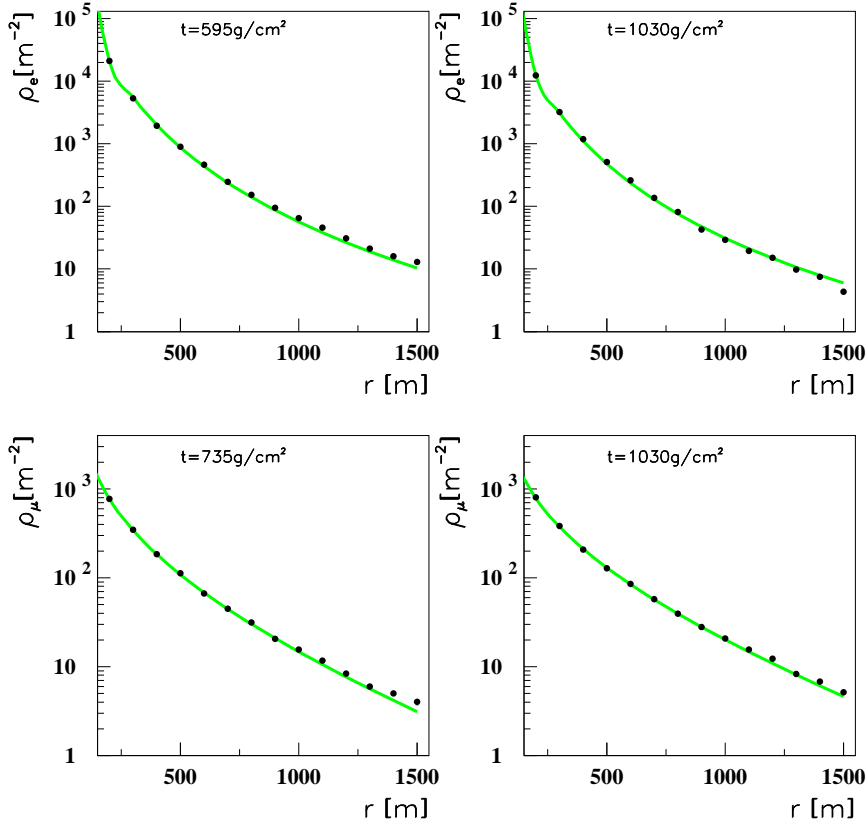}{0.70}
\caption{Electron (top) and muon (bottom) lateral distributions 
of a $10^{10}$~GeV vertical proton shower at different atmospheric 
altitudes.  The solid lines are fits to the data using the NKG-like
parametrization.  The error bars are, in all cases, 
smaller than the points~\cite{Dova:2001jy}.}
\label{ldf} 
\end{figure}

The NKG formula may also be extended to describe showers initiated
by baryons~\cite{Dova:2001jy}.  In such an extension, one finds 
a deviation of behavior of the Moli\`ere radius 
described in Eq.~(\ref{moliere}) when using a 
value of the age parameter which is derived from theoretical 
predictions for pure electromagnetic cascades.
The need for a different age parameter to reproduce the 
electromagnetic component of hadronic induced showers has been addressed 
experimentally~\cite{Linsley:denver,Aguirre,Porter,Kawaguchi,Nagano:zt,Yoshida:1994jf,Glushkov,Coy} 
and extensively studied by several authors~\cite{Coy,Bourdeau:yf,Plyasheshnikov}.
It is possible to generalize the NKG formula for the electromagnetic component of 
baryon-induced showers by modifying the exponents in Eq.~(\ref{nkg}).
From simulations, fits to lateral distribution functions (LDF) 
of electrons and positrons as a function of depth, $t$, yield an age parameter given by
\begin{eqnarray}   
s  = 3 \, \left(1+ \frac{2 \,\beta}{t}\right)^{-1} \,\,,\label{age2}
\end{eqnarray}
where the floating parameter $\beta$ takes into account the above mentioned deviations from the 
theoretical value $s_{_{\rm NKG}}$~\cite{Dova:2001jy}.

The modified NKG formula provides a good description of 
the $e^{+}e^{-}$ lateral distribution 
at all stages of shower development for values of $r$
sufficiently far from the hadronic core.
Fortunately, this is the experimentally interesting region, since typical
ground arrays can only measure densities at 
$r > 100 $~m from the shower axis, where detectors are not saturated.

To illustrate the validity of this parametrization, 
we show in Fig.~\ref{ldf} (top) the Monte Carlo $e^+ e^- $ density 
distributions  corresponding to 
a single $10^{10}$~GeV proton shower at selected  atmospheric depths.
The total number of electrons obtained from the fit to each 
single shower is slightly lower 
than the true value due to the invalidity of the parametrization 
close to the shower core. It should be mentioned that an NKG-like formula can be used to 
parametrize the total particle's density observed in baryon-induced showers~\cite{Roth:2003rs}.

In the case of inclined showers, one normally analyzes particle densities
in the plane perpendicular to the shower axis.  Simply projecting distributions
measured at the ground into this plane is a reasonable approach for near-vertical showers, but
is not sufficient for inclined showers.   In the latter case, 
additionally asymmetry is introduced because of both 
unequal attenuation of the electromagnetic components arriving at the ground earlier than and 
later than the core~\cite{Dova:2001jy}, and geometrical effects which also reduce the
early compared to the late 
flux~\cite{Bertou:GAP}. Moreover, deflections on the geomagnetic field become important for showers 
inclined by more than about $70^\circ.$

In the framework of cascade theory, any effect coming from the influence of the
atmosphere should be accounted as a function of the slant depth 
$t$~\cite{Kamata}. Following this idea, a LDF valid 
at all zenith angles $\theta < 70^{\circ}$ can be determined by considering
\begin{equation}
t'(\theta,\zeta)  =  t\,\sec\theta\,(1+ {\cal K}\,\cos\zeta)^{-1}\,,
\end{equation}
where $\zeta$ is the azimuthal angle in the shower plane, 
${\cal K} =  {\cal K}_0\,\tan\theta,$ and ${\cal K}_0$ is a constant extracted 
from the fit~\cite{Dova:2001jy,Dova:dq}.  Then, the  particle lateral distributions 
for inclined showers $\rho(r,t')$ are  given by 
the corresponding vertical LDF $\rho (r,t)$ but evaluated at slant 
depth $t'(\theta,\zeta)$ where the dependence on the azimuthal angle is evident.

\begin{figure} [t]
\postscript{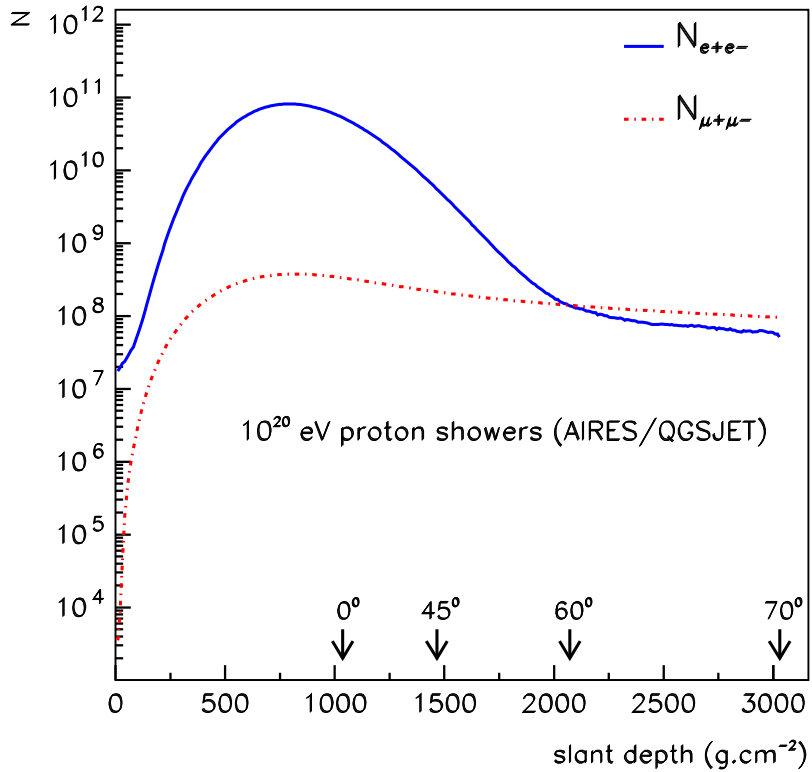}{0.70}
\caption{Longitudinal development of  muons and electrons as a function of the slant depth 
for $10^{11}$~GeV proton-induced showers.}
\label{longi} 
\end{figure}

For zenith angles $\theta > 70^{\circ}$, the surviving electromagnetic component at ground is 
mainly due to muon decay and, to a much smaller extent, hadronic interactions, pair production 
and bremsstrahlung. As a result the lateral distribution follows that of the muon rather closely.
In Fig.~\ref{longi} the longitudinal development of the muon and electron components are shown. It is 
evident from the figure that  for very inclined showers  
the electromagnetic development is due mostly to muon decay~\cite{Ave:2000dd,Cillis:ij}. 

The consequences of the LPM effect and pair production in the geomagnetic field on the longitudinal cascade 
distribution 
initiated by photons were already discussed in this section. Since the lateral distribution of particles is 
strongly correlated with the development of the shower in the atmosphere, the LPM effect has consequences
for the observed LDF at ground level.  In particular, unconverted photons result in  
large fluctuations and  steeper lateral profiles than nuclei~\cite{Bertou}.

In summary, the growth of the electromagnetic cascade is governed by 
bremsstrahlung and pair production.  The mean free path for interactions via
these processes depends on energy and atmospheric depth.  Below $10^{10}$~GeV,
each particle sees screened nuclei, while at higher energy collective effects
suppress the cross section.  On top of that, ultra high energy gamma ray
interactions in the geomagnetic field also come into play, reducing the 
importance of the LPM cross section suppression.  

The lateral distribution of the electromagnetic component of a shower can be effectively 
parametrized.  The well-known NKG lateral distribution function, which strictly applies to only 
purely electromagnetic showers, can be extended to describe not only the electromagnetic portion 
of baryon-induced showers but also the signal produced by all particles reaching ground level. This 
provides a handle on one of the most 
useful shower observables available to surface arrays.

\subsection{The muon component}

The muonic component of EAS differs from the electromagnetic component for two main reasons.
First, muons are generated through the decay of cooled 
($E_{\pi^\pm} \alt 1$~TeV) charged pions, and thus the muon content is sensitive to the initial
baryonic content of the primary particle.  Furthermore, since there is no ``muonic cascade'', the number
of muons reaching the ground is much smaller than the number of electrons. Specifically, there are
about $5\times 10^{8}$ muons above 10~MeV at ground level for a vertical $10^{11}$~GeV proton 
induced shower.
Second,  the muon has a much smaller cross section for radiation and pair production than the electron, 
and so the muonic component of EAS develops differently than does the electromagnetic component.
The smaller multiple scattering suffered by muons leads to 
earlier arrival times at the ground for muons than for the electromagnetic component.  

The ratio
of electrons to muons depends strongly on the distance from the core;
for example, the $e^+  e^-$ to  $\mu^+  \mu^-$ ratio for a $10^{11}$~GeV vertical proton 
shower varies from 17 to 1 at 200~m from the core to 1 to 1 at 2000~m.
The ratio between the electromagnetic and muonic shower components behaves somewhat differently
in the case of inclined showers.  For zenith angles greater than $60^{\circ}$, the 
$e^+  e^-$/$\mu^+  \mu^-$  ratio  
remains roughly constant at a given distance from the core.  As the zenith angle grows beyond $60^\circ$,
this ratio decreases, until at $\theta = 75^{\circ}$, it is 400 times smaller than for a vertical shower.
Another difference between inclined and vertical showers is that the average muon energy at ground changes 
dramatically.  For horizontal showers, the lower energy muons are filtered out by a combination of 
energy loss mechanisms and the finite muon lifetime: for vertical showers, the average muon energy 
is 1~GeV, while for horizontal showers it is about 2 orders of magnitude greater.
 The muon densities obtained
in shower simulations using {\sc sibyll} 2.1 fall more rapidly with lateral 
distance to the shower core
than those obtained using {\sc qgsjet} 01. 
This can be understood as a 
manifestation of the enhanced leading
particle effect in {\sc sibyll}, which can be traced to the relative 
hardness of the electromagnetic form factor profile function. 
The curvature of the distribution
$(d^2\rho_{\mu}/dr^2)$ is measurably different in the two cases, and,
with sufficient statistics, could 
possibly serve as a discriminator between hadronic interaction models, 
provided the primary species can be determined from 
some independent observable(s)~\cite{Anchordoqui:2003gm}.

\begin{figure} [t]
\postscript{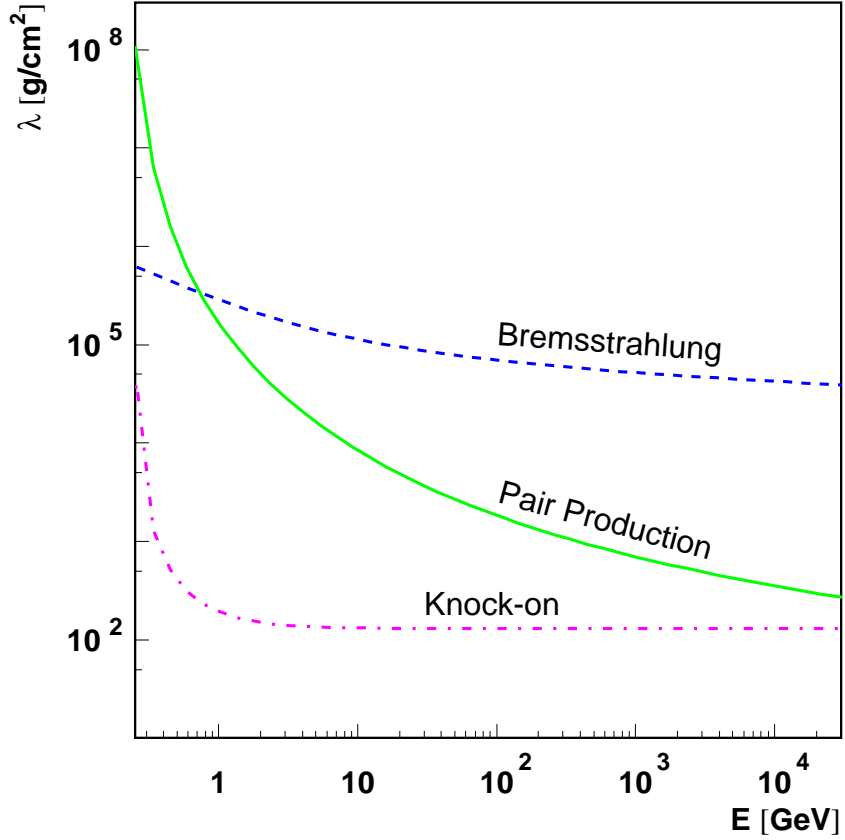}{0.70}
\caption{Mean free path in air for the different muonic interactions as a function of the initial kinetic 
energy. This figure is courtesy of Sergio Sciutto.} 
\label{sergio}
\end{figure}

High energy muons lose energy through $e^{+}e^{-}$ pair production, 
muon-nucleus 
interaction, bremsstrahlung, and knock-on electron ($\delta$-ray) 
production~\cite{Cillis:2000xc}.  The first three processes 
are discrete in the sense that they are characterized by high inelasticity and a large mean free path.  
On the other hand,  because of its 
short mean free path and its small inelasticity, knock-on electron production can be 
considered a continuous process. The muon bremsstrahlung cross section is suppressed by a factor of 
$(m_e / m_\mu)^2$ with respect to electron bremsstrahlung, see Eq.~(\ref{brem}).  Since the radiation 
length for 
air is about $36.7~\mathrm{g}/\mathrm{cm}^2$, and the vertical atmospheric depth is 1000~g/cm$^2$, 
muon bremsstrahlung is of negligible importance for vertical air shower development.  Energy loss due to 
muon-nucleus interactions is somewhat smaller than muon bremsstrahlung.    
As can be seen in Fig.~\ref{sergio}, energy  loss by pair production is slightly more important than 
bremsstrahlung at about 1~GeV, and becomes increasingly dominant with energy.  
Finally, knock-on electrons have a 
very small mean free path (see Fig.~\ref{sergio}), but also a very small inelasticity, so that this 
contribution to the energy loss is comparable to that from the hard processes.

In addition to muon production through charged pion decay, photons can directly generate muon pairs,
or produce hadron pairs which in turn decay to muons. In the case of direct pair production, the large 
muon mass leads to a higher threshold for this process than for electron pair production. 
Furthermore, QED predicts that $\mu^+ \mu^-$ production is suppressed by a factor 
$(m_{e}/m_{\mu})^2$ compared the Bethe-Heitler cross section. The cross section for hadron production by 
photons is much less certain, since it involves the {\it hadronic structure} of the photon. This has been 
measured at HERA for photon energies corresponding to $E_{ \mathrm{lab}} = 2 
\times 10^{4}$~GeV~\cite{Adloff:1997mf,Derrick:1992kk}.  This energy is still well below the energies of 
the highest energy cosmic rays, but nonetheless, these data do constrain the extrapolation of the cross 
sections to high energies.  To give an idea of the  rates, at 100~GeV
the cross section for $\gamma \rightarrow e^{+}e^{-}$ is $\approx $~650~mb, {\it i.e.} much larger than 
the cross sections for hadronic interaction ($\approx $ 1.4 mb) or for muon pair production 
($\approx $ 0.015 mb).

\subsubsection{Muon lateral distribution function}

Now we consider the lateral distribution of the muon component of an extensive air shower.
Unlike the electrons and photons, muons are relatively unaffected by multiple Coulomb
scattering, and so their lateral distribution function (LDF) retains some information about the parent 
pion trajectories.
In what follows, we first discuss a parameterization 
characterizing the muon LDF which is motivated only by the muon genealogy.
After that, we add to the discussion the effect of the geomagnetic field on the evolution 
of the lateral distribution.

One of the earliest parameterizations of the muon LDF in vertical showers was
empirically derived by Greisen~\cite{Greisen}.  This LDF was inspired by the
NKG parametrization, and is factorized into two terms,
\begin{equation}  
\rho_{\mu}(r) =  N_{\mu}(t) \,\,\,f_{\mu}(r) \,\,,
\end{equation}
where $N_{\mu}(t)$ gives the normalization as a function of depth $t$,
\begin{equation} 
\label{sf1}
f_{\mu}(r)  \approx   \left( \frac{r}{r_{\rm G}} \right)^{-0.75} \,\,\left( 
1+\frac{r}{r_{\rm G}} \right)^{-2.5}  
\end{equation}
is a structure function describing the lateral shape of the shower, and $r_{\rm G}=320$~m
is analogous to the Moli\`ere radius.

Later, Vernov and collaborators
proposed a semi-analytical form of the structure function~\cite{Vernov},
\begin{equation}  
\label{sf2}
f_{\mu}(r)  \approx \,\left(\frac{r}{r_0}\right)^{-\Gamma} \,\,\,\exp \left(-\frac{r}{r_0}\right)\,\,,
\end{equation}
with $\Gamma=0.4$ and $r_0=80$~m. Similar approaches were also suggested by Hillas' group 
at the University of Leeds~\cite{Hillas:buda} and by the SUGAR Collaboration~\cite{Brownlee}.
The slopes from Eqs.~(\ref{sf1}) and (\ref{sf2}) are in very good agreement with each other 
at intermediate distance, but Eq.~(\ref{sf2}) predicts a distribution which is flatter close
to the shower core and more strongly damped at large distances.
These LDFs have been used to fit 
experimental data. However, neither function reproduces
the whole radial range of an extensive air shower. This is a consequence
of neglecting the shower age in formulating the structure functions.
Very recently, the KASCADE Collaboration has used an NKG formula 
to fit muon density distributions~\cite{Antoni:2000mg}. The fits were performed 
close to the shower core ($r<200$~m) with non-conventional 
values of $r_{\rm M}$ and age parameter, $s$.

One expects there to be a dependence of the LDF parameters on the shower age.
However, in contrast to electrons, muons in an air shower are less attenuated and little
affected by Coulomb scattering, so the dependence of the LDF on 
the shower age is not the same as that exhibited by
the electromagnetic component.  The lateral growth of the 
shower is largely determined by the direction 
of emission of the parent particle and hence increases while the shower 
propagates downward. Two approaches for including shower age-dependence in the muon structure function have
been discussed in the literature~\cite{Alessio:je,Dova:2001jy}.  Here, we
consider the more recent treatment, in which a Vernov-like approach is used taking a
slope dependence on atmospheric depth, $\Gamma =  2 - s,$ with $s$ as given in Eq.~(\ref{age2}).
  
It is easily seen~\cite{Bosia:fw,Bergamasco:qm} that, if the parent particles are 
created with a $p_{_{T}}$ distribution,  $p_{{_T}}/
p_0 \exp(- p_{_{T}}/p_0) dp_{_{T}}/p_0 $, then  
the Vernov distribution at ground level has a value of $r_0$ given by
\begin{equation}  
r_0  =  \frac{2}{3}\,\, \langle h_p \rangle\,\,\frac{\langle p^{\mu}_{_{T}}\rangle }{\langle E_{\mu} 
\rangle},
\label{rmu}
\end{equation}
where $\langle p_{_{T}} \rangle= 2p_0$ 
is the mean transverse momentum, $\langle E_{\mu} \rangle$  the mean energy of muons
and  $\langle h_p \rangle$ the mean height of muon production. 
These approximate expressions can 
serve to calculate the variation with depth of the parameters characterizing 
the lateral spread.
The ratio $\langle p^{\mu}_{\perp}\rangle /\langle E_{\mu} \rangle$ can be considered constant while 
the shower develops~\cite{Ave:2000xs} and the variation of $r_0$ with altitude
is determined only by the dependence of $\langle h_p \rangle$ on depth, $t$. 

Muons are produced in every pion generation and their energy
distribution follows that of their parents. There are three
phenomena contributing to the 
behavior of $\langle h_p \rangle$ as a function of $t$.
The first is simply the dependence of the atmospheric density
on height and temperature. For an isothermal atmosphere of scale $h_0,$ one obtains 
$\langle h_p\rangle \propto h_0\, \ln(t/t_p)$.
The second phenomenon is the ``pionization'' process: the competition between 
pion production and decay.
The last contribution to the behavior of $\langle h_p \rangle$ is associated with
systematics induced by hadronic interaction models. In what follows, we leave aside the issue of  
systematic errors and as an example adopt {\sc qgsjet} 98 as the hadronic interaction model. 
Combining all these considerations, the characteristic radius
$r_0(t)$ becomes,
\begin{equation}  
r_0(t)  =  \frac{2}{3}\, \frac{\langle p^{\mu}_{_{T}} \rangle}{\langle E_{\mu} \rangle} \,\,\, 
h_0\,\,\, \frac{t_{\rm GL}}{t}\,\,\,\ln \left[\frac{t}{t_p} \right]\,\,,
\label{rmu2}
\end{equation}
where the subscript GL indicates 
that a quantity is given at ground level. For  $t = t_{\rm GL}$, Eq.~(\ref{rmu}) is recovered. 

In Fig.~\ref{ldf} we show the  $\mu^+ \mu^- $ 
density distributions for a single $10^{10}$~GeV proton shower 
at various depths.  Fits to the Vernov-like distribution are overlaid
on the simulation results, indicating validity of the parametrization~\cite{Dova:2001jy}.
Furthermore, the total number of muons $N_{\mu}$ from the fits agrees quite well 
with the corresponding values predicted by the Monte Carlo simulation, 
even though the fits are performed at core distances $r>100$~m.

Muons can travel long distances without interacting with the medium, and consequently the ground density 
profiles are significantly modified by the Earth's magnetic field. 
The global shower observables, like longitudinal 
and lateral distributions, are not affected by the geomagnetic field for zenith angles  
$\theta < 70^{\circ}$~\cite{Cillis:ij}. 
However, for the case of very inclined showers, which are dominated by muons, the density at ground 
is rendered quite asymmetric by the geomagnetic field.
In the remainder of this section we describe these effects quantitatively~\cite{Ave:2000xs}.

Consider a highly relativistic muon of energy $E_\mu \approx c p$ and transverse momentum $p_{_{T}}$ 
that travels a distance $d$ to reach ground.  
This muon suffers a deviation $r$ from the shower axis given by
\begin{equation}  
r \simeq  \frac{c\,\, p_{_{T}}\,\,d}{E_\mu} \,\,.
\label{ave1}
\end{equation}
Now, it is easily seen that if the energy spectrum of muons is taken as $\phi(E) = A\,E_\mu^{-\gamma}$,
the muon density is given by~\cite{Ave:2000xs}
\begin{equation}
\rho_\mu (r) = \frac{A}{2\pi}\,\, (c\,p_{_{T}}\,d)^{1-\gamma}\,\, r^{-3+ \gamma} \,\,.
\end{equation}

To take into account the effect of the geomagnetic field, define a cartesian coordinate, $(x,y),$ 
in the plane transverse to the shower axis, with $y$ aligned to $\vec{B}_{\perp}$. The circular 
symmetry of the shower 
is distorted depending on 
$\vec{B}_{\perp}$, the distance traveled by the muons, and 
their energy distribution. For very large zenith angles the pattern results in two lobes on each side of  
the shower axis, one for the negatively and one for the positively charged muon components. 
The magnetic deviation $\delta x$ experienced by muons of different charges
is~\cite{Ave:2000xs}
\begin{equation}  
\delta x =  \frac{e\,\, B_\perp\, d^2}{2p}\,\, ,
\label{ave2}
\end{equation}
where $e$ is the electron charge. Combining this with Eq.~(\ref{ave1}) 
we obtain,
\begin{equation}  
\delta x =  0.15\, \left(\frac{B_\perp}{{\rm T}}\right)\, \left(\frac{d}{{\rm m}}\right) \,
\left(\frac{p_{_{T}}}{{\rm GeV}} \right)^{-1}\,\, \overline r 
= \alpha \,\,\overline r \,\,,
\label{ave3}
\end{equation}
where $\overline r$ corresponds to the muon deviation in the transverse plane in the absence of a magnetic 
field, and $\alpha$ measures the ratio of displacement in the transverse plane due to $p_{_{T}}$ as 
well as the displacement due to the magnetic field. The density of muons in the transverse plane can be 
obtained by making the transformation
\begin{equation}  
x =   \overline x + \alpha \sqrt{\bar x^2 +\bar y^2}\,, \,\,\,\,\,\,\,\,
y =   \overline y \,\,,
\label{ave4}
\end{equation}
where the barred and unbarred coordinates indicate the position of the muon in the transverse plane in 
the absence and presence of the geomagnetic field, respectively.
The muon number density reads
\begin{equation}
\rho_\mu (x,y) = \overline \rho_\mu 
(\overline x,\overline y) \,\,\left[\frac{ \partial(\overline x\overline y)}{\partial(x y)}\right]\,\,,
\label{transf}
\end{equation}
where $\overline \rho_\mu (\overline x,\overline y)$ is the density 
at a distance $\overline r = (\overline x^2 + \overline y^2)^{1/2}$ in the case $\vec B = 0$ and the 
last factor is the Jacobian of the transformation.

\begin{figure} [t]
\postscript{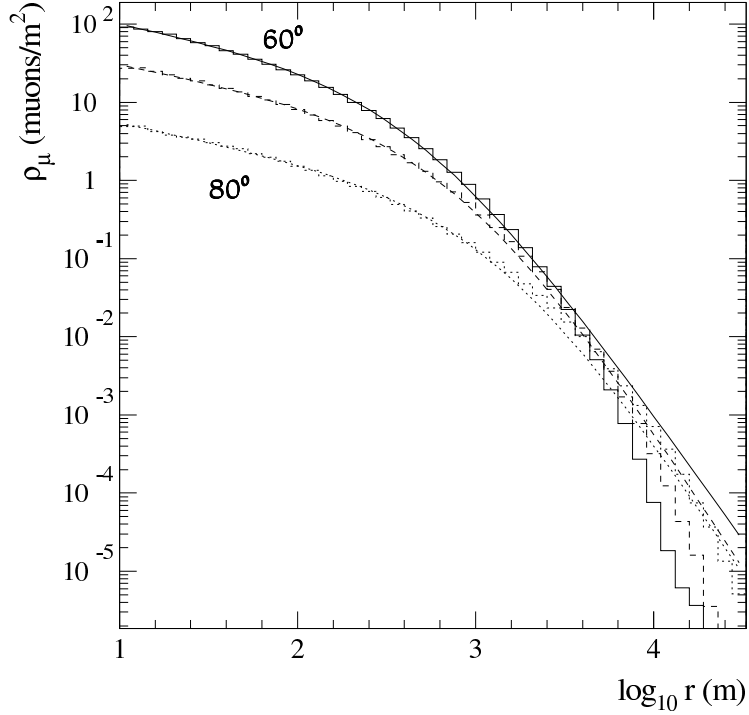}{0.60}
\caption{Lateral distribution of muons from {\sc aires} simulations superimposed over the best
fits obtained using Eq.~(\ref{14}). From top to bottom the curves correspond to $60^\circ$, 
$70^\circ$ and $80^\circ$~\cite{Ave:2000xs}.} 
\label{gzk2}
\end{figure}

In a realistic situation the transverse position of the muon $\overline r$
is affected by both multiple scattering and the transverse position of the parent 
pions. Following~\cite{Ave:2000xs}, to account for this effect we use Eq.(\ref{ave1}), 
setting $d$ to a constant and we 
assume that at a given $\overline r$ there is an energy distribution.  
For convenience one introduces the variable $\epsilon= \log_{10}E_\mu,$ such that
\begin{equation}
\langle \epsilon \rangle = A - \gamma  \log_{10} \overline r \,.
\label{epsilon}
\end{equation}
The muon density is taken to be 
\begin{equation}  
\rho_\mu (\overline r, \epsilon)   =  
P(\epsilon;\langle \epsilon \rangle, \sigma)\,\, \rho_\mu(\overline r) \,\,,
\end{equation}
where $P$ is a distribution of mean $\langle \epsilon \rangle$ and standard deviation $\sigma$. 
Now, one obtains the muon number density in the coordinate system $(x,y),$ by using Eq.~(\ref{transf}), 
\begin{equation}  
\rho_\mu (x,y)    =  
\int d \epsilon \,\,P(\epsilon;\langle \epsilon \rangle, \sigma) \,\,\rho_\mu(\overline r) \,\,,
\label{ave5}
\end{equation}
where
\begin{equation}  
\bar r  =  \left[\left(x- \frac{e\, B_\perp\, d^2 \,c}{2E_\mu}\right)^2 + y^2 \right]^{1/2}\,\,.
\end{equation}
The muon number density given in Eq.~(\ref{ave5}) depends on 3 quantities: 
{\it (i)} the distribution of $\epsilon$ that hereafter is taken as a Gaussian with mean 
given by Eq.~(\ref{epsilon})  and $\sigma \sim 0.4,$
{\it (ii)} the effective distance to the production point $d$, and
{\it (iii)} the lateral distribution function of the muons in the transverse plane. 
Figure~\ref{gzk2} shows fits~\cite{Ave:2000xs} to the lateral distributions at different zenith angles 
using the NKG-like LDF, 
\begin{equation}
\rho(\overline r) = {\cal N}\, \overline r^{-\psi}\,\,
\left[ 1 + \frac{\overline r}{{\cal R}}\right]^{-\kappa}\,\,,
\label{14}
\end{equation}
with $N,$ $\psi,$ $\kappa,$ ${\cal R}$ as given in Table~\ref{t2}, and $d$  taken as 16~km, 32~km, and 
88~km, for $\theta$ = $60^\circ$, $70^\circ,$ and $80^\circ$, respectively. 
One can see from the figure that the parametrization reproduce the simulation 
quite well up to a core distance of 1~km.

\begin{table}
\caption{Best values for the parameters in Eqs.~(\ref{epsilon}) and (\ref{14}) as obtained from fits 
to Monte Carlo simulations, using {\sc sibyll} 1.6 to process the hadronic interactions.}
\begin{center}
\begin{tabular}{ccccccc}
\hline \hline
$\theta$  & \hspace{1cm} $A$ \hspace{1cm}   & \hspace{1cm} $\gamma$ \hspace{1cm}  & \hspace{1cm} ${\cal N}$ \hspace{1cm}  & 
\hspace{1cm} $\psi$ \hspace{1cm} & \hspace{1cm} $\kappa$ \hspace{1cm} &  ${\cal R}$ \\ \hline
$60^\circ$ & $2.67 \pm 0.23$ & $0.75 \pm 0.1$ & 569.9 & 0.52  & 4.05 & 782.8~m \\
$70^\circ$ & $4.04 \pm 0.20$ & $0.73 \pm 0.06$ & 227.1 & 0.49  & 4.35 & 1010.0~m \\
$80^\circ$ & $3.63 \pm 0.20$ & $0.81 \pm 0.07$ & 78.4 & 0.52  & 4.49 & 1513.0~m \\
\hline \hline
\end{tabular}
\end{center}
\label{t2}
\end{table}

\subsubsection{Muon content of the shower tail}

As discussed in the previous section, once the shower particle energies fall below 
$\epsilon_0$, ionization losses take over from other 
electromagnetic processes, and the number of electrons and photons in the shower begins to 
decrease, while the number of muons remains more-or-less the same.
We will refer to this region of the shower as the ``tail.''  Most 
ground arrays are located below the altitude at which $X_{\rm max}$ occurs, even in the case of vertical 
showers induced by ultra high energy primaries.  This means that ground arrays observe predominantly shower
tails. In this section we describe the variation of the shower tail's muon content with energy, and 
compare the original calculations of Hillas from the early 1970's with more recent detailed Monte Carlo
simulations.

\begin{figure}[tbp]
\begin{minipage}[t]{0.49\textwidth}
\postscript{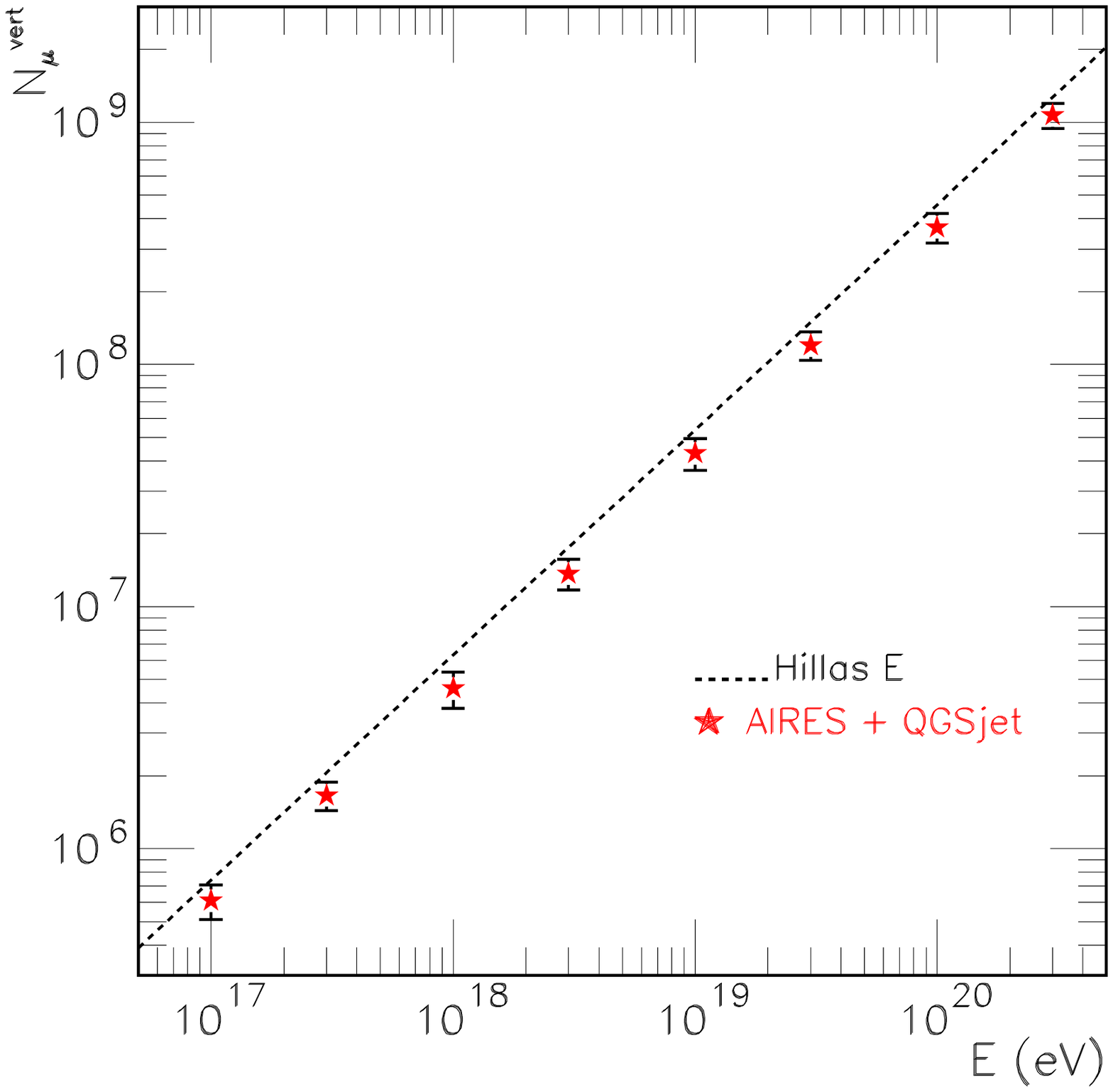}{0.99}
\end{minipage}
\hfill
\begin{minipage}[t]{0.49\textwidth}
\postscript{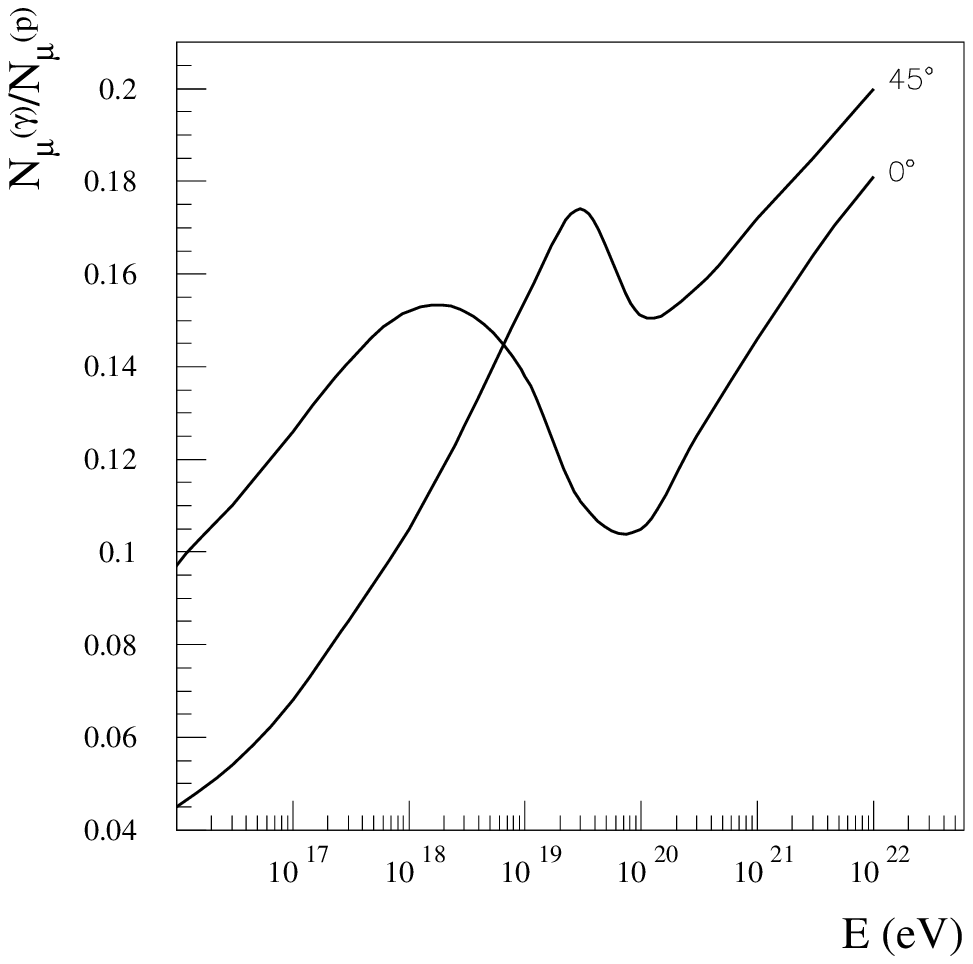}{1.}
\end{minipage}
\caption{Left panel:
Total number of muons at ground level as a function of the shower energy. 
 The dashed line indicates the Hillas parametrization for model ''E'',  
with a threshold energy set to that of the SUGAR experiment~\cite{Winn:un}.
(For vertical showers considering the SUGAR's energy threshold, one obtains an exponent 
$0.93$ rather than the $0.94$ used in the text.)
The stars illustrate the results obtained from
simulations carried out with {\sc aires + qgsjet} 01, assuming proton
primaries. The particles were injected vertically and the observation level was placed at 
250~m above sea level. Muons
with energies below the threshold 0.75~GeV are not taken into
account in the simulations~\cite{Anchordoqui:2003gm}.  If the hadronic 
interactions are modeled with {\sc sibyll} 1.6 rather than {\sc qgsjet} 01, an
exponent of $0.88$ best fits the simulation~\cite{Ave:2001xn}.
Right panel: Ratio of the muon content for EAS produced by primary gammas and protons. 
The geomagnetic 
field is set to the PAO Southern site~\cite{Plyasheshnikov:2001xw}.}
\label{hillas}
\end{figure}

Muons are produced when a shower has cooled sufficiently to allow
pions to decay before they interact (recall that the probability for decay of a 1~TeV pion in the atmosphere
is somewhat less than 10\%).  
Hillas' group at the University of Leeds reviewed various models for this cooling process and
analyzed their consistency with data from emulsion experiments as well as cosmic air shower 
observations~\cite{Hillas:1,Hillas:2}.
They found the data at ground level to be best reproduced by the model (so-called ``E'') which  
predicted that for $\theta = 14^\circ,$ the number of muons in proton showers scales as 
\begin{equation} \label{hs}
N^p_\mu \propto E^{0.94} \,\,.  
\end{equation}
Interestingly, this result differs only by an offset in the normalization when compared to 
the prediction from full-blown modern-day Monte Carlo simulations, as shown in Fig.~\ref{hillas}. 
Of course, the exponent in Eq.~(\ref{hs}) varies with zenith angle.
It is possible to take into account the zenith angle dependence either 
through the ``constant intensity cut'' method~\cite{Alvarez-Muniz:2002xs},
or by simply
determining the behavior of the exponent as a function of zenith angle~\cite{Ave:2000dd}.
Furthermore, it has been recently noted that this exponent is more accurately
taken to have a logarithmic energy dependence~\cite{Alvarez-Muniz:2002ne}.  

The muon content of EAS at ground level $N_\mu,$ as well as the ratio $N_{\mu}/N_{e}$, 
are sensitive to primary composition (here, $N_e$ is the electron content at ground level).
To estimate the ratio of the  muon content of nucleus induced to proton 
induced showers, we can resort again to the principle of superposition. 
Using Eq.~(\ref{hs}) we find that the total 
number of muons produced by the superposition of $A$ individual proton showers 
is, $N_{\mu}^A \propto A (E_{_A}/A)^{0.94}$. Consequently, in a vertical shower, 
one expects a cosmic ray nucleus to produce about $A^{0.06}$  more 
muons than a proton.  This implies that a shower initiated by 
an iron nucleus produces about 27\% more muons than a proton shower.  Note, however,
that a change in the hadronic interaction model could produce a much larger effect than a 
change in the primary species.   For example, replacing {\sc qgsjet} 01 with {\sc sibyll} 1.6 
as the hadronic interaction model leads to a prediction of 60\% more muons for an 
iron shower than for a proton shower~\cite{Ave:2000dd}.

The situation for gamma-induced showers is a bit different.  In this case
the muon component of the shower does not simply follow
Eq.~(\ref{hs}) because of the LPM and geomagnetic field effects~\cite{Plyasheshnikov:2001xw}. 
Competition between the two processes leads to a complex behavior in $N_{\mu}^{\gamma}/N_{\mu}^p$, as 
shown in Fig.~\ref{hillas}.

In this section we have described the four main energy loss mechanisms for muons en route through the 
atmosphere.  The rate of energy attenuation for muons is much smaller than it is for electrons, 
and the energy loss processes are only really of interest in the case of extremely inclined showers 
for which the original electromagnetic component is mostly absorbed.  In such cases, small electromagnetic
sub-showers can still arise from bremsstrahlung, pair production and knock-on electrons.
In addition, muon-nucleus interactions induce hadronic sub-showers.  We also discussed the
effect of different energy loss mechanisms on the electron and muon distributions in time and space.
Because they are less subject to multiple scattering, muons tend to arrive at the ground
earlier and more compressed in time.  The ratio of muons to electrons far from the core
is much greater than it is near the core, and this effect is more pronounced at higher zenith angles.

The muon content of the shower tail is quite sensitive to unknown details of 
hadronic physics.  This implies that attempts to extract composition information from measurements
of muon content at ground level tend to be systematics dominated. 
The muon LDF is mostly determined by the distribution in phase space of the parent
pions.  However, the pionization process together with muon deflection in the geomagnetic field obscures 
the distribution of the first generation of pions. 
A combination of detailed simulations, high
statistics measurements of the muon LDF  and
identification of the primary species using uncorrelated observables 
could shed light on hadronic interaction models. 

\section{Fingerprints of primary species in EAS}
\label{discrimination}

A determination of primary composition is invaluable in revealing the origin of 
cosmic rays as this information would provide important bounds on sources and on possible production 
and acceleration mechanisms. 
In addition, a proper interpretation of anisotropy information requires knowledge of the primary mass 
due to the influence on propagation of the galactic and intergalactic magnetic fields.
Attempting to determine the primary composition of cosmic rays by measuring various shower parameters 
is fraught with systematic uncertainty. Furthermore, because of the stochastic nature of the 
extensive air showers, there are inherent shower-to-shower fluctuations in measured shower observables
that cannot be attributed to experimental systematic error alone.
Therefore, the determination of primary composition on an event-by-event basis 
is an intractable problem. Nevertheless, statistical analyses 
of shower observables known to correlate with the primary composition are possible. 
Based on the general signatures of the EAS described in previous sections, 
we provide a summary of the observables that help to separate primary species.

\subsection{Photon showers}

In this section we provide an overview of how the EAS characteristics described in the previous
sections allow one to distinguish photon primaries from other species. 
As discussed in section~\ref{EM} photon-induced showers are expected 
to generate fewer muons than baryon-induced showers. This clear signature can be 
exploited by surface arrays which are equipped with dedicated muon detectors or 
are capable of distinguishing muons using shower observables sensitive to the muon content.
The AGASA Collaboration~\cite{Shinozaki,Shinozaki:ve} has used the muon content of the detected EAS to 
set bounds on the percentage of photon primaries present in the observed flux. AGASA comprises 111 stations covering an area of 100 km$^2$. Detectors of $2.8 - 10~{\rm m}^2$ area, capable of measuring muon 
densities up to $\approx 10$ m$^{-2}$, were deployed in 27 stations.   
The analysis of the AGASA Collaboration, which takes into account the LPM effect and 
conversion in the geomagnetic field, shows that at the 95\% CL, the fraction 
of $\gamma$-rays above $10^{10}$~GeV, $10^{10.25}$~GeV, and $10^{10.5}$~GeV is less than 
34\%, 59\%, and 63\%, respectively.  Of course, these 
bounds depend on the hadronic interaction models used to simulate the showers.
Several models were used in this analysis, and the reported limits are the
least restrictive ones.

\begin{figure} [t]
\postscript{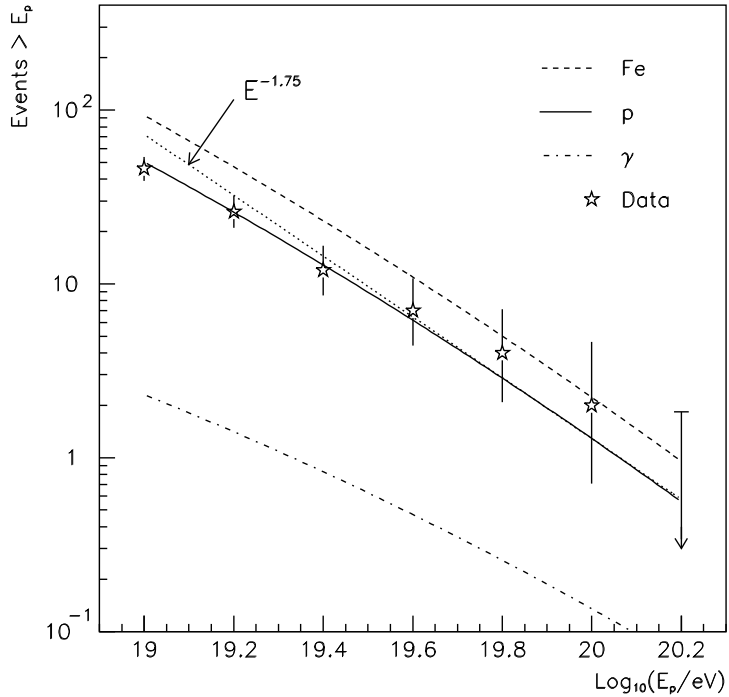}{0.60}
\caption{Integral number of inclined events as a function of energy for the Haverah Park data set 
compared with the predictions for iron, proton, and photon primaries. Here the energy is calculated 
assuming a proton primary. The slope of the assumed primary spectrum $\propto E^{-1.75}$ is shown 
to illustrate the increase of triggering efficiency with energy~\cite{Ave:2000nd}.}
\label{rate-nw}
\end{figure}

 Another powerful tool for discriminating between photons and baryons using data collected by surface detectors 
relies on comparing the flux of vertical showers to that of inclined
showers, a technique which
exploits the attenuation of the electromagnetic shower
component for large slant depths. 
As an illustration of this technique, we describe below
the constraints on the gamma ray flux obtained from 
Haverah Park measurements~\cite{Ave:2000nd}. 
The first crucial ingredient in the analysis is the vertical flux normalization.  This should 
be determined in a way which is free from systematic uncertainties associated with 
the primary composition.  Fluorescence detectors, which record ``calorimetric''
measurements,  provide the best tool to attain this normalization, and in the 
analysis described here the data from Fly's Eye~\cite{Bird:wp} were used~\cite{Ave:2001sd}. 
From this known vertical spectrum and Monte Carlo simulations of
shower propagation and detector response, 
a prediction can be made for the expected rate of inclined
events for each type of primary. 
For inclined showers in the zenith angle range $60^\circ < \theta < 80^\circ$, the 
Haverah Park experiment collected 46 
events with energy above $10^{10}$~GeV and 7 events above $10^{10.6}$~GeV. 
A Comparison of these observations to the results extracted from simulations is shown
Fig.~\ref{rate-nw}. If one assumes the primary spectrum comprises a mixture 
of protons and photons, then the Haverah Park data imply that 
above $10^{10}$~GeV, less 
than 30\% of this admixture can be photons,  and 
above $10^{10.6}$~GeV less than 55\%  can be 
photons. Both of these statements are made at the 95\% CL~\cite{Ave:2000nd}. 
Even though Fly's Eye provides a flux measurement which is independent of 
the mass composition, one should be aware of 
the inherent systematic uncertainties in aperture 
estimates of fluorescence detectors.  A separate normalization technique
using both fluorescence and surface array data leads to bounds within 
20\% of the previous estimates~\cite{Ave:2001xn}.   

The sensitivity of PAO for isolating gamma ray primaries
using this method was estimated in~\cite{Ave:2002ve}.
Given the huge statistics -- above $60^\circ$ the aperture
of the observatory is increased by almost a factor 2 -- PAO will
place severe constraints 
on the photon content of the observed flux. 
Additionally, hybrid techniques available to PAO will facilitate
independent cross-checks on the systematic uncertainties.

The effect of the geomagnetic field on 
photons also leads to an energy dependence and characteristic anisotropy  in the directional 
distribution of the fraction of events with abnormally deep profiles which are not compatible 
with proton or nuclei primaries. This technique has not yet been implemented in the analysis of real data, 
but the potential for the HiRes and PAO experiments has been evaluated ~\cite{Vankov:2002cb,Bertou}.

\subsection{Hadronic primaries} 

\label{HPries}

\begin{figure} [t]
\postscript{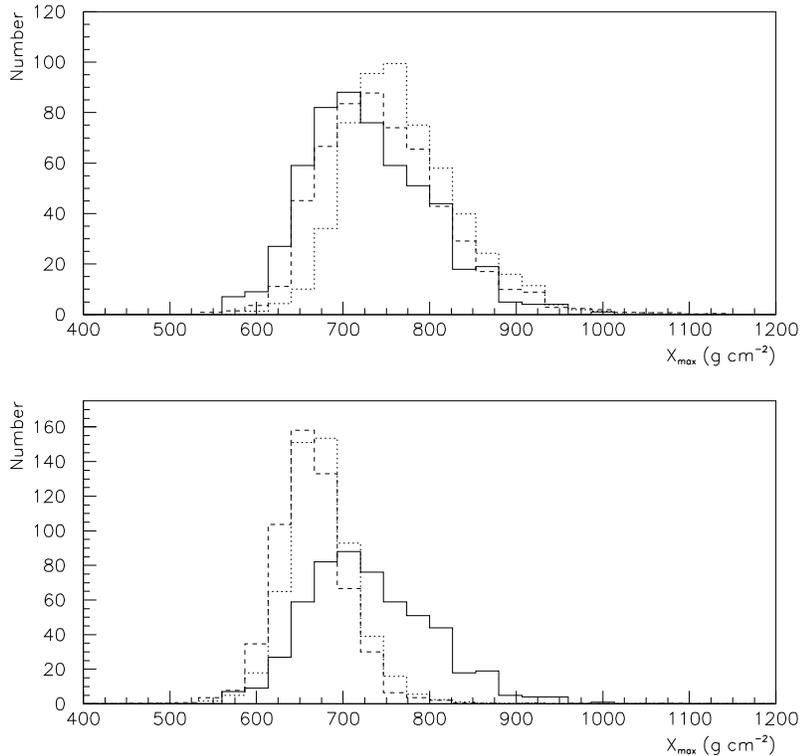}{0.70}
\caption{Distribution of observed $X_{max}$ from HiRes stereo data for showers in the
energy range $10^{9.5} - 10^{10.4}$~GeV (solid histogram). The predictions for proton (upper figure) 
and for iron (lower figure) are given for {\sc qgsjet} 01 (dashed histogram) and {\sc sibyll} 2.1 
(dotted histogram) 
This figure is courtesy of Greg Archbold.}. 
\label{hires-xmax}
\end{figure}

We now discuss how baryonic species may, to some extent, be distinguished by the
signatures they produce in the atmosphere.  As in the previous section, we consider both surface array
and fluorescence detector observables.

As mentioned in Sec.~\ref{EM}, the main purpose of fluorescence detectors is to measure the properties of the longitudinal development. The shower longitudinal profile can be parameterized using the Gaisser-Hillas 
function~\cite{Gaisser:icrc}
\begin{equation}
N_e (X) = N_{e,{\rm max}} \left(
\frac{X - X_1}{X_{\rm max}-X_1}\right)^{[(X_{\rm max} - X_1)/\lambda]}\,\,
\exp\left\{\frac{X_{\rm max} - X}{\lambda}\right\},
\,\,\, X \geq X_1,
\end{equation}
where $N_{e,{\rm max}}$ is the size at the maximum, $X_1$ is the depth of the
first observed interaction, and $\lambda$ is a floating parameter in the fit, 
generally fixed to $70~{\rm g/cm^2}$. Using this parametrization, fluorescence detectors can measure
$X_{\rm max}$ with a statistical precision typically around $30~{\rm g} / {\rm cm}^2$. 

The speed of shower development is the clearest indicator of the primary composition. It was 
shown in Sec.~\ref{EM} using the superposition model that  there is a difference between the 
depth of maximum in proton and iron induced showers. In fact, nucleus-induced showers develop faster, 
having $X_{\rm max}$ higher in the atmosphere. From Monte Carlo simulations, one finds
that the difference between the 
average $X_{\rm max}$ for protons and iron nuclei is about 90 -- 100~g/cm$^2$. However, 
because of shower-to-shower fluctuations, it is not possible to obtain meaningful 
composition estimates from $X_{\rm max}$ on a shower-by-shower basis, though one can derive composition
information from the magnitude of the fluctuations themselves.    
For protons, the depth of first interaction fluctuates more than it does for iron, and 
consequently the fluctuations of $X_{\rm max}$ are larger for protons as well.  Specifically,
the standard deviation $\sigma (X_{\rm max})$ is  $53~{\rm g} / {\rm cm}^2$ for protons and
$22~{\rm g} / {\rm cm}^2$ for iron~\cite{Cronin:2004ye}. These fluctuations depend only weakly on the
choice of interaction model.    
The HiRes Collaboration has recently analyzed their stereo data sample in 
the energy range $E = 10^{9.5} - 10^{10.4}$~GeV~\cite{Archbold}.  The results 
are shown in Fig.~\ref{hires-xmax},
together with the expected distributions using several hadronic interaction models.  While
agreement between data and Monte Carlo is not perfect for any of the models, the data do 
qualitatively suggest a proton dominated composition.

\begin{figure}[tbp]
\begin{minipage}[t]{0.49\textwidth}
\postscript{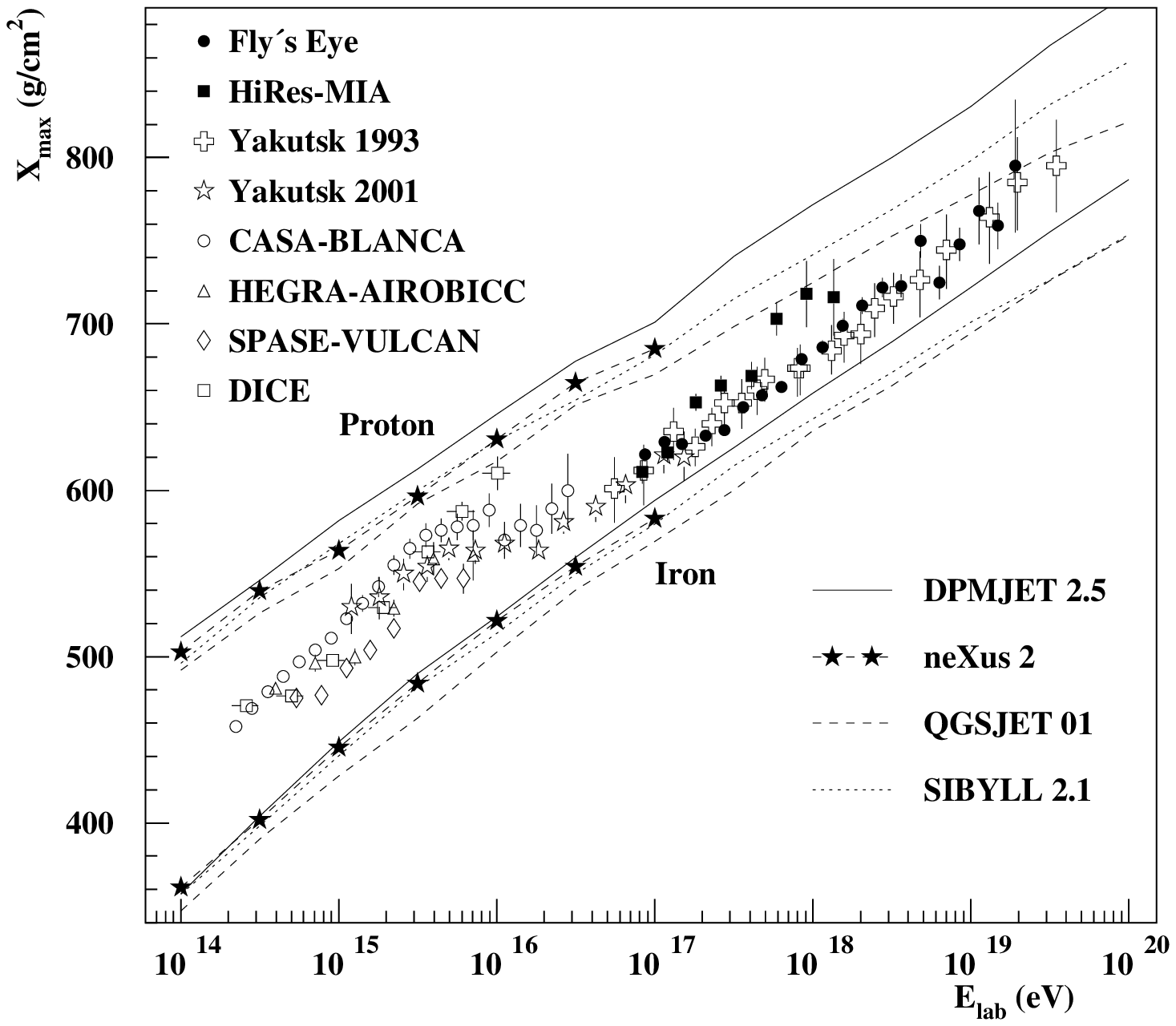}{0.99}
\end{minipage} 
\hfill
\begin{minipage}[t]{0.49\textwidth}
\postscript{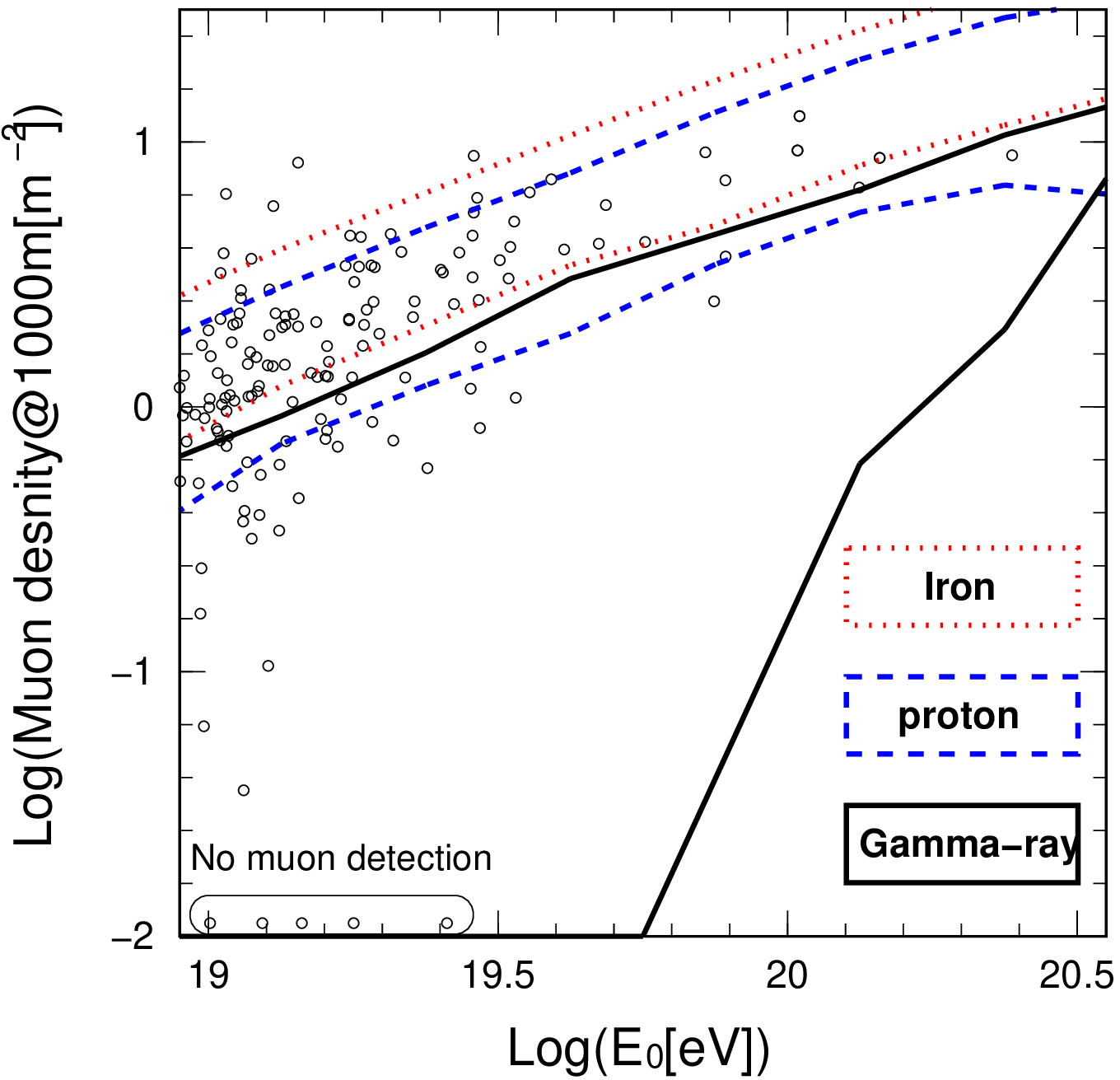}{0.90}
\end{minipage}
\caption{Left panel: Variation of $X_{\rm max}$ with energy as seen by different experiments:
Fly's Eye~\cite{Bird:yi}, HiRes-MIA~\cite{Abu-Zayyad:1999xa,Abu-Zayyad:2000ay}, 
HEGRA-AIROBICC~\cite{Arqueros:1999uq}, CASA-BLANCA~\cite{Fowler:2000si}, DICE~\cite{Swordy:1999um}, 
SPASE-VULCAN~\cite{Dickinson}, and YAKUTSK~\cite{Knurenko}. The rising curves indicate simulated 
results for proton and iron primaries using various hadronic interaction 
models~\cite{Knapp:2002vs}. Right panel: The circles indicate the experimental measurements of
$\log \rho_{\mu}$ at 1000 m from the core {\em vs}. logarithm of the primary energy. The 
lines indicate the $1\sigma$ band for iron, proton, and photon predictions~\cite{Shinozaki}.} 
\label{elong}
\end{figure}

Changes in the mean mass composition of the cosmic ray flux as a function of energy will 
manifest as changes in the mean values of $X_{\rm max}$. This change of $X_{\rm max}$ 
with energy is commonly known as the elongation rate theorem~\cite{Linsley:P3}:
\begin{equation} 
D_e = \frac{\delta X_{\rm max}}{\delta \ln E} \,\,.
\end{equation}
For purely electromagnetic showers, $X_{\rm max}(E) \approx  X_0\, \ln(E/\epsilon_0)$ and
then the elongation rate is $D_e \approx X_0$. For proton primaries, the multiplicity rises 
with energy, and thus
the resulting elongation rate becomes smaller.  This can be understood by noting that, on average,
the first interaction is determined by the proton mean free path in the atmosphere, $\lambda_N$.
In this first interaction the incoming proton splits into $\langle n(E) \rangle$ secondary particles, 
each carrying an average energy $E/\langle n(E) \rangle$. Assuming that $X_{\rm max}(E)$ 
depends logarithmically on energy,
as we found with the Heitler model described in Sec.~\ref{EM}, it follows that, 
\begin{equation}
X_{\rm max}(E)=\lambda_N + X_0\, \ln[E/\langle n(E) \rangle]\,\, .
\end{equation}
If we assume a multiplicity dependence $\langle n(E) \rangle \approx n_0 E^{\Delta}$, then  
the elongation rate becomes,
\begin{equation} 
\frac{\delta X_{\rm max}}{\delta \ln E}= X_0\,\left[1-\frac{\delta \ln \langle n(E) 
\rangle}{\delta \ln E} \right] + \frac{\delta \lambda_{N}}{\delta \ln E}
\end{equation}
which corresponds to the form given by Linsley and Watson ~\cite{Linsley:gh},
\begin{equation} 
D_e = X_0 \,\left[ 1-\frac{\delta \ln \langle n(E) \rangle}{\delta \ln E} + 
\frac{\lambda_{N}}{X_0} \frac{\delta \ln(\lambda_{N})}{\delta \ln E} \right] =  X_0\,(1-B) \,\,.
\label{ERLW}
\end{equation}
Using the superposition model introduced in Sec.~\ref{EM} and assuming that 
\begin{equation}
B \equiv \Delta - \frac{\lambda_{N}}{X_0} \,\,\frac{\delta \ln \lambda_{N}}{\delta \ln E}
\label{theB}
\end{equation} 
is not changing with energy, one obtains for mixed primary 
composition~\cite{Linsley:gh}
\begin{equation}  
D_e =\, X_0\,(1-B)\,
\left[1 - \frac{\partial \langle \ln A \rangle }{\partial \ln E} \right]\, .
\label{er}
\end{equation}
Thus, the elongation rate provides a measurement of the change of the 
mean logarithmic mass with energy.  One caveat of the procedure discussed above is 
that Eq.~(\ref{ERLW}) accounts for the energy dependence of the cross section and violation of Feynman 
scaling only for the first interaction. Note that subsequent interactions are assumed to be 
characterized by Feynman scaling and constant interaction cross sections, see Eq.~(\ref{theB}).  
Above $10^{7}$~GeV, these secondary interactions play a more important role, and thus the predictions of 
Eq.~(\ref{er}), depending on the hadronic interaction model assumed, may
vary by up to 20\%~\cite{Alvarez-Muniz:2002ne}.

For convenience, the elongation rate is often written in terms of energy decades,
$D_{10} = \partial \langle X_{\rm max} \rangle/\partial \log E$, where $D_{10} = 2.3 D_e.$ 
In Fig.~\ref{elong} we show the variation of $\langle X_{\rm max} \rangle$ with primary energy 
as measured by several experiments together with predictions from a variety of hadronic interaction models. 
For protons and iron, 
Monte Carlo simulations indicate $D_{10} \approx 55~{\rm g}/{\rm cm}^2$ and for photons
$D_{10} \approx 84~{\rm g}/{\rm cm}^2$~\cite{Cronin:2004ye}. 
Recent results presented by the HiRes Collaboration~\cite{Archbold} using stereo data favours 
a light component above $10^{9}$~GeV, and the reported 
variation of $\langle X_{\rm max} \rangle$ with logarithm of primary energy is 
$D_{10} = 54.5 \pm 6.5$, consistent with a constant or slowly changing composition between  
$10^{9}$~GeV and $10^{10.4}$~GeV.

As an attentive reader should know by now, the muon content of the shower is 
strongly sensitive to the nature of the primary.
The AGASA Collaboration used measurements of the muon component to discern
the primary composition~\cite{Shinozaki}. For events 
with estimated energy $> 10^{10}$~GeV, and zenith $\leq 36^\circ$, the muon density at 
1000~m from the shower core was used to estimate the primary mass.  Expected muon densities for 
iron and proton primaries were estimated from Monte Carlo simulations, and comparison of the 
expected to observed densities suggests a dominance of light composition. Specifically,
above $10^{10}$~GeV the average fraction of iron
is $14^{+16}_{-14}$\%, rising to  $30^{+7}_{-6}$\% above $10^{10.25}$~GeV, and  
above $10^{10.5}$~GeV the fraction is found to be below 66\% at the $1 \sigma$ level. In  Fig.~\ref{elong} 
we show the distributions of muon density at 1000~m from the core as a function of primary energy 
as reported by the AGASA Collaboration, together with the predictions for $\pm 1 \sigma $ bounds for 
iron, proton and photon induced showers.

\begin{figure}[tbp]
\begin{minipage}[t]{0.32\textwidth}
\postscript{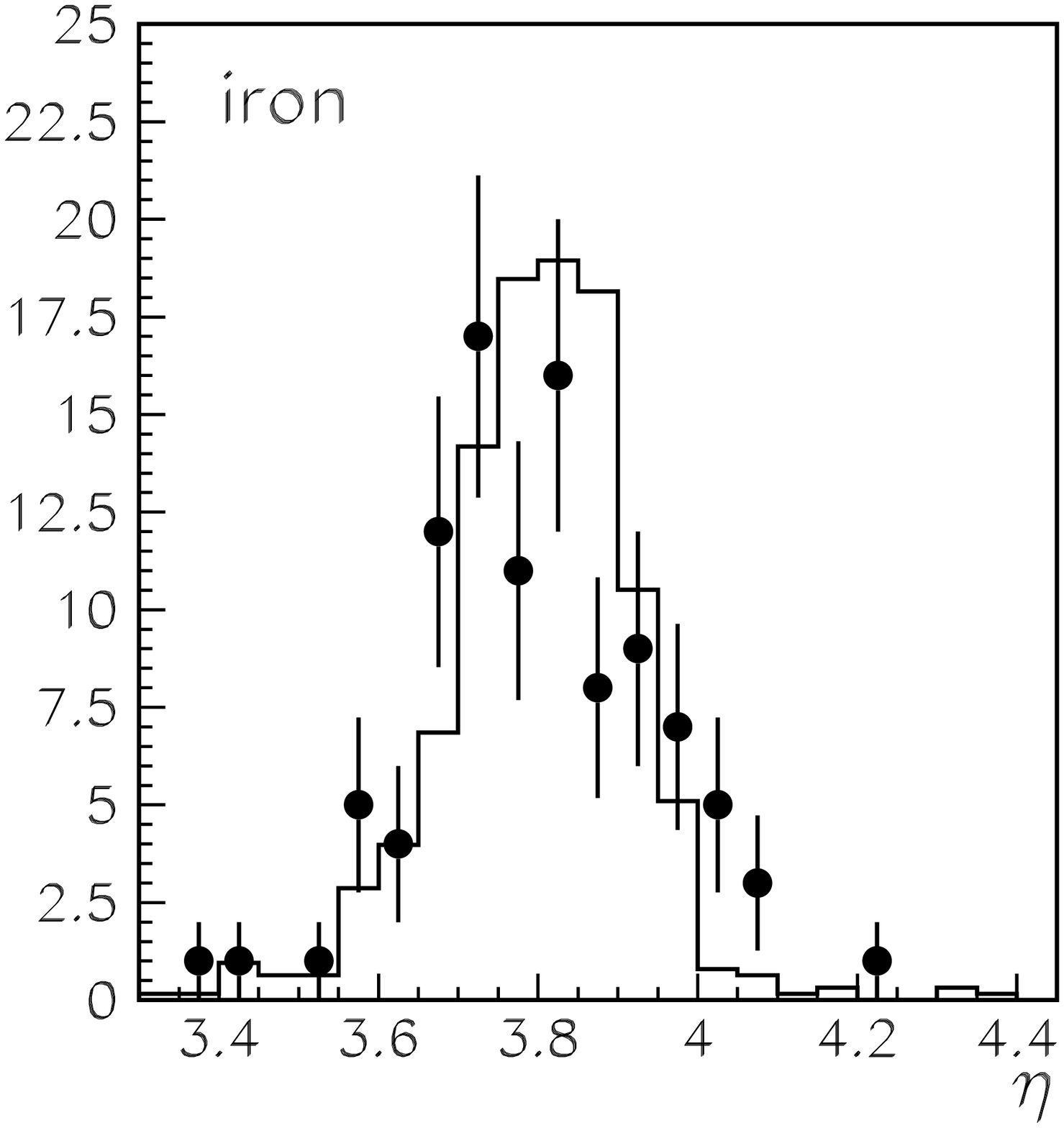}{0.99}
\end{minipage}
\begin{minipage}[t]{0.32\textwidth}
\postscript{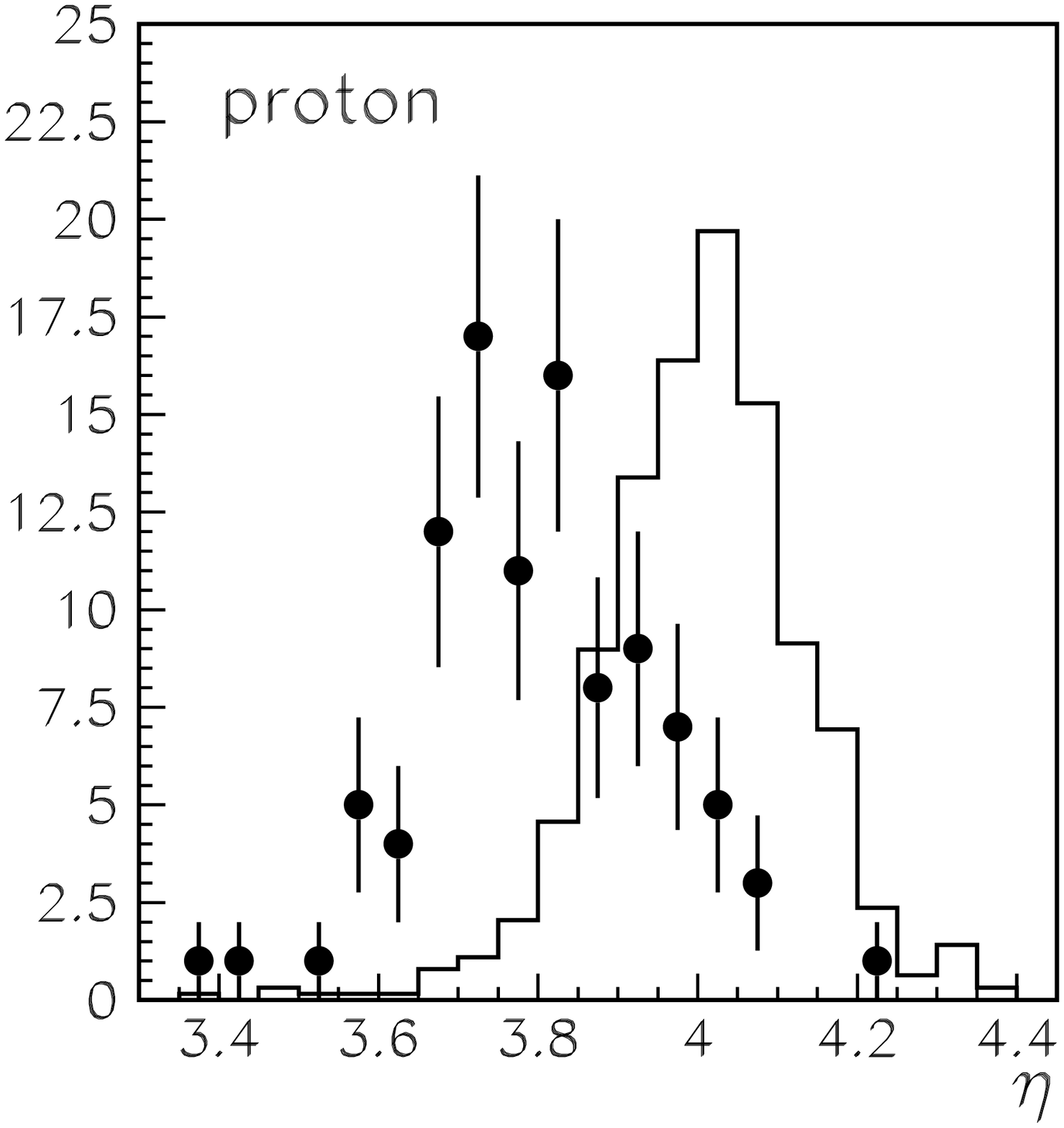}{0.99}
\end{minipage}
\begin{minipage}[t]{0.32\textwidth}
\postscript{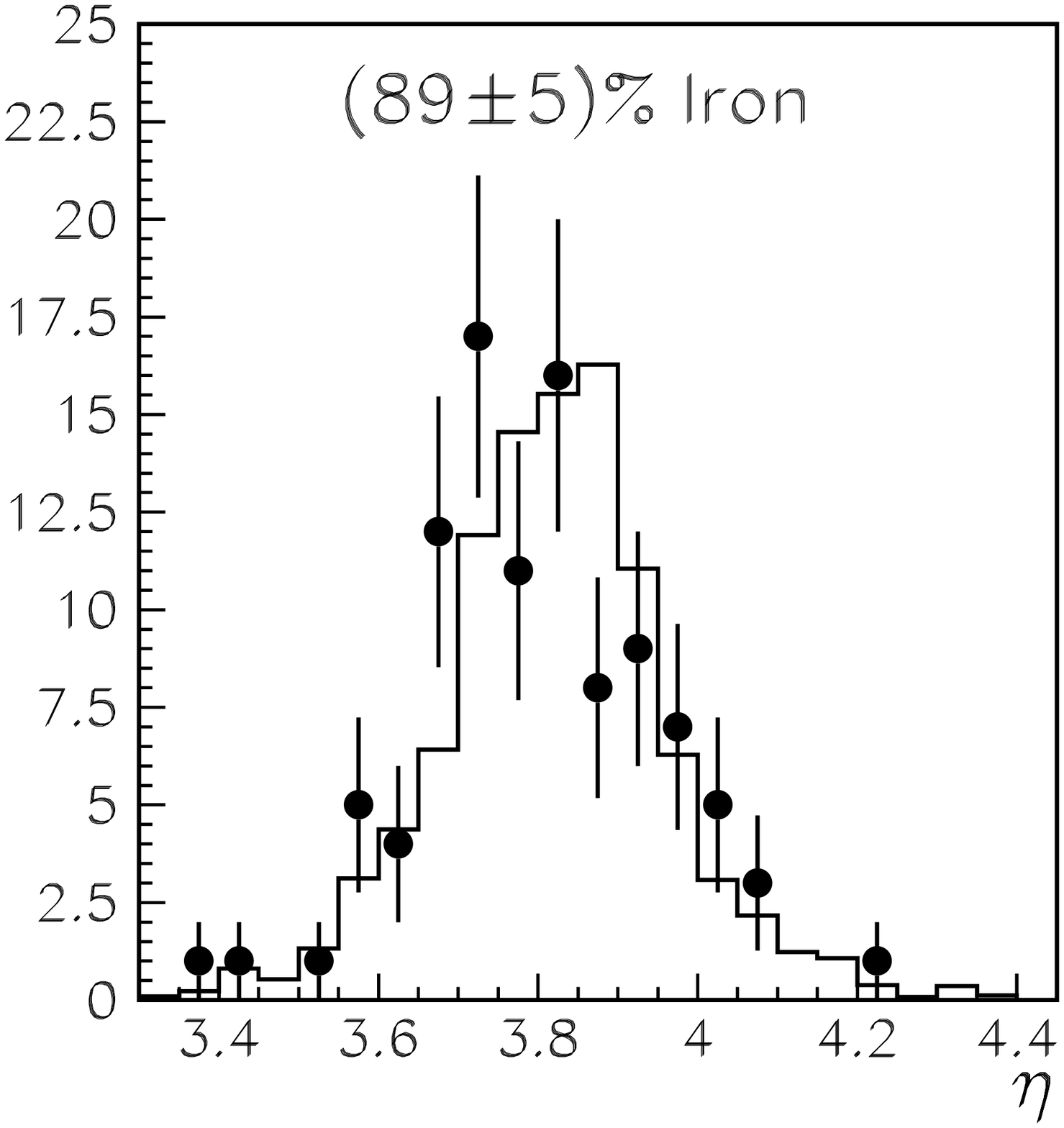}{0.99}
\end{minipage}
\caption{Comparison of the simulated distribution of $\eta$ (histogram) with measured distributions (points)
for iron (left) and proton (middle). One can see that an iron-dominated composition best fits the data, 
but that the addition of a lighter component is needed to fit the distribution at large values of $\eta$.
The right panel shows the best composition fit to the measured distribution of $\eta$ (points) from 
maximum likelihood analysis.}
\label{protonfe}
\end{figure}

The steepness of the lateral distribution of particles at ground level also correlates with the depth of 
shower maximum, and
thus carries information about the primary species.  For instance, a proton-initiated 
shower would have a steeper average 
lateral distribution, since the shower develops deeper in the atmosphere than an 
iron-initiated shower of the same energy. Recently, this approach has been used in 
conjunction with  the latest shower 
and detector simulation codes to re-interpret the lateral distribution
measurements from Haverah Park~\cite{Ave:2002gc} and Volcano 
Ranch~\cite{Dova:2003an,Dova:2003ng,Dova:2002yt}.
In the case of the Volcano Ranch array, 80 scintillators were laid out in a grid 
with a separation of 147 m, facilitating a very fine-grained measurement of the lateral distribution. 
Recall that the signal measured by plastic scintillators is the average energy loss in the scintillator of 
electrons, muons and photons in units of the energy loss of vertically penetrating muons. The correpsonding 
lateral distribution was parameterized by an NKG-like formula~\cite{Linsley:denver},
\begin{equation}
S_{\rm VR}(r) = {\frac{N}{r_{\rm M}^{2}}} \frac{\Gamma (\eta - \alpha)}{2\pi \,\,\Gamma(2 - \alpha)\,\, 
\Gamma(\eta -2)} \,\,\left( {\frac{r}{r_{\rm M}}} 
\right)^{-\alpha} \left( 1 + {\frac{r}{r_{\rm M}}} \right)^{-(\eta-\alpha)},
\end{equation}
where $N$ is the shower size, $\eta$ and $\alpha$ describe the logarithmic slope,
and $r_{\rm M} \simeq 100$~m at Volcano Ranch.
For events at median energy $10^{9}$ GeV and shower 
sizes $N = 4 \times 10^{7} - 6 \times 10^{9}$, measurements of $\eta$ 
(with $\alpha$ fixed to 1) were analyzed
in the range of zenith angle $1.0 < \sec \theta < 1.1$~\cite{Linsley:P1,Linsley:P2}. 
These measurements are shown in Fig.~\ref{protonfe} along with the recently simulated results 
for purely proton and iron primaries using {\sc qgsjet} 98 as the hadronic interaction model. 
One can clearly see the dependence on primary composition reflected in the distribution of $\eta$. 
To quantify this dependence, a maximum likelihood technique was employed assuming a bimodal
composition of proton and iron. The cosmic ray mass composition, deduced from Volcano Ranch data, 
resulted in a mean fraction (89 $\pm$ 5(stat) $\pm$ 12(sys)) \% of iron 
in the whole energy range $10^{8.7}$~GeV to $10^{10}$~GeV; mean energy  $10^{9}$~GeV.  
The resulting admixture is also shown in Fig.~\ref{protonfe}. Systematic uncertainties associated with the 
hadronic interaction model are important in this analysis. If {\sc qgsjet} 98 is replaced 
by {\sc qgsjet} 01, 
then the fraction of iron
changes from $(89 \pm 5)\%$ to $(75 \pm 5)\%$. 

\begin{figure}[tbp]
\begin{minipage}[t]{0.49\textwidth}
\postscript{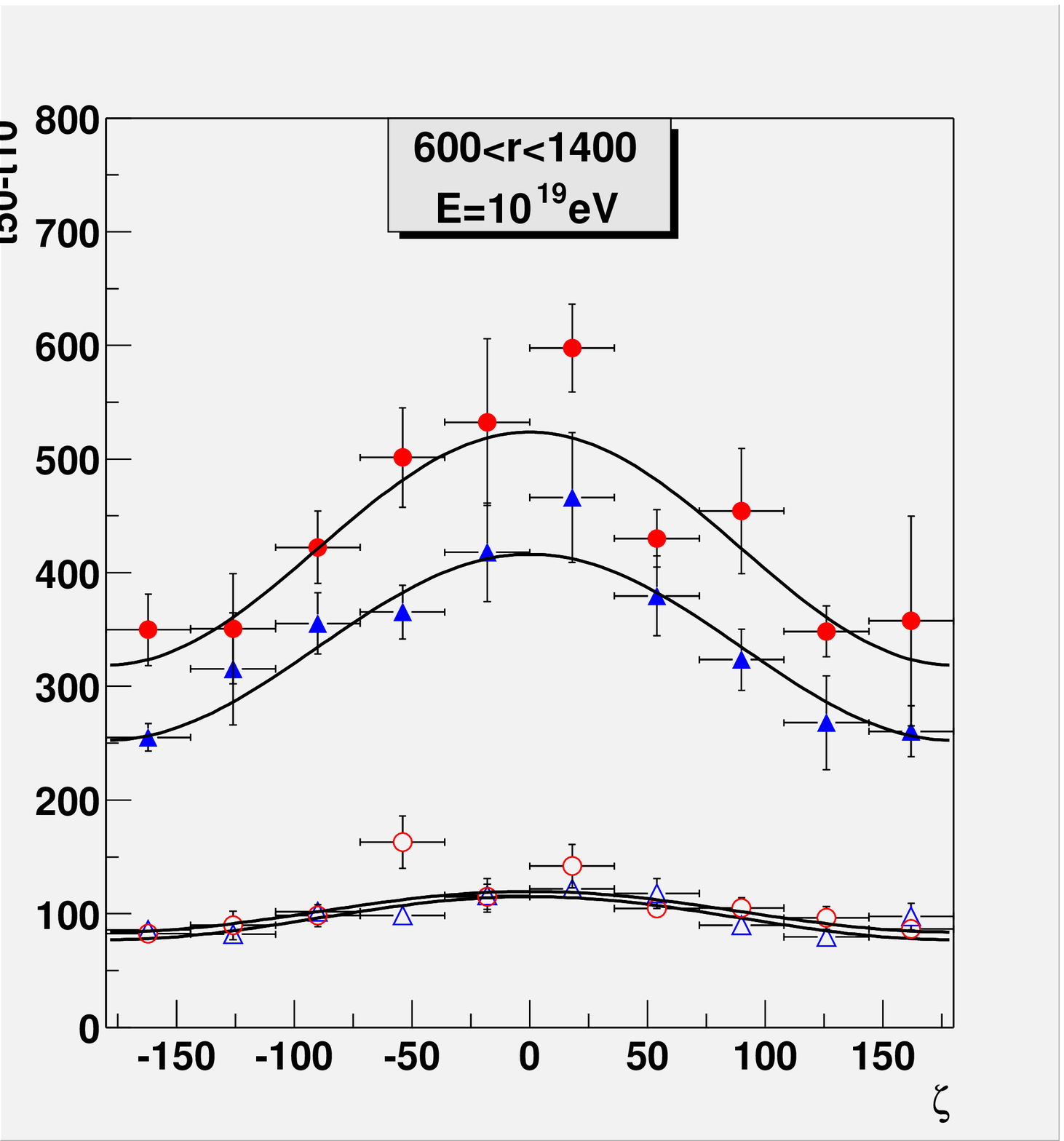}{0.99}
\end{minipage}
\begin{minipage}[t]{0.49\textwidth}
\postscript{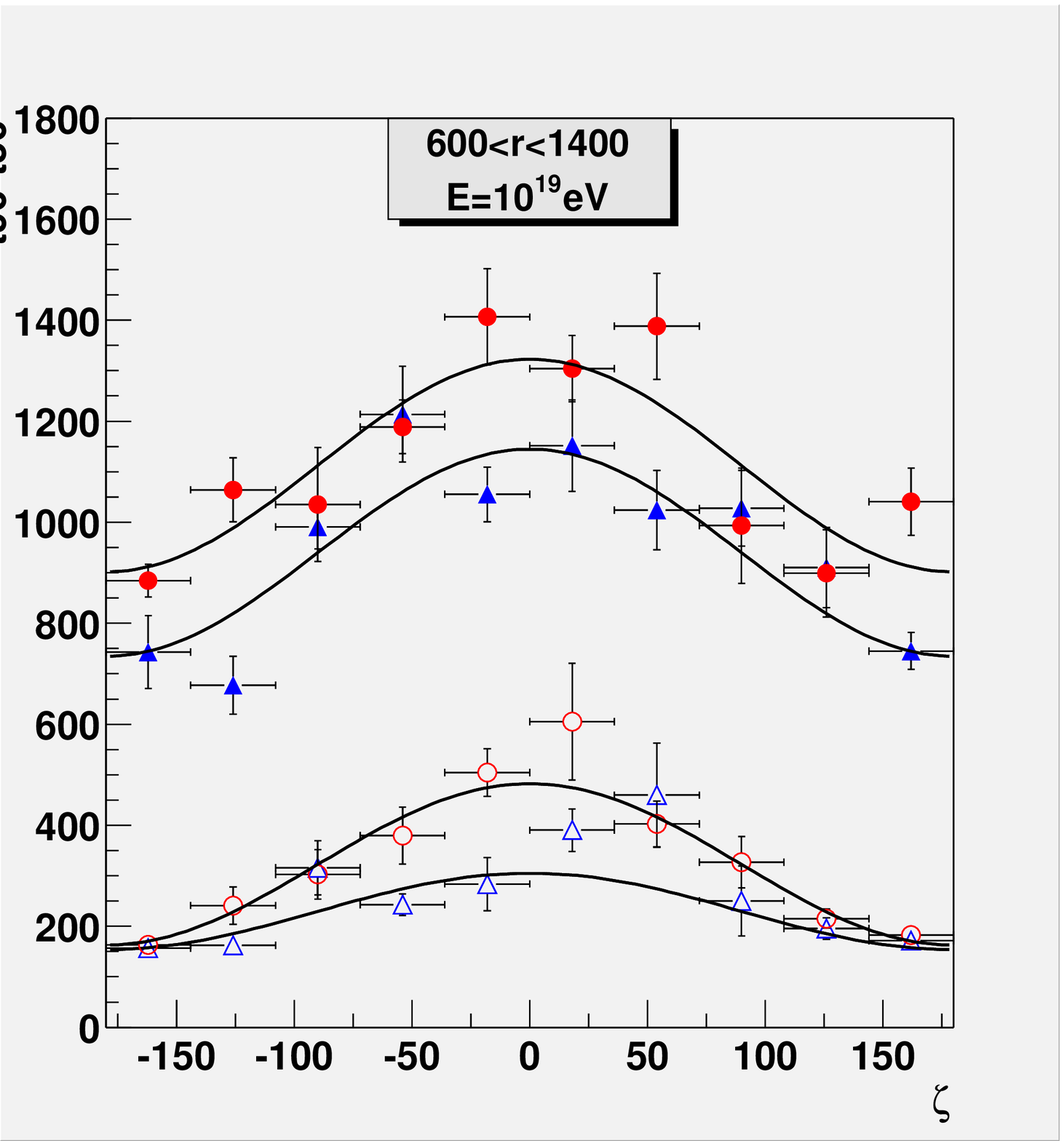}{0.99}
\end{minipage}
\caption{The triangles (iron) and circles (proton) indicate 
the rise time (left) and fall time (right) as a function of  $\zeta$.
Solid symbols correspond to a primary zenith angle of $35^\circ$, while the
open symbols correspond to $60^\circ$~\cite{Dova:2003rz}.}
\label{tere}
\end{figure}

\begin{figure}[tbp]
\begin{minipage}[t]{0.49\textwidth}
\postscript{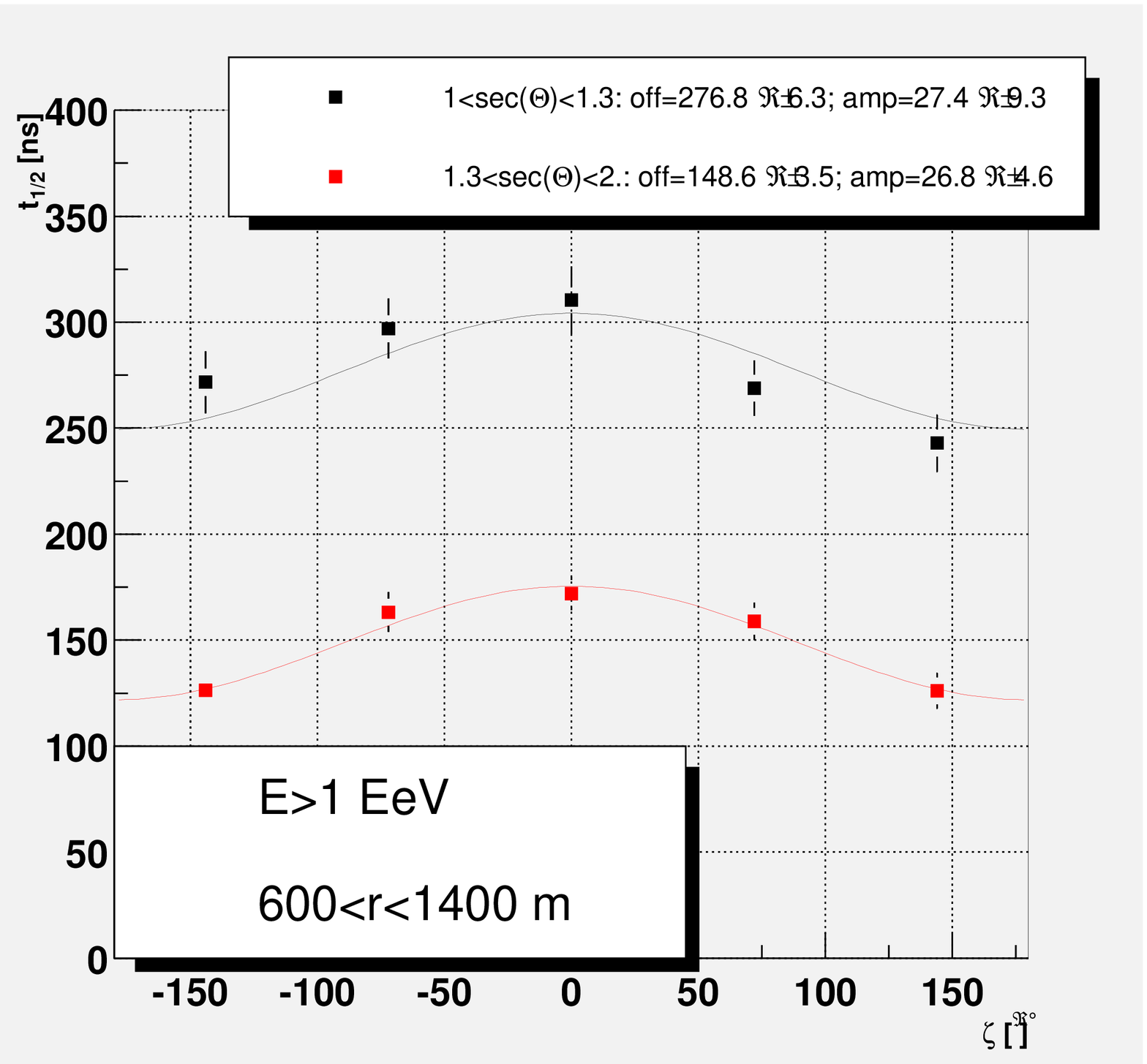}{0.99}
\end{minipage}
\begin{minipage}[t]{0.49\textwidth}
\postscript{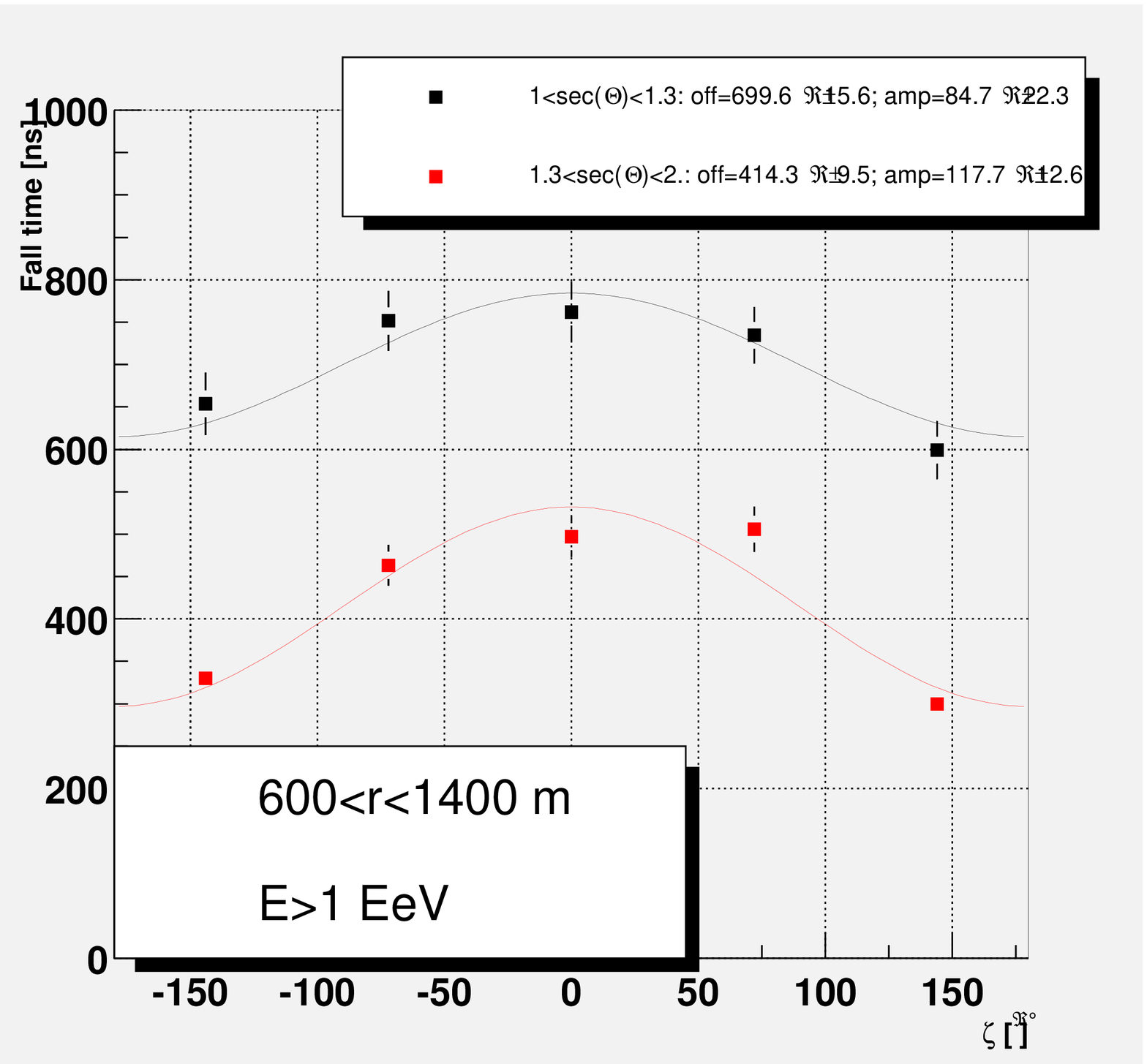}{0.99}
\end{minipage}
\caption{Time distributions from data collected by the PAO during the period of
May to November 2002 in the radial range $600 - 1400$~m~\cite{Dova:2003rz}.
In each plot the upper points corresponds to $1 < \sec \theta  < 1.3$ while the lower
corresponds to $1.3 < \sec \theta < 2$.  A fit to a linear cosine is overlayed on 
the points, where the fitted parameter ``off'' is the mean value and ``amp'' is
the amplitude of the asymmetry~\cite{Dova:2003rz}.}
\label{tere2}
\end{figure}

As described in Sec.~\ref{EM} the electromagnetic component of an EAS suffers more scattering and energy loss than the muonic component and consequently, muons tend to arrive earlier and over a shorter period of time. This means that parameters characterizing the time structure of the EAS 
will be correlated with $X_{\rm max}$ and hence with primary mass.  An early study of the shower signal 
observed in water \v{C}erenkov detectors at the Haverah Park 
array~\cite{Watson:ja} established the utility of a 
shower property known as rise time in estimating the primary composition. 
Specifically, the rise time, $t_{1/2}$, is defined as the time for the 
signal to rise from 10\% to 50\% of the full signal. 
Interestingly, the  relation between rise time and depth of maximum also implies a relation
between the rise time and the elongation rate. As suggested by Linsley \cite{Linsley:P3} if $P$ represents the average
value of some shower parameter, such as the rise time, which does not depend
explicitly on primary energy but depends on the
depth of observation, $X$, and the depth of shower maximum $X_{\rm max}$, then
the elongation rate can be derived from the following expression:
\begin{equation}
\left( \frac{\delta P}{\delta \ln E} \right)_{X} = - F D_{e} \left( \frac{\delta
 P}{\delta X} \right)_{E}
\end{equation}
where $F$ depends on the depth dependence of $P$. For a depth dependence of the form $f(X/X_{max})$, 
$F=X/X_{\rm max},$ whereas for  $f(X-X_{\rm max})$, $F=1$.

This alternative approach was applied to Haverah Park data using
the rise time  $t_{1/2}$ of the signals~\cite{Walker:sw}
to produce a measurement of the elongation rate from
$10^{8.3}$~GeV to $10^{11}$~GeV. By means of an experimentally determined value for the dependence of $t_{1/2}$ with depth, they obtained  an elongation rate of
 $70 \pm 5$~g/cm${^2}$, averaged over the previously mentioned energy range.
Their result is suggestive of an evolution to lighter species with rising energy.

Recently, an extension of the work on the shower front thickness using Haverah Park data was performed, 
focusing on the highest energy events~\cite{Avejapan}. In this analysis the averaged values of the rise 
time at a large distance from the core were compared with Monte Carlo ({\sc corsika/qgsjet} 01) predictions 
for proton and iron values. The result indicates a large fraction ($\approx 80\%$) of iron nuclei 
at 10$^{10}$~GeV~\cite{Watson:2003ba}.

Azimuthal asymmetries in the size~\cite{England} and time
structure of signals at the ground~\cite{Dova:2003rz} 
have been observed in non-vertical showers. The origin of 
these asymmetries has been discussed in Sec.~\ref{EM}.  The AUGER Collaboration has found that the 
asymmetry in time distributions offers a new possibility for the determination of  mass
composition, because its magnitude is strongly dependent on the muon to electromagnetic ratio at the 
observation level.  A preliminary study of the timing information of EAS using simulated proton and iron 
events was used to estimate the sensitivity of the PAO in discrimination of baryonic primaries~\cite{Dova:2003rz}.  The following observables were analyzed:  the rise 
time (time between 10\% and 50\% of the integrated signal), fall time (time between 50\% and 90\%) and 
the time between 10\% and 90\% of the signal. The timing variables as a function of the azimuth angle in 
the shower plane, $\zeta$, at fixed range of core distances for proton and iron showers  are shown 
in Fig.~\ref{tere}. The incoming direction of the shower 
corresponds to  $\zeta=0$.  As one can see from the figure, these distributions tend to flatten 
with increasing  primary mass.  A first analysis seems to indicate that the fall time would
be a better discriminating tool. 
One expects stronger asymmetries at intermediate core distances, where the electromagnetic component
dominates. In Fig.~\ref{tere2} we show the mean rise time and fall time as a function of  
$\zeta$, for events with energy above $10^{9}$~GeV collected by the PAO 
in the radial range $600 - 1400$~m. These results, while preliminary, indicate the promise of 
this method for composition studies, once large statistics samples become available.

\begin{figure} [t]
\postscript{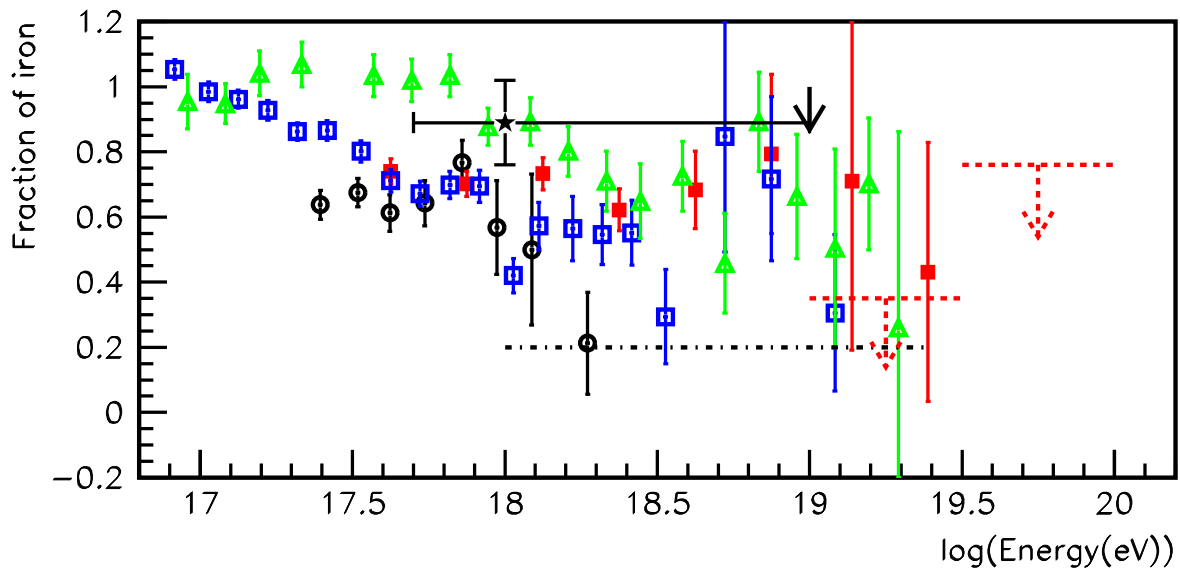}{1.0}
\caption{Iron fraction from various experiments: Fly's Eye ($\triangle$), AGASA A100 ({\tiny $\blacksquare$}), AGASA A1 ({\tiny $\square$}) using {\sc sibyll 1.6}~(\cite{Dawson:1998kk} and references therein) 
and Haverah Park~\cite{Ave:2002gc}, using {\sc qgsjet98} ({\large $\circ$})). The mean composition 
determined 
in~\cite{Dova:2003an} with the corresponding error for the Volcano Ranch energy range using 
{\sc qgsjet98} ({$\star$}) is shown. The solid line arrow indicates the recent result using rise time
measurements from Haverah Park~\cite{Avejapan}. The dashed arrow lines represents upper limits obtained 
by the AGASA Collaboration with {\sc qgsjet98}~\cite{Shinozaki}. The dot dashed horizontal line 
corresponds to results reported by the HiRes Collaboration~\cite{Archbold} } 
\label{ana}
\end{figure}

In summary, the primary composition has been studied over various
energy ranges using several experimental techniques.
The variety of results is summarised in Fig.~\ref{ana},
which shows the fraction of iron as a function of energy.
Surface arrays such as Haverah Park, Volcano Ranch and Akeno-AGASA infer $X_{\rm max}$, and hence the
overall composition, from properties of secondary particle distributions at the ground.
In this case, the primary source of systematic error arises from uncertainties
in the hadronic interaction models.
Fluorescence detectors such as Fly's Eye and HiRes observe an image of the longitudinal shower 
profile and extract $X_{\rm max}$ directly.
Consequently such measurements do not suffer from uncertainties in 
hadronic event generators, though the data analysis still faces the challenge of 
assessing atmospheric properties as a function of time.
Future stereo data from HiRes and hybrid PAO observations will provide a higher 
statistics sample with a better control of the systematic uncertainties and will 
certainly provide clues to the nature of ultra high energy cosmic rays.  

\section{Deeply Penetrating Showers} 
\label{nus}

 Even at large zenith angles, the mean free path of a
 neutrino in the atmosphere is much larger than 
 the atmosphere's slant depth. However, nearly horizontal
 showers are especially interesting since in this case
 the likelihood of interaction is maximized, and furthermore,
 neutrino induced showers can be easily distinguished
 from those induced by hadrons high in the atmosphere.
 In this section we first consider strategies 
 for detecting these kinds of signatures. After that,
 we focus attention on recent novel scenarios
 with TeV-scale quantum gravity and discuss the observables 
 of neutrino showers associated with both sub-planckian and 
 trans-planckian physics.

\subsection{Everyday weakly interacting neutrinos}

Neutrinos are unique and thus far relatively untapped astronomical
messengers~\cite{Gaisser:1994yf}.  Up to now, the only directly observed extraterrestrial
neutrinos are those from the Sun~\cite{Hirata:ub} and from supernova 
SN1987A~\cite{Hirata:1987hu,Bionta:1987qt};  these
are low energy (MeV-range) neutrinos.  Higher energy neutrinos 
should be generated by 
cosmic ``beam dumps'' in which baryonic particles collide with interstellar media.  
These neutrinos are particularly appealing for astronomy since they are undeflected by
magnetic fields and  they can travel cosmological distances without
interacting~\cite{Halzen:km,Halzen:2002pg}. In addition to providing a 
new window for astronomy, cosmic neutrino observations may also
enlighten our understanding of fundamental physics. For instance, it will be possible 
to test Lorentz invariance at extremely high energies and
to hunt for exotic processes such as neutrino decay, CPT violation, and 
small $\delta m ^2$ oscillations into sterile neutrinos 
(see {\it e.g.,}~\cite{Bahcall:2002zh,Bahcall:2002ia,Beacom:2002vi,Beacom:2003nh,Beacom:2003eu,Beacom:2003zg,Anchordoqui:2003vc}).
Since neutrinos interact only weakly, very large detector volumes are 
required, ideally on the order of $1 {\rm km}^3$we~\cite{Resvanis:ab}.  As 
discussed in the introduction the PAO overlooks $15 {\rm km}^3$we~\cite{Capelle:1998zz}, which 
is sufficient to collect a statistically significant sample of 
neutrino showers, provided they can be separated from the 
hadron and photon-induced showers.
In what follows we discuss the characteristics of neutrino-induced EAS, and comment
on the qualities which may provide a handle to separate these showers
from background.
 
In a typical collision in the Earth's atmosphere the neutrino (with energy $E_\nu$) scatters off a proton 
either via the charged current, $(\nu_l,\overline \nu_l) N \to (l^-, l^+)$ + anything,
or the neutral current, $(\nu_l,\overline \nu_l) N \to (\nu_l, \overline \nu_l)$ + 
anything.\footnote{The rate of interaction of $\nu_e,$ $\nu_\mu,$ $\nu_\tau,$ $\overline\nu_\mu,$ 
$\overline\nu_\tau,$ with atmospheric electrons is negligible compared to interactions with nucleons. 
The case of $\overline \nu_e e$ interactions is exceptional because of the intermediate Glashow resonance 
formed via $W$ boson production at $E_{\overline \nu_e} \approx 
10^{6.8}$~GeV~\cite{Glashow:W}.}
The kinematics of these reactions can be characterized by 
the inelasticity parameter $y = (1 -\cos \vartheta^*)/2$ and the 
4 momentum fraction of the proton carried by the struck quark 
$x = Q^2/(ys),$ where $-Q^2$ is the invariant momentum 
transfer between the incident neutrino and the outgoing lepton.
For a given $Q^2$ the lowest $x$ is achieved when $y=1$ and the lowest $y$ when $x=1$. Thus, kinematically 
the small values of $x$ are associated with large values of $y$ and vice versa.

The charged current differential cross section of a neutrino 
scattering on an isoscalar nucleon $N \equiv (n+p)/2$ of mass $M$ is given by
\begin{eqnarray}
\frac{d^2 \sigma_{\nu N}^{\rm CC}}{dx \, dy} & = & \frac{G_F^2\, M\, E_\nu}{\pi} \left(\frac{m_W^2}{Q^2 + m_W^2}\right)^2 \left[
\frac{1+ (1 -y)^2}{2} F_2^{\rm CC} (x, Q^2) \right. \nonumber \\
 & - & \left. \frac{y^2}{2} F_L^{\rm CC} (x, Q^2) + y \left( 1- \frac{y}{2} \right) x F_3^{\rm CC} 
(x,Q^2) \right]
\label{DIS}
\end{eqnarray}
where $m_W \simeq 80.423$~GeV denotes the $W$ boson mass, $G_F = 1.16639 \times 10^{-5}$~GeV$^2$ is
the Fermi coupling constant and the structure functions $F_i$ in terms of the quark distribution 
functions of the nucleon $q_i(x, Q^2)$ read
\begin{equation}
F_2^{\rm CC} = 2x \, \left(\frac{d_v +u_v}{2} + d_s +\overline u_s + s_s + \overline c_s + b_s 
+ \overline t_s\right)\,\,,
\end{equation}
\begin{equation}
x F_3^{\rm CC} = 2x \, (d_s + s_s - \overline u_s - \overline c_s)\,\,,
\end{equation}
and
\begin{equation}
F_L^{\rm CC} = F_2^{\rm CC} - 2 x F_1^{\rm CC} \,\,.
\end{equation}
Here the subscripts $v$ and $s$ label valence and sea contributions, and $u$, $d$, $c$, $s$, $t$, 
$b$ denote the distributions for various quark flavors in a proton. 
In the deep inelastic factorization 
scheme, Eq.~(\ref{DIS}) can be re-written in terms of quark distributions 
as~\cite{Gandhi:1995za,Gandhi:1995tf} 
\begin{equation}
\frac{d^2 \sigma_{\nu N}^{\rm CC}}{dxdy} = \frac{2G_F^2\, M\, E_\nu}{\pi} \left(\frac{m_W^2}{Q^2 + m_W^2}\right)^2 \left[
xq^{\rm CC}(x,Q^2) + (1-y)^2 x \overline q^{\rm CC} (x,Q^2)\right]\,\,,
\label{DIS2}
\end{equation}
where
\begin{equation}
q^{\rm CC}(x, Q^2) = \frac{u_v(x,Q^2) + d_v(x,Q^2)}{2} + \frac{u_s(x, Q^2) + d_s(x,Q^2)}{2} + 
s_s(x, Q^2) + b_s(x,Q^2)\,\,,
\end{equation}
and
\begin{equation}
\overline q^{\rm CC}(x, Q^2) = \frac{u_s(x,Q^2) + d_s(x,Q^2)}{2} +  c_s(x, Q^2) + t_s(x,Q^2)\,\,.
\end{equation}
In  Eq.~(\ref{DIS2}) we have omitted perturbative QCD corrections, which for $\sqrt{s} > 10^{3.4}$~GeV 
are insignificant. The average energy loss of this process is $\langle y \rangle \approx 0.2.$ Duplicating this procedure, 
one can straightforwardly obtain the cross section for the 
neutral current neutrino-nucleon interaction, 
\begin{equation}
\frac{d^2 \sigma_{\nu N}^{\rm NC}}{dx\,dy} = \frac{G_F^2\, M\, E_\nu}{2\pi} \left(\frac{m_Z^2}{Q^2 + m_Z^2}\right)^2 \left[
xq^{\rm NC}(x,Q^2) + (1-y)^2 x \overline q^{\rm NC} (x,Q^2)\right]\,\,,
\end{equation}
where the quark densities are given by
\begin{eqnarray}
q^{\rm NC}(x,Q^2) & = & \left[\frac{u_v(x,Q^2)+d_v(x,Q^2)}{2}\right] \,\left[
(g_V^d +g_A^d)^2 + (g_V^u + g_A^u)^2\right] \nonumber \\
 & + & 2 \left[\frac{u_s(x,Q^2)+d_s(x,Q^2)}{2}\right] \,\left[(g_V^d)^2 + (g_A^d)^2 + (g_V^u)^2 
+ (g_A^u)^2 \right] \nonumber \\
 & + & 2 [s_s(x,Q^2) + b_s(x,Q^2)]\, [(g_V^d)^2 + (g_A^d)] \nonumber \\
 & + & 2 [c_s(x,Q^2) + t_s(x,Q^2)]\, [(g_V^u)^2 + (g_A^u)^2] \,\,,
\end{eqnarray}
and 
\begin{eqnarray}
\overline q^{NC}(x,Q^2) & = & \left[\frac{u_v(x,Q^2)+d_v(x,Q^2)}{2}\right] \,\left[
(g_V^d -g_A^d)^2 + (g_V^u - g_A^u)^2\right] \nonumber \\
 & + & 2 \left[\frac{u_s(x,Q^2)+d_s(x,Q^2)}{2}\right] \,\left[(g_V^d)^2 + (g_A^d)^2 + (g_V^u)^2 
+ (g_A^u)^2 \right] \nonumber \\
 & + & 2 [s_s(x,Q^2) + b_s(x,Q^2)]\, [(g_V^d)^2 + (g_A^d)] \nonumber \\
 & + & 2 [c_s(x,Q^2) + t_s(x,Q^2)]\, [(g_V^u)^2 + (g_A^u)^2] \,.
\end{eqnarray}
Here, $m_Z\simeq 91.187$~GeV is the mass of the neutral intermediate boson and
\begin{equation}
g_V^d = -\frac{1}{2} + \frac{2}{3} \sin^2 \theta_{\rm W}\,, \hspace{2cm} g_A^d = -\frac{1}{2} \,,\end{equation}
\begin{equation}
g_V^u = \frac{1}{2} - \frac{4}{3} \sin^2 \theta_{\rm W} \,, \hspace{2cm} g_A^u = \frac{1}{2}\,,
\end{equation}
are the vector and axial-vector couplings for down-- and up--type quarks, respectively; with  
$\sin^2 \theta_{\rm W} \simeq 0.226$ the weak mixing parameter~\cite{Hagiwara:fs}.
Similar calculations lead to the cross sections for $\overline \nu N$ scattering.
For further details see {\it e.g.,}~\cite{Huitzu}.

The $x-Q^2$ region probed by ultra high energy neutrinos, 
\begin{equation}
x \sim \frac{m_W^2}{s \,\langle y \rangle} \sim 3.2 \times 10^4 \,\, 
\left(\frac{s}{{\rm GeV}^2} \right)^{-1} \,\sim 10^{-7}\,,
\end{equation}
is well beyond the region accesible by the HERA experiments (see Fig.~\ref{hera}). As we 
discussed in Sec.~\ref{hadronic}, in the renormalization group-improved parton 
model,  the structure functions are extrapolated to very low $x$ considering leading order (LO), 
next to leading order (NLO), and/or double--leading--logarithmic evolution of DGLAP equations. Using the CTEQ4 pdf's~\cite{Lai:1996mg} 
one finds~\cite{Gandhi:1998ri}: 
\begin{equation}
\sigma_{\nu N}^{\rm CC} = 5.53   \left(\frac{E_\nu}{{\rm GeV}}\right)^{0.363}~{\rm pb}\,,
\label{sigmaCC}
\end{equation}
\begin{equation}
\sigma_{\nu N}^{\rm NC} = 2.31 \left(\frac{E_\nu}{{\rm GeV}}\right)^{0.363}~{\rm pb}\,,
\end{equation}
\begin{equation}
\sigma_{\overline \nu N}^{\rm CC} = 5.52 \left(\frac{E_\nu}{{\rm GeV}}\right)^{0.363}~{\rm pb}\,,
\end{equation}
and
\begin{equation}
\sigma_{\overline \nu N}^{\rm CC} = 2.29   \left(\frac{E_\nu}{{\rm GeV}}\right)^{0.363}~{\rm pb}\,.
\end{equation}
For $10^{7}~{\rm GeV} \alt E_\nu \alt 10^{12}~{\rm GeV},$ these cross sections are correct
to about 10\%, which is smaller than the systematic uncertainties that cosmic ray experiments
typically contend with.  Note that the reason this uncertainty is small compared to the uncertainty in 
the cross section shown in Fig.~\ref{sigmahera} is a consequence of the existence of two viable 
models for cross section extrapolation in the case of $pp$ interactions.  
Neutrino interaction lengths
\begin{equation}
L=1.7\times 10^7~\kmwe~\left({\pb\over\sigma_{(\overline\nu \nu) N}^{\rm (CN)C}}\right)
\end{equation}
are therefore far larger than the Earth's atmospheric depth, which is only
$0.36~\kmwe$ even when traversed horizontally.  As a consequence, neutrino showers,
unlike baryon or photon induced showers, can begin deep in the atmosphere.
So, to obtain a clean signal of neutrino-induced showers one should be able to identify
deeply developing cascades in the whole  sample of EAS. 

For large zenith angles ($\theta > 70^\circ$), 
an air shower initiated by a neutrino can be distinguished from that of
an ordinary hadron by its shape. As discussed in Sec.~\ref{EMP}, baryons interact high
in the atmosphere. Consequently, at ground level the
electromagnetic part of these inclined showers is totally extinguished (more
than 6 equivalent vertical atmospheres were gone through) and only
the muon channel survives. Besides, the shower front is extremely
flat (radius of curvature $>$ 100 km) and the particle time spread is very
narrow ($\Delta t < 50$ ns). Since neutrinos can interact deeply in the atmosphere, 
they can initiate showers in the volume of air immediately above the detector. 
Therefore, quasi-horizontal showers initiated by neutrinos would still present a curved 
front (radius of curvature of a few km), with particles well spread over time, 
${\cal O} (\mu{\rm s})$,  allowing a good characterization of the cascade
against background.

If the primaries are mainly electronic and muonic neutrinos, as
expected from pion decays, two types of neutrino showers can be
distinguished: ``mixed'' (with full energy) or ``pure hadronic''
(with reduced energy), respectively~\cite{Billoir:nq}. In the charged current
interaction of a $\nu_e$, an ultra high energy electron having about
80\% of the $\nu_e$ energy is
produced and initiates a large electromagnetic cascade parallel
to the hadronic cascade.  In contrast, the charged current
interaction of a $\nu_\mu$ produces a muon which is not
easily detectable by existing experiments. In the presence of maximal $\nu_{\mu}/\nu_{\tau}$-mixing,
$\nu_\tau$-showers must also be considered. The $\tau$
mean flight distance is $\sim 50 E_\tau/(10^9$~GeV)~km, whereas the distance between the 
position of first impact and ground is $\sim$ 30 km, hence only
$\tau$'s with energy $\alt 10^{8.9}$~GeV will decay before reaching
the ground.  Since the $\tau$ is produced with about 80\% of the 
$\nu_\tau$ energy, showers initiated by ultra high energy $\nu_\tau$'s will
be indistinguishable from  $\nu_{\mu}$ showers.

Another interesting category of neutrino-related showers comprises 
events in which a neutrino skims the Earth, traveling at a low 
angle along a chord with length of order its interaction 
length. Some of 
these Earth-skimmers may be converted into charged leptons in the 
Earth's crust. Unlike electrons that do not escape from rocks, at the energies of interest, 
muons and tau leptons travel up to ${\cal O} (10~{\rm km})$ inside the Earth. Although muons
do not produce any visible signal in the atmosphere, taus can  produce clear signals for both 
fluorescence eyes~\cite{Feng:2001ue} and surface arrays~\cite{Bertou:2001vm} if 
they decay above the detector. The phenomenon would thus increase the $\nu$-event rate
and enhance the sensitivity of the PAO to neutrino fluxes.

Up to now we have only considered signals one might expect from Standard Model (SM) processes.
Many scenarii with new physics beyond the electroweak scale, $M_{\rm EW},$ have been proposed, 
some of which increase the weak interaction cross section 
(see {\it e.g.},~\cite{Domokos:1986qy,Bordes:1997bt,Domokos:1998ry,Fodor:2003bn}), and hence would 
have observable implications. As an example of such SM extensions, we consider in the last part of this 
review a scenario which has generated a great deal of recent interest.

\subsection{Neutrino interactions mediated by gravity}

A promising route towards reconciling the apparent mismatch of the
fundamental scales of particle physics and gravity is to modify the 
short distance behavior of gravity at scales much larger
than the Planck length. This can be accomplished in a straightforward
manner~\cite{Arkani-Hamed:1998rs,Antoniadis:1998ig,Randall:1999ee} if 
one assumes that the SM fields are confined to a 4-dimensional world 
(corresponding to our apparent universe), while gravity lives in a higher 
dimensional space. One virtue of this assumption is that very large extra dimensions are
allowed without conflicting with current experimental
bounds, leading to a fundamental Planck mass much lower than its 
effective 4-dimensional value. In particular, if spacetime is taken as a direct product of a
4-dimensional spacetime and a flat spatial $n$-dimensional torus $T^{n}$
(of common linear size
$2\pi r_c$), one obtains a
definite representation of this picture in which the effective 4-dimensional
Planck scale,
$M_{\rm Pl} \sim 10^{19}$~GeV, is related to the fundamental scale of
gravity, $M_D$, according to $M^2_{\rm Pl} = 8 \pi \, M_D^{n+2} \,
r_c^n$~\cite{Arkani-Hamed:1998rs}. Now, a straightforward calculation shows that, for $n=1$, low-scale 
gravity within toroidal compactifications is excluded, as gravity would be modified at the 
scale of our solar system. Astrophysical constraints require
$M_D \gg 10$~TeV for $n= 2,3$ and $M_D \agt 4$~TeV for $n=4$~\cite{Hannestad:2003yd}. For 
$n \geq 5,$ however, $M_D$ may be as low as a 
TeV~\cite{Acciarri:1999jy,Acciarri:1999bz,Abbott:2000zb,Abazov:2003gp,Anchordoqui:2003jr}.

From our 4-dimensional point of view the massless graviton appears as an infinite tower of 
Kaluza--Klein (KK) modes, of which the lowest is the massless graviton itself, 
but the others are massive. The mass squared of each KK graviton mode reads,
$m^2 = \sum_{i=1}^n \,\ell_i^2 / r_c^{2},$
where the mode numbers are $\ell_i \in \mathbb{Z}.$ Note that the weakness of the gravitational 
interaction is partially compensated 
by the tower of KK modes that are exchanged: the square coupling $M^{-2}_{\rm Pl}$ 
of the graviton vertex is exactly cancelled by the large multiplicity of KK excitations 
$\sim s^{n/2} \,r_c^n,$ so that the final product is $\sim s^{n/2}/M_D^{2+n}.$ 
Indeed, if one includes in the interaction 
the brane Goldstone modes, a form factor $\sim e^{-m^2/M_D^2}$ is introduced at each graviton 
vertex~\cite{Bando:1999di}.
This exponential suppression, which parametrizes the effects of a finite brane tension, 
provides a dynamical cutoff in the (otherwise divergent) sum over all KK contributions 
to a given scattering amplitude. Altogether, one may wonder 
whether the rapid growth of the cross section with energy in neutrino-nucleon reactions mediated by 
spin 2 particles carries with it observable deviations from SM 
predictions.\footnote{In what follows we only take into account 
KK excitations on the gravity sector without considering string recurrences of any other field. 
For a treatment of the latter see {\it e.g.,}~\cite{Domokos:1998ry}.}

A simple Born approximation to the elastic
neutrino-parton cross section (which underlies the
total neutrino-proton cross section) leads, without modification, to
$\hat \sigma_{\rm el} \sim \hat s^2$~\cite{Nussinov:1998jt,Jain:2000pu}. Unmodified, 
this behavior by itself eventually
violates unitarity. This  may be seen either by examining the
partial waves of this amplitude, or by studying the high energy
Regge behavior of an amplitude 
$A_R (\hat s,\hat t) \propto \,\hat s^{\alpha(\hat t)}$
with spin-2 Regge pole, {\it viz.,} intercept $\alpha(0)=2$. For the latter,
the elastic
cross section is given by
\begin{equation}
\frac{d\hat\sigma_{\rm el}}{d\hat t}\, \sim\, \frac{|A_R(\hat s, \hat t)|^2}{\hat s^2}\, \sim 
\hat s^{2\alpha(0)-2}\,\sim \hat s^2,
\end{equation} 
whereas the total cross section reads as
\begin{equation}
\hat \sigma_{\rm tot}\, \sim \frac{\Im {\rm m} [A_R(\hat s,0)]}{\hat s}\,\sim \hat s^{\alpha(0)-1}\,\sim 
\hat s,
\end{equation}
so that eventually, $\hat \sigma_{\rm el}>\hat\sigma_{\rm tot}$~\cite{Anchordoqui:2000uh}.  Eikonal
unitarization schemes modify these behaviors. Specifically, for large impact parameter,
a single Regge pole exchange amplitude yields 
$\hat \sigma_{\rm tot} \sim \ln^2(\hat s/s_0)$~\cite{Kachelriess:2000cb,Kisselev:2003rz}. 
For short impact parameters, partial wave unitarity is a tall order as corrections to 
the eikonal amplitude are expected to become important. Note that graviton self interactions carry 
factors of $\hat t$ associated to the vertices, and thus as $\hat t$ increases, so too does the attraction 
among the scattered particles.  Eventually it is expected that gravitational collapse to a black
hole (BH) will take place, absorbing the initial 
state in such a way that short distance effects will be screend by the appearance of a 
horizon~\cite{Banks:1999gd,Giddings:2001bu,Dimopoulos:2001hw,Feng:2001ib}.\footnote{This paper does 
not purport to be a comprehensive review of TeV-scale gravity BHs; for 
an up-to-date and detailed discussion of the topic, the reader is 
referred to~\cite{Kanti:2004nr}.}

According to the Thorne's hoop conjecture~\cite{Thorne:ji}, 
a BH forms in a two-particle collision when and only when the impact parameter is smaller than the 
radius of a Schwarzschild BH of mass equal to $\sqrt{\hat s} \equiv \sqrt{xs}$. The conjecture
thus predicts a total cross section for BH production proportional 
to the area subtended by a ``hoop'' 
\begin{equation}
\hat\sigma_{BH} = F(n)\,\pi r_s^2(\sqrt{\hat{s}}) 
\label{hoopsigma}
\end{equation} 
of radius~\cite{Myers:un,Argyres:1998qn}
\begin{equation}
\label{schwarz}
r_s(\sqrt{\hat s}) =
\frac{1}{\md}
\left[ \frac{\sqrt{\hat{s}}}{\md} \right]^{\frac{1}{1+n}}
\left[ \frac{2^n \pi^{(n-3)/2}\Gamma({n+3\over 2})}{n+2}
\right]^{\frac{1}{1+n}}\,,
\end{equation}
where $F(n)$ is a constant of order unity. 
Recent progress has confirmed the validity of Eq.~(\ref{hoopsigma}) and evaluated the
dimension-dependent constant $F(n)$, analytically in four
dimensions~\cite{Eardley:2002re} and numerically in higher
dimensions~\cite{Yoshino:2002tx}. Of course, this work is purely in
the framework of classical general relativity, and is expected to be 
valid for energies far above the Planck scale,  for which curvature 
is small outside the horizon and strong quantum effects 
are hidden behind the horizon. Extending it to the  
planckian regime of center-of-mass energies close to $\md$ requires a
better understanding of quantum gravity than we now possess. Thus it
is important to impose a cutoff on the mass
of microscopic BHs for which Eq.~(\ref{hoopsigma}) can reasonably be
expected to hold. 

\begin{figure} [t]
\postscript{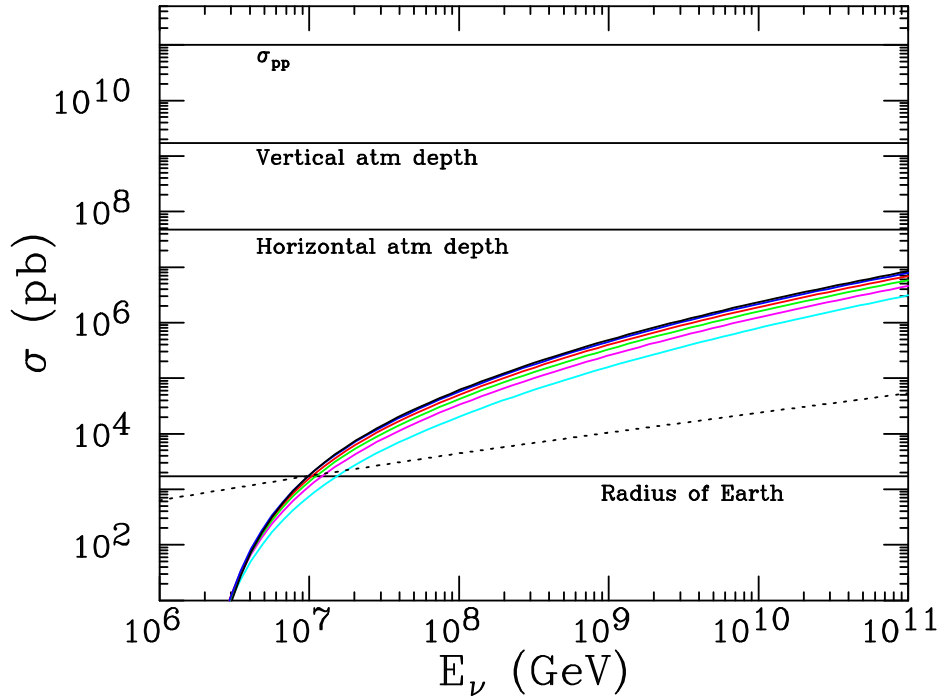}{0.75}
\caption{Lower bounds on BH production cross sections for $n=2,\ldots,
7$ from below, assuming $\md = 1~\tev$ and $\xmin=1$.  Energy loss has
been included according to Eq.~\eqref{sigma}.  The SM cross section
$\sigma_{\nu N}^{\rm CC},$ as given in Eq.~(\ref{sigmaCC}), is indicated by the dotted line. The 
typical range of $pp$ interactions, as well as cross sections required for shower triggering in 
different characteristic targets are also shown for comparison. This figure is courtesy of Jonathan Feng.}
\label{bh}
\end{figure}

In the course of collapse a certain  amount of energy is 
radiated in gravitational waves by the
multipole moments of the incoming shock waves, leaving only
a fraction $y \equiv \mbh/\sh$ available to Hawking
evaporation, where $\mbh$ is a {\it lower bound}
on the final mass of the BH.  This ratio depends on the classical impact parameter $b,$ 
and so the inclusive production of BHs proceeds through different final
states for different values of $b$. These final 
states are characterized by the fraction $y(z)$ of the initial parton center-of-mass
energy which is trapped within
the horizon. Here, $z= b/b_{\rm max},$ and $b_{\rm max}/r_s(\sqrt{\hat s})$ is given in 
Table~\ref{bhtable}~\cite{Yoshino:2002tx}.
With a lower cutoff $M_{\rm BH,min}$
on the BH mass required for the validity of the semi-classical description,
this implies a joint constraint
\begin{equation}
 y(z)\,\,\sqrt{\hat s} \ge M_{\rm BH,min}
\label{constraint}
\end{equation}
on the parameters $x$ and $z$. Because of the monotonically decreasing nature
of $y(z)$, Eq.~(\ref{constraint}) sets an {\it upper}
bound $\bar z(x)$ on the impact parameter for
fixed $x.$ The corresponding parton-parton BH cross section
is $\hat \sigma_{_{\rm BH}} (x) = \pi \bar b^2(x),$
where $\bar b=\bar z b_{\rm max}.$  The total BH production cross section
is then~\cite{Anchordoqui:2003jr}
\begin{equation}
\sigma_{_{\nu N \to {\rm BH}}}(E_\nu,M_{\rm BH,min},M_{D}) \equiv \int_{\frac{M_{\rm BH,min}^2}{
y^2(0) s}}^1 \, dx
\,\sum_i f_i(x,Q) \ \hat \sigma_{_{\rm BH}}(x) \,\,,
\label{sigma}
\end{equation}
where $i$ labels parton species and the $f_i(x,Q)$ are pdf's (to derived the BH production 
cross sections shown in Fig.~\ref{bh} we used the CTEQ6M pdf's~\cite{Pumplin:2002vw,Stump:2003yu}).
The momentum scale $Q$ is taken as $r_s^{-1},$ which is a typical
momentum transfer during the gravitational collapse~\cite{Emparan:2001kf}. 
In contrast to SM processes, BH production is not supressed by perturbative couplings and is 
enhanced by the sum over all partons, particularly the gluon.

\begin{table}
\caption{Gravitational collapse parameters.}
\begin{center}
\begin{tabular}{c|cccccccc}
\hline \hline
$n$ & 2 & 3 & 4 & 5 & 6 & 7 \\ \hline
$b_{\rm max}/r_s(\sqrt{\hat s})$  & $\,\,\,1.052\,\,\,$ & $\,\,\,1.118\,\,\,$ & $\,\,\, 1.166 \,\,\,$ & 
$\,\,\, 1.206 \,\,\,$ & $\,\,\, 1.238 \,\,\,$ & $\,\,\, 1.264 \,\,\,$ \\
$F(n)$ & 1.341 & 1.515 & 1.642 & 1.741 & 1.819 & 1.883 \\
\hline \hline
\end{tabular}
\end{center}
\label{bhtable}
\end{table}

Subsequent to formation, the BH proceeds to decay 
dominantly through radiation of standard SM particles on the 3-brane~\cite{Emparan:2000rs}.  
This occurs because the SM particles live on the brane, so that the
relevant phase space for BH decay into these fields is governed by the
4-dimensional projection of the horizon area of the
$(4+n)$-dimensional BH, and by the Hawking temperature which is common
to bulk and brane. Modulo some grey body 
factors~\cite{Kanti:2002nr,Kanti:2002ge,Harris:2003eg}, the dominance of the SM radiation is a 
result of the much larger number of degrees of freedom.

The choice of a lower limit of integration in Eq.~(\ref{sigma}) requires additional
explanation. This limit determines the minimum mass for BHs
that will be included in the calculation. The semiclassical description outlined above is only 
reliable when the energy of the emitted particle is small compared to the BH
mass, because it is only under this condition that both the gravitational
field of the brane and the back reaction of the metric during the emission
process can be safely neglected~\cite{Preskill:1991tb}. Since the total number 
of particles emitted by the BH is roughly equal to its entropy, 
\begin{equation}
S = \frac{4 \pi\, \mbh\,r_s(\mbh)}{n+2} \,\,,
\end{equation}
the criterion we employ is to assume that the BH has an entropy $S \gg 1.$ 
For $\xmin \equiv M_{\rm BH,min}/M_D > 3$ and $n\ge 5$, one finds $S > 10,$ so that most
of the decay process can be well described within the semiclassical approximation.  
Moreover, the string cross section derived~\cite{Dimopoulos:2001qe} from 
the Virasoro-Shapiro amplitude is expected to be 
considerably larger than that given in Eq.~(\ref{hoopsigma}), and so it may be reasonable 
even to take $\xmin$ as
low as 1, for which $S \agt 3$~\cite{Anchordoqui:2003ur}.

Although the BH production cross section, ${\cal O} (M_{\rm EW}^{-1})$, is about 5
orders of magnitude smaller than QCD cross sections, ${\cal O}
(\Lambda_{\rm QCD}^{-1}),$ one of the most startling prdictions of TeV-scale 
gravity theories is that at the LHC, BH events could be filtered out of the QCD background,  both 
in $pp$~\cite{Giddings:2001bu,Dimopoulos:2001hw} (see 
also~\cite{Cheung:2001ue,Anchordoqui:2002cp,Mocioiu:2003gi,Chamblin:2003wg,Anchordoqui:2003ug,Chamblin:2004zg}) as well as in PbPb~\cite{Chamblin:2002ad} 
collisions. At energies of 
interest, however, the cosmic ray luminosity, $ L \sim 10^{-24}$ cm$^{-2}$ s$^{-1},$ 
is about 50 orders of magnitude smaller than the LHC luminosity, 
thus making it futile to hunt for BHs in baryonic cosmic rays.  On the other hand, 
as can be seen in Fig.~\ref{bh}, although greatly reduced by the cross section for BH production,
neutrino interaction lengths are still far larger than the Earth's atmospheric depth. Neutrinos therefore
would produce BHs with roughly equal probability at any point in the atmosphere.
As a result, the light descendants of the BH
may initiate low-altitude, quasi-horizontal showers at rates significantly higher than SM
predictions. Because of these considerations the atmosphere provides a
buffer against contamination by mismeasured baryons (for which the electromagnetic channel 
is filtered out) allowing a good
characterization of BH-induced showers when $S \gg
1$~\cite{Anchordoqui:2001ei,Ringwald:2001vk,Dutta:2002ca,Anchordoqui:2002vb,Cardoso:2004zi}.\footnote{Additionally, 
neutrinos that traverse
the atmosphere unscathed may produce BHs through interactions in the
ice or water and be detected by neutrino 
telescopes~\cite{Kowalski:2002gb,Alvarez-Muniz:2002ga,Jain:2002kz}.} Furthermore, a similar technique 
to that employed
in discriminating between photon and hadron showers can be applied to isolate BH mediated showers 
from neutrino SM events~\cite{Anchordoqui:2001cg}. Specifically, if an anomalously 
large quasi-horizontal deep 
shower rate is found, it may be ascribed to either an enhancement of the incoming neutrino flux, or an 
enhancement in the neutrino-nucleon cross section. However, these two possibilities may be 
distinguished by separately binning events which arrive at very
small angles to the horizontal, the so-called ``Earth-skimming'' 
events. An enhanced flux will increase 
both quasi-horizontal and Earth-skimming event rates, whereas a large BH 
cross section suppresses the latter, because the hadronic decay products of 
BH evaporation do not escape the Earth's crust.

In summary, the signal for ultra high energy neutrinos is quasi-horizontal giant air showers 
initiated deep in the atmosphere: showers with large electromagnetic components, 
curved fronts, and signals well spread over time. These shower characteristics are easily 
differentiated from 
EAS initiated by baryons or photons. The low target density for neutrino interactions provided by 
the atmosphere must be compensated by monitoring large areas at the Earth's surface. In particular, 
the PAO will have an acceptance exceeding 1~km$^3$ of water for $E_\nu \agt 10^8$~GeV, and thus
will be able to search for extraterrestrial sources of ultra high energy neutrinos. Moreover, this 
observatory holds great promise for probing physics beyond the SM. An optimist might even imagine the 
discovery of microscopic BHs, the telltale signature of the universe's unseen dimensions.

\section{EAS in a Nutshell} 

In this article, we have reviewed the general properties and techniques for 
modelling air showers initiated 
by ultra high energy particles interacting in the Earth's atmosphere.

The incidence of a single high energy particle on the upper atmosphere 
gives rise to a roughly conical cascade of particles which 
reaches the Earth in the form of a giant ``saucer'' traveling
at nearly the speed of light. The number of secondaries in the cascade readily increases through 
subsequent generations of particle interactions. Because of the prompt decay of neutral 
pions, about 30\% of the energy in each generation is transferred to an electromagnetic 
cascade. Roughly speaking, at $10^{11}$~GeV, baryons and charged pions have interaction lengths of 
the order of $40~{\rm g}/{\rm cm}^2,$ increasing to about $60~{\rm g}/{\rm cm}^2$ at $10^7$~GeV. 
Additionally, 
below $10^{10}$~GeV, photons, electrons, and positrons have mean interaction lengths of
$37~{\rm g}/{\rm cm}^2,$ whereas above this critical energy the competing LPM and geomagnetic effects lead
to interaction lengths between $45-60~{\rm g}/{\rm cm}^2$. 
Altogether, the atmosphere acts as a natural colorimeter with variable density, providing a vertical 
thickness of 26 radiation lengths and about 15 interaction lengths. Amusingly, this is not too different 
from the number of radiation and interaction lengths at the LHC detectors.\footnote{For example, the 
CMS electromagnetic calorimeter is $\agt 25$ radiation lengths deep, and the hadron calorimeter constitutes 
11 interaction lengths.}

The number of muons does not increase linearly with energy, because of the previously mentioned pionization 
process: at higher energy more generations are required to cool the pions to the point where they are 
likely to decay before interaction. Production of extra generations results in a larger fraction of the 
energy being lost to the electromagnetic cascade, and hence a smaller fraction 
of the original energy being delivered to the $\pi^{\pm}$. Ultimately, about 90\% of 
the primary particle's energy is dissipated in the electromagnetic cascade. The 
remaining energy is carried by hadrons, as well as muons and neutrinos produced in $\pi^{\pm}$ decays. 

By the time they reach the ground, relatively vertical showers have evolved 
fronts with a curvature radius of a few km, and far from the shower core their constituent
particles are well spread over time, typically of the order of a few 
microseconds.  For such a shower both the muon component and a large portion of
the  electromagnetic component survive to reach the ground, and their 
lateral distributions can be accurately parametrized.  
Although the lateral distribution function depends on the experiment,
surface measurements of both gamma-- and baryon--induced 
showers can be fitted with  NKG-like formulae.  From this 
distribution the primary energy can be determined.

For inclined showers the electromagnetic component is absorbed long before reaching the 
ground, as it has passed through the equivalent of several vertical atmospheres: 2 at a zenith angle 
of $60^\circ$, 3 at $70^\circ$, and 6 at $80^\circ$. In these showers, only high energy muons created in 
the first few generations of particles survive past 2 equivalent vertical atmospheres. The rate of energy 
attenuation for muons is much smaller than it is for electrons, thus the shape of the resulting shower
front is  very flat (with curvature radius above $100~\km$), and its time extension is very short (less 
than $50~\ns$).  The damping of the electromagnetic component of the shower provides a means to 
search for both photon and neutrino primaries.  In particular 
quasi-horizontal showers with electromagnetic components at the ground would suggest a deeply penetrating 
primary, such as the elusive ultra high energy neutrino.

As we have seen, the chief uncertainty in shower modelling arises from
lack of definitive knowledge about the nature of hadronic interactions. 
This is because the center-of-mass energies involved in cosmic ray 
collisions are orders of magnitude beyond that achievable in present and foreseeable future
experiments. Moreover, man-made accelerators are designed to probe QCD physics in 
the high tranverse momentum region, and air shower physics is driven by interactions in the very forward 
direction. The analysis of extensive air showers then requires the extrapolation of hadronic 
interaction models more than 2 orders of magnitude in  center-of-mass energy beyond the 
highest accelerator energies ($\sqrt{s} = 1.8$~TeV) to date. In fact, the required extrapolation is 
much greater than this because air showers involve nuclei as well as single hadrons both as targets 
and projectiles. Efforts towards
improving our understanding of soft and semi-hard processes are clearly required.

The muon content of the shower tail is quite sensitive to the unknown details of 
hadronic physics at ultra high energies.  
This implies that attempts to extract composition information from measurements
of muon content at ground level tend to be systematics dominated.  There are, however, 
complimentary methods for uncovering the primary species which are less dependent
on knowledge of the hadronic physics. 
One well-established method involves using fluorescence telescopes to determine the 
energy dependence of the depth of shower maximum, the so-called elongation rate, which
is sensitive to the evolution of the primary composition with energy.  Unfortunately, fluorescence
detection has its own set of systematic uncertainties associated with the knowledge of atmospheric
properties as a function of time. Future hybrid experiments, such as the PAO, will record 
events with simultaneous observation of particles reaching the ground
and the shower profile in the atmosphere, and thus provide 
a new arsenal of data for controlling the systematic errors.
Furthermore, the giant aperture of the array will generate an extensive air shower
sample of unprecedented size, ushering in a golden age of cosmic ray physics.

\begin{acknowledgments}
We would like to thank  M\'aximo Ave, Jim Cronin, Luis Epele, Jonathan Feng, 
Haim Goldberg, Francis Halzen, Dieter Heck, Johannes Knapp, M\'onica Mance\~nido, Markus Risse, 
Markus Roth, Sergio Sciutto, Al Shapere, Alan Watson, Tom Weiler, and Tokonatsu Yamamoto for 
fruitful collaborations. We also acknowledge Albert de Roeck and our colleagues of the AUGER, L3, 
and D0 collaborations for many illuminating discussions. 
We are indebted to several collaborators mentioned above as well as 
Felix Aharonian, Greg Archbold, Karsten Eggert, Max Klein, Kenji Shinozaki, 
Hristofor Vankov, Enrique Zas, and the HiRes and AGASA collaborations for allowing 
us to use various figures from their papers in this review. This work has been supported, in part, 
by the US National Science Foundation (under grant No. PHY--0140407) and IFLP--CONICET Argentina.
\end{acknowledgments}

\end{document}